\documentclass[sn-mathphys-num]{sn-jnl}

\usepackage{graphicx}%
\usepackage{multirow}%
\usepackage{amsmath,amssymb,amsfonts}%
\usepackage{amsthm}%
\usepackage{mathrsfs}%
\usepackage[title]{appendix}%
\usepackage{xcolor}%
\usepackage{textcomp}%
\usepackage{manyfoot}%
\usepackage{booktabs}%
\usepackage{algorithm}%
\usepackage{algorithmicx}%
\usepackage{algpseudocode}%
\usepackage{listings}%

\theoremstyle{thmstyleone}%
\theoremstyle{thmstyletwo}%

\theoremstyle{thmstylethree}%

\begin{document}

\title[Article Title]{Boundary conditions and electromagnetic effects on the phase transition of a zero spin bosonic system}


\author*[1]{\fnm{Emerson} \sur{Corrêa}}\email{ecorreae@ufpa.br}

\author[2]{\fnm{Michelli} \sur{Sarges}}\email{michellisarges@yahoo.com.br}
\equalcont{These authors contributed equally to this work.}


\affil*[1]{\orgdiv{Faculdade de Física}, \orgname{Universidade Federal do Pará}, 
\postcode{66075-110},
\city{Belém}, \state{Pará}, \country{Brazil}}


\affil[2]{\orgdiv{Independent researcher},  \city{Belém}, \state{Pará}, \country{Brazil}}



\abstract{In the following paper, we will study a charged scalar field under an electromagnetic external field taking into account the spatial confining of the system. We shall use the Coleman-Weinberg method in one-loop approximation to obtain the effective potential of the model in the proper time representation. Through generalized Matsubara formalism, we applied several kinds of boundary conditions on the frontier of the system. The regularization of the model is performed by a scheme independent of the external electromagnetic applied field. The model presents phase transition and we carry out its analysis by the free energy density functional of the bosonic system. The findings show magnetic catalysis, electric catalysis, and inverse electric catalysis phenomena, all of them depending on the thickness of the system.}




\maketitle


\section{Introduction}\label{sec1}


\hspace{0.53cm}It is speculated that microseconds after the Universe's evolution, there was a phase transition during the system cooling, from the quark-gluon plasma phase (QGP)  to the hadronic phase. It is further argued that transition would have created powerful magnetic fields in the primordial Universe and therefore such external fields would have interacted with QCD matter~\cite{Grasso,DiagramaQCDMag,PRD50,DiagramaQCDMagItalianos}. 

Among the wide variety of phenomena that magnetic fields provide when pervading a system, we can list the chiral magnetic effect stands out (the magnetic field separates the electrical charges of the medium along the direction of the external field)~\cite{Kharzeev,MagChiral,MagChiralRev}, magnetic catalysis (MC) (magnetic fields reinforce the effective mass of the fermion)~\cite{Gusynin,Debora,Italianos1,Indianos,Igor,MiranskyPR2015,Kojo,PRD2019} and inverse magnetic catalysis (IMC) (the magnetic field contributes to restoring chiral symmetry, through the decrease of chiral condensate while the magnetic background gets an augmenting)~\cite{IMC-PRD,IMC1,IMC2,IMC3,IMC4,IMC5,IMC6}. This IMC effect was discovered in the context of lattice QCD in 2012~\cite{IMC-JHEP}. After that, effective models for QCD were revisited to reobtain the IMC phenomenon with the coupling constant dependent on the magnetic field field~\cite{IMC7,IMC8,IMC9,IMC10,IMC11,IMC12,IMC13,EPJA-Emerson}.

In contrast, considering the existence of an external electric field, QCD on the lattice once again brought us another interesting result: an increase in the transition temperature of the system while the electric field gets higher values (electrical catalysis)~\cite{EndrodiPRD2024}. This result does not agree with what was found by effective QCD models, i.e., a decrease in the pseudocritical temperature
as a function of the electric field (inverse electrical catalysis) for a growing intensity of electric field~\cite{Tavares,Tavares1}.

Beyond the cosmological context, magnetic fields are produced in non-central collisions of relativistic
heavy ions. Indeed, since the 2000s the Large Hadron Collider (LHC) and later the Relativistic Heavy Ion Collider (RHIC) mimic in the terrestrial scenario similar conditions to those present immediately after the Big Bang, through the creation of the QGP. These very intense magnetic fields are on
average transverse to the nuclear collision plane. For example, in LHC, Pb+Pb collision with center of mass energy $\sqrt{s}=2.76\,\mathrm{TeV}$ can created magnetic fields of intensity $\sim 10^{16}\,\mathrm{Tesla}$. On the other hand, in RHIC, Au+Au collisions at $\sqrt{s}=0.2\,\mathrm{TeV}$ gives magnetic field on the order of $10^{15}\,\mathrm{Tesla}$ (notice that $eB \sim m^{2}_{\pi} \sim 10^{14}\,\mathrm{Tesla}$)~\cite{Kharzeev,Ions,IonsPesados}. 

Having discussed the QCD under extreme conditions, let us move on to the scalar field study.

The scalar field plays a fundamental role in Quantum Field Theory (QFT) due to the Higgs field. The celebrated Higgs mechanism breaks the electroweak symmetry and is responsible for the mass generation of particles in the Standard Model (SM). The particle associated with the fluctuations of this scalar field that permeates all space (the Higgs boson), was detected in 2012~\cite{PLBHiggs2012-1,PLBHiggs2012-2}. From a fundamental point of view, we can say that the zero-spin field is the cornerstone of the SM and is the simplest case of a QFT~\cite{Zee}. 

The spinless field immersed in an external magnetic field in conjunction with a thermal bath was considered in the QED context in Ref.~\cite{PRDElmfors1995}. The authors showed that when thermal magnetization changes from diamagnetic to paramagnetic behavior, the vacuum contribution is greater than the thermal one. Recently, in Ref.~\cite{RBEFEmerson2024}, the mass parameter of the charged scalar field under a constant external magnetic field was studied in an infinite volume through a connection between Ritus' and Schwinger's methods to obtain the propagator function of the system.  

Moreover, finite size effects in scalar models were considered in Refs.~\cite{PRDAdolfo2012,PRDAdolfo2013} through the application of two-particle irreducible formalism (2PI) in the study of symmetry breaking/restoration, without and with an external magnetic field. First- and second-order phase transitions for the spinless field were analyzed in Refs.~\cite{PLAEmerson2013} and \cite{EPJCErich2017}, respectively, using some Landau levels and not only the lowest Landau level $(\ell = 0)$ as is normally done.

Despite its importance, it is surprising that an analysis involving the charged scalar field embedded in an electromagnetic background in interaction with a thermal reservoir was done, albeit approximately, only newly~\cite{Zamora} (up to our knowledge). The authors of this Ref. studied the phase structure of a self-interaction scalar field under an orthogonal external electromagnetic field in two distinct sectors: weak and strong electric fields, no chemical potential, and without considering the important effects that come from the finite size of the system. 

In addition to theoretical interest, as discussed above, the inclusion of electromagnetic fields in physical systems reveals a variety of interesting phenomena, and we can expand the discussion and ask ourselves: what are the boundary conditions that systems with spatial restrictions obey? Strong electromagnetic fields in the range $[3-10~\mathrm{GeV}]$ are created even in intermediate energy heavy-ion collisions. Such intensities exceed the Schwinger critical field, $eE_{cr}\sim m_{e}^{2} \sim (0.5 \cdot10^{-3}\,\mathrm{GeV})^{2}$, which one generates non-linear effects on vaccum~\cite{PRC-Ion}. The created electromagnetic fields in heavy-ion collisions are perpendicular to the collision plane~\cite{PRX-Ion}, but we have no knowledge about what kind of boundary conditions the scalar fields are subject with the finite volume. This manuscript is in a speculative direction, by simulating various types of boundary conditions that a bosonic system constrained on one spatial direct and immersed in an electromagnetic field can assume. The system analised also can simulate the behaviour of matter under extreme conditions, as produced in heavy-ion collisions, taking into account the effects due the smallness volume and the differents boundary conditions allowed in the compacted dimension.

The main purpose of this manuscript is to calculate the Feynman propagator of a zero-spin field under an external electromagnetic background along $z$-direction considering several kinds of spatial boundary conditions. We will use this propagator to compute radiative corrections on the mass parameter of a self-interacting Bose field under the external fields $\textbf{E}$ and $\textbf{B}$. This approach encompasses multiple phenomena and is a natural extension of the research started in Ref.~\cite{PRDEmerson2023} in which the Dirac field was considered.

The setup used here can be found in \cite{NuclPhysBAdolfo2002,PhysRepAdolfo2014}. These references draw attention to quantum field theory defined on a toroidal topology and will be our starting point in this investigation. Initially, we have a bosonic self-interaction system defined in $D$ dimensions. By application of Matsubara generalized formalism we include the finite temperature as well as one spatial restriction, which is represented by topology $\mathrm{\Gamma}_{D}^{2}=S^{1}\times S^{1}\times\mathrm{R}^{D-2}$, being $S^{1}$ the representation of one compact dimension. Such a system retains $(D-2)$ dimensions without any spatial restriction, which corresponds to the bosonic heated system confined between two parallel planes orthogonal to $z$-direction with separation $L_{z}$, i.e., the system geometry assumes thin film form.

In this perspective, we shall consider the quantum effects due to the finite size system on the $z$-coordinate using periodic, quasiperiodic, and antiperiodic boundary conditions, in this compactified spatial coordinate. Some findings include system thickness-dependent magnetic catalysis, inverse electric catalysis, and electric catalysis. Furthermore, we find the phenomenon of inverse electric catalysis simultaneously with electric catalysis for different field regimes, considering periodic boundary conditions. This new result is strongly dependent on $L_{z}$ and was reported in a fermionic system in Ref.~\cite{Tavares}, and considering the weak electric field and strong electric field limits in the scalar propagator at unlimited space in Ref.~\cite{Zamoa-Eext}. Here this peculiar behavior is obtained without any approximations in the propagator.

This article is organized as follows: In Sec. II we define the model and compute the Feynman propagator of the scalar field under the parallel fields $\textbf{E}$ and $\textbf{B}$, in light of a hybrid method, namely: from Ritus' eigenfunctions~\cite{Ritus,ElizaldeRitus,Mexicanos,RBEFEmerson2015,CMEmerson2022} for a charged scalar field, we sum over all Landau levels for obtain a closed expression in terms of Schwinger's proper time~\cite{Schwinger,Lawrie}. In Sec. III we applied the external field independent regularization prescription established in~\cite{Avancini-MFIR} to get finite results on proper time integrals. The phase structure of the model is addressed in Sec. IV. Conclusions and some remarks are drawn in Sec.V. 
Throughout this paper, we will use $\hbar=c=k_{B}=1$ and an Euclidean space.

\section{The model}\label{sec2}

\hspace{0.53cm}The model deal with is defined by the Lagrangian density written in a free $D$-dimensional Euclidean space
\begin{eqnarray}
\mathcal{L} = \frac{1}{2}\left|\partial_{\mu} \Phi\right|^2 + \frac{m_{0}^{2}}{2}\,\left|\Phi\right|^2 + \frac{\lambda_{0}}{4!}\,\left|\Phi\right|^{4},
\end{eqnarray}
where $m_{0}$ and $\lambda_{0}$ are the bare mass and coupling constant system, respectively both at zero temperature.

Let us use the Coleman-Weinberg method to calculate the effective potential in the loop expansion without counterterms (see Ref.~\cite{Coleman-Weinberg})
\begin{eqnarray}
V = V_{0}+V_{1}+V_{2}+\cdots,
\end{eqnarray}
being $V_{0}$, $V_{1}$, $V_{2}$, the contribution at zero-loop (tree approximation), one-loop, two-loop, and so on. In this paper, we restrict our attention to one-loop order.

The classical potential is
\begin{eqnarray}
V_{0} = \frac{m_{0}^{2}}{2}\,\varphi_{c}^{2} + \frac{\lambda_{0}}{4!}\,\varphi_{c}^{4},
\end{eqnarray}
where the classical field is defined by $ \varphi_{c} \equiv\langle 0\,|\sqrt{\Phi^{*}\Phi} \,|0\rangle$. The one-loop contribution is given by the series
\begin{eqnarray}
V_{1} = \int \,\frac{d^Dp}{(2\pi)^D} \, \sum_{n=1}^{+\infty} \, \frac{1}{2n} \,\left[\frac{(\lambda_{0}/2) \varphi_{c}^{2}}{p^2+m_{0}^{2}}\right]^n,
\end{eqnarray}
where $n$ represents the number of vertices in the graphs. Within one-loop approximation, we shall consider just $n=1$, i.e., we shall compute the tadpole diagram. In Fig.~$1$, we observe the zero- and one-loop diagrams. 
\begin{figure}
\centering
\includegraphics[{width=5.0cm}]{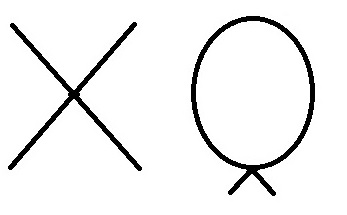}
\caption{~Vertex and one-loop contributions for the effective potential considered in this paper.}
\label{Fig1}
\end{figure}

The physical mass of the system is obtained by the condition~\cite{Dolan}
\begin{eqnarray*}
m^2 \,=\, \frac{d^{2}V}{d\varphi_{c}^2}\left.\frac{}{}\right|_{\varphi_{c} \,=\, 0},
\end{eqnarray*}
which results, in our approximation,
\begin{eqnarray*}
m^2 \,=\, m_{0}^{2}+\frac{\lambda_{0}}{2} \,G_{F},
\end{eqnarray*}
where $G_{F}$ is the propagation function
\begin{eqnarray}
G_{F} = \int \,\frac{d^Dp}{(2\pi)^D} \, \frac{1}{p^2+m_{0}^{2}}.
\label{free}
\end{eqnarray}
In the next subsection, we calculate the Feynman propagator of the scalar field under parallel external fields.
\subsection{Charged scalar field propagator under electromagnetic effects}

\hspace{0.53cm}The Klein-Gordon equation under an electromagnetic background written in $D$-dimensional Euclidean space is given by
\begin{eqnarray*}
\left(-D_{\mu}^{2} + m_{0}^{2}\right)\Phi\left(u\right) = 0,
\label{Eq.K-GcomCampo}
\end{eqnarray*}
where $u\equiv (\tau,x,y,z,\vec{r}\,)$, being $\vec{r}$ a $(D-4)$-dimensional vector and the covariant derivative is defined by $D_{\mu} ={\partial}_{\mu} - i e A_{\mu}$, being $e$ the charge of Bose field and $A$ the $D$-potential.

The Feynman propagator of the spin-zero field satisfies the equation
\begin{eqnarray*}
\left(D_{\mu}^{2} - m_{0}^{2}\right)G\left(u,u^{\prime},A\right)= - \,\delta^{4}(u-u^{\prime}).
\label{Prop K-GcomCampo}
\end{eqnarray*}

We choose the gauge $A \equiv (iz E,0,x B,0,\vec{0}\,)$, which means spinless field $\Phi\left(u\right)$ is subject to an electromagnetic field $\textbf{E}$ and $\textbf{B}$ along $z$-direction. This propagator has been calculated by Ritus' method in~\cite{CJPEmerson2022}, and its expression on the momentum space in the limit $u \rightarrow u^{\prime}$ is given by
\begin{eqnarray}
G_{F}\left(A\right) &=&  \int \,\frac{d^{D-4}p}{(2\pi)^{D-4}} \, \frac{i\omega_{E}}{2\pi}\frac{\omega_{B}}{2\pi}\,\sum_{\ell,\,\ell^{\,\prime} \,=\, 0}^{+\infty} \nonumber \\
&\times&\frac{1}{i\omega_{E}(2\ell+1)+\omega_{B}(2\ell^{\,\prime}+1) + \vec{p}^{\,2}+m_{0}^2}.
\label{magPhi}
\end{eqnarray}
The discrete variables $\ell$ and $\ell^{\,\prime}$ denote the Landau levels, $\vec{p}~$ is a $(D-4)$-dimensional vector, in addition we used the notation $\omega_{ E} \equiv |e|\mathrm{E} $ e $\omega_{B} \equiv |e|\mathrm{B}$. Below, we will obtain a closed expression for the scalar propagator.

The proper time representation and the identity, 
\begin{eqnarray*}
\mathcal{A}^{-1} = \int_{0}^{+\infty} \, dS \exp{\left(-S \, \mathcal{A}\right)},
\end{eqnarray*}
implies that
\begin{eqnarray}
G_{F}\left(A\right) &=& \frac{i\omega_{E}\,\omega_{B}}{4\pi^{2}} \, \int_{0}^{+\infty} {dS} \,  \int \,\frac{d^{D-4}p}{(2\pi)^{D-4}} \,\exp\left[-S\left(\vec{p}^{\,2}+m_{0}^{2}\right)\right] \nonumber \\
&\times& \,\sum_{\ell,\,\ell^{\,\prime} \,=\, 0}^{+\infty} \, \exp\left\{-S\left[\frac{}{}i\omega_{E}(2\ell+1)+\omega_{B}(2\ell^{\,\prime}+1)\right]\right\}. 
\label{prop}
\end{eqnarray}

In Eq.~(\ref{prop}), we have two geometric series on the Landau levels $\ell$ and $\ell^{\,\prime}$. After performing the summation, we get
\begin{eqnarray}
G_{F}\left(A\right) &=& \frac{i\omega_{E}\,\omega_{B}}{4\pi^{2}} \, \int_{0}^{+\infty} {dS} \,  \int \,\frac{d^{D-4}p}{(2\pi)^{D-4}} \,\exp\left[-S\left(\vec{p}^{\,2}+m_{0}^{2}\right)\right] \nonumber \\
&\times&\left[\frac{\exp\left(i\omega_{E}S\right)}{-1+\exp\left({2i\omega_{E}S}\right)}\right]\left[\frac{\exp\left(\omega_{B}S\right)}{-1+\exp\left({2\omega_{B}S}\right)}\right]. 
\label{cot30}
\end{eqnarray}
Rearranging the exponentials that carry the electromagnetic field in Eq.~(\ref{cot30}), we have
\begin{eqnarray}
G_{F}\left(A\right) &=& \frac{\omega_{E}\,\omega_{B}}{16\pi^{2}} \, \int_{0}^{+\infty} {dS} \,  \int \,\frac{d^{D-4}p}{(2\pi)^{D-4}} \,\exp\left[-S\left(\vec{p}^{\,2}+m_{0}^{2}\right)\right] \nonumber \\
&\times&\left[\frac{1}{\sin{(\omega_{E}S)}}\right]\left[\frac{1}{\sinh{(\omega_{B}S})}\right].
\label{cotJaca}
\end{eqnarray}

Due to the gauge we have chosen, the variables $\tau$ and $z$ are coupling, as well as $x$ and $y$. The dimensional reduction $D \, \rightarrow \,(D-2) \, \rightarrow \,(D-4)$ is a natural consequence of these coupled variables. In other words, the presence of external fields $\textbf{E}$ and $\textbf{B}$ in the propagator hidden four dimensions of $D$-dimensional system~\cite{CJPEmerson2022,Gusynin,MiranskyPR2015}. Notwithstanding, we can restore these four dimensions obscured by the external fields using a simple trick. The following integrals enable us to express the electromagnetic effects contained in Eq.~(\ref{cotJaca}) as 
\begin{eqnarray*}
 \int\,\int\, \frac{dp_{\tau}}{2\pi}\frac{dp_{z}}{2\pi} \exp\left[-\left(p_{\tau}^{2}+p_{z}^{2}\right)\frac{\tan{(\omega_{E}S)}}{\omega_{E}}\right] = \frac{\omega_{E}}{4\pi}\cot{(\omega_{E}S)},
\end{eqnarray*}
and
\begin{eqnarray*}
 \int\,\int\, \frac{dp_{x}}{2\pi}\frac{dp_{y}}{2\pi} \exp\left[-\left(p_{x}^{2}+p_{y}^{2}\right)\frac{\tanh{(\omega_{B}S)}}{\omega_{B}}\right] = \frac{\omega_{B}}{4\pi}\coth{(\omega_{B}S)}.
\end{eqnarray*}

Taking into account these integrals, the scalar propagator under electric and magnetic backgrounds becomes
\begin{eqnarray}
G_{F}\left(A\right) &=& \int_{0}^{+\infty} {dS} \,  \int \,\frac{d^{D-4}p}{(2\pi)^{D-4}} \,\exp\left[-S\left(\vec{p}^{\,2}+m_{0}^{2}\right)\right]\left[\frac{1}{\cos{(\omega_{E}S)}}\right]\left[\frac{1}{\cosh{(\omega_{B}S})}\right] \nonumber \\
&\times& \int \frac{dp_{\tau}}{2\pi}\frac{dp_{z}}{2\pi} \frac{dp_{x}}{2\pi}\frac{dp_{y}}{2\pi} \, 
\exp\left\{-S\left[(p_{\tau}^{2}+p_{z}^{2})\frac{\tan{(\omega_{E}S)}}{\omega_{E}S}+(p_{x}^{2}+p_{y}^{2})\frac{\tanh{(\omega_{B}S)}}{\omega_{B}S}\right]\right\}. \nonumber \\
\label{Jaca2}
\end{eqnarray}

Finally, focusing on the $(D-4)$ uncoupling momentum coordinates, it is easy to show that the spinless field propagator under the external electromagnetic field given by Eq.~(\ref{Jaca2}) becomes, after $(D-4)$ Gaussian integrals
\begin{eqnarray}
G_{F}\left(A\right) &=&\frac{1}{(4\pi)^{(D-4)/2}} \, \int_{0}^{+\infty} \frac{dS}{S^{(D-4)/2}} \,  \exp\left(-S \,m_{0}^{2}\right) \left[\frac{1}{\cos{(\omega_{E}S)}}\right]\left[\frac{1}{\cosh{(\omega_{B}S})}\right]\nonumber \\
&\times& \int \frac{dp_{\tau}}{2\pi}\frac{dp_{z}}{2\pi} \frac{dp_{x}}{2\pi}\frac{dp_{y}}{2\pi}  \,\exp\left\{-S\left[(p_{\tau}^{2}+p_{z}^{2})\frac{\tan{(\omega_{E}S)}}{\omega_{E}S}+(p_{x}^{2}+p_{y}^{2})\frac{\tanh{(\omega_{B}S)}}{\omega_{B}S}\right]\right\}. \nonumber \\
\label{EeB1}
\end{eqnarray}

Let us take the limit for pure magnetic field. In this case, we have
\begin{eqnarray*}
\lim_{\omega_{E}\,\rightarrow \,0}\left[\frac{1}{\cos{(\omega_{E}S)}}\right] \,\exp\left\{-S\left[(p_{\tau}^{2}+p_{z}^{2})\frac{\tan{(\omega_{E}S)}}{\omega_{E}S}\right]\right\} = \exp\left[-S\left(p_{\tau}^{2}+p_{z}^{2}\right)\right]
\end{eqnarray*}
Thus,

\begin{eqnarray}
G_{F}\left(B\right) &=&\frac{1}{(4\pi)^{(D-4)/2}} \, \int_{0}^{+\infty} \frac{dS}{S^{(D-4)/2}}  \, \exp\left(-S \,m_{0}^{2}\right)\left[\frac{1}{\cosh{(\omega_{B}S})}\right] \nonumber \\
&\times& \int \frac{dp_{\tau}}{2\pi}\frac{dp_{z}}{2\pi} \frac{dp_{x}}{2\pi}\frac{dp_{y}}{2\pi}  \,\exp\left\{-S\left[(p_{\tau}^{2}+p_{z}^{2})+(p_{x}^{2}+p_{y}^{2})\frac{\tanh{(\omega_{B}S)}}{\omega_{B}S}\right]\right\}. \nonumber \\
\label{justB1}
\end{eqnarray}
Eq.~(\ref{justB1}) is in agreement with that obtained in \cite{RBEFEmerson2024}, where the spinless field propagator was considered under an external magnetic field in coordinate space.

Analogously, the scalar propagator under just an electrical external field, reads
\begin{eqnarray}
G_{F}\left(E\right) &=&\frac{1}{(4\pi)^{(D-4)/2}} \, \int_{0}^{+\infty} \frac{dS}{S^{(D-4)/2}} \,  \exp\left(-S \,m_{0}^{2}\right) \left[\frac{1}{\cos{(\omega_{E}S)}}\right]\nonumber \\
&\times& \int \frac{dp_{\tau}}{2\pi}\frac{dp_{z}}{2\pi} \frac{dp_{x}}{2\pi}\frac{dp_{y}}{2\pi}  \,\exp\left\{-S\left[(p_{\tau}^{2}+p_{z}^{2})\frac{\tan{(\omega_{E}S)}}{\omega_{E}S}+(p_{x}^{2}+p_{y}^{2})\right]\right\}. \nonumber \\
\label{JustE1}
\end{eqnarray}
Both Eqs.~(\ref{justB1}) and (\ref{JustE1}) in the free space ($B \,\rightarrow \,0$ or $E \,\rightarrow \,0$, respectively), give correctly the Klein-Gordon propagator in $D$-dimensional Euclidean space, i.e.,
\begin{eqnarray}
G_{F} &=& \frac{1}{(4\pi)^{(D-4)/2}} \, \int_{0}^{+\infty} \frac{dS}{S^{(D-4)/2}} \,  \int \frac{d^4 p}{(2\pi)^4}  \,\exp\left[-S\left(p_{\tau}^{2}+p_{x}^{2}+p_{y}^{2}+p_{z}^{2}+m_{0}^{2}\right)\right]. \nonumber \\
\label{freeS}
\end{eqnarray}
The Eq.~(\ref{freeS}) corresponds to the Eq.~(\ref{free}) in the proper time representation. From now on, we will set $D=4$.
\subsection{System at zero temperature and spatially unlimited under $\textbf{E}$ and $\textbf{B}$}

\hspace{0.53cm}The Euclidean Lagrangian density of the charged scalar field embedded in an external electromagnetic field is given by~\cite{PRDAdolfo2013}
\begin{eqnarray*}
\mathcal{L} = \frac{1}{2}\left|D_{\mu} \Phi\right|^2 + \frac{m_{0}^{2}}{2}\,\left|\Phi\right|^2 + \frac{\lambda_{0}}{4!}\,\left|\Phi\right|^{4}.
\end{eqnarray*}

We are interested in the system under the electromagnetic background. In this case, we have
\begin{eqnarray*}
m^2 = m_{0}^{2} + \frac{\lambda_{0}}{2} \,G_{F}({E},{B}),
\end{eqnarray*}
with $G_{F}({E},{B})$ expressed in the Eq.~$(\ref{EeB1})$.

\subsection{System at finite temperature and with different types of boundary conditions under $\textbf{E}$ and $\textbf{B}$} 

\hspace{0.53cm}We assume the thermal equilibrium of the system with a thermal reservoir and we shall explore the effects that come from one compacted spatial direction 
simultaneously with the application of constant and external electric or magnetic fields. This will be performed by including the thermodynamic variables on the propagation function, considering different types of boundary conditions in $z$-coordinate. In this case, the physical mass reads
\begin{eqnarray}
m^{2} = m^{2}_{0}+\frac{\lambda_{0}}{2} \,G_{F}(T,\mu,L_{z},\mu_{z},{E},{B}).
\label{massacorrigida}
\end{eqnarray}

The thermal effects on the scalar field are included by Matsubara formalism. This approach uses the imaginary-time coordinate $\tau$ to take into account the finite temperature in the system. Then, through the so-called Kubo-Martin-Schwinger conditions (KMS) for bosons, we have to use periodic boundary conditions for the field
\begin{eqnarray*}
\Phi \, ({\tau},x,y,z,\vec{r}\,) &=&  \, \Phi \,({\tau}+\beta,x,y,z,\vec{r}\,).
\end{eqnarray*}
This allows establishing a discrete set of values on the temporal momenta
\begin{eqnarray}
p_{\tau} \rightarrow  {\omega}_{n_\tau} \equiv \frac{\pi}{\beta}
\left(2  n_{\tau}\right)&-&i\mu, \,\,\,\,\,\,\, n_{\tau} = 0,\pm 1 , \pm 2, \cdots. \nonumber\\
 \int \,\frac{dp_{\tau}}{2\pi}\,g\left(p_{\tau},p_{x},p_{y},p_{z},\vec{p}\,\right) & \rightarrow & \frac{1}{\beta }\,\sum_{ n_{\tau}=-\infty}^{\infty} \,g \left( {\omega}_{n_{\tau}},p_{x},p_{y},p_{z} ,\vec{p}\,\right),
 \label{Mat1}
\end{eqnarray}
where $T=\beta^{-1}$ is the temperature of the system and $\mu$ its chemical potential. 

The geometry of our system is a film of thickness $L_{z}$ (since $x$ and $y$ directions remain unlimited). In virtue of this, is necessary to obtain the analogous of Eq.~(\ref{Mat1}) in the spatial $z$-direction, i.e., let us evaluate the finite size effects on the same direction of $\mathbf{E}$ and $\mathbf{B}$. However, there is no reason to use periodic boundary conditions in spatial coordinates. In other words, we do not have to use the KMS conditions for spatial coordinates as well.

In this sense, finite size effects give us freedom in choosing boundary conditions in the spatial coordinates. This characteristic can be expressed as
\begin{eqnarray*}
\Phi \, (\tau,x,y,z,\vec{r}\,) &=& e^{-i\,\pi\,\alpha_{z}} \, \Phi (\tau,x,y,z+L_{z},\vec{r}\,).
\end{eqnarray*}
The parameter $\alpha_{z}$ represents periodic boundary conditions (PBC), antiperiodic boundary conditions (ABC), and quasiperiodic boundary conditions (also called twisted boundary conditions) (QBC). The value $\alpha_{z} \,=\, 0$ denote the PBC case. The ABC case is represented by $\alpha_{z} \,=\, 1$ and the range $0 \, < \alpha_{z} < \, 1$\, establishes QBC on the system~\cite{twisted,Erich,Inagaki2022,NuclPhysAEmerson2023}.

In terms of parameter $\alpha_{z}$, the Bose field assume the discrete frequencies ${\omega}_{n_z}$ such that
\begin{eqnarray}
p_{z} \rightarrow {\omega}_{n_z} \,\,\equiv\,\, \frac{\pi}{L_z}
\left(2 \, n_{z}\right)&-&i\,\mu_{z} \,\,\,\,;\,\,\,\,  \mu_{z} = i \frac{\pi}{L_z}\alpha_{z} \nonumber      \\
 \int\,\frac{dp_{z}}{2\pi}\,g(p_{\tau},p_{x},p_{y},p_{z},\vec{p}\,) & \rightarrow & \frac{1}{ L_{z}}\,\sum_{ n_{z}=-\infty}^{\infty} g \left(p_{\tau},p_{x},p_{y} ,{\omega}_{n_{z}},\vec{p}\,\right), 
 \label{Mat2}
\end{eqnarray}
where $n_{z} \in \mathrm{Z}$ and $L_z$ corresponds to the film thickness. Eqs.~(\ref{Mat1}) and (\ref{Mat2}) define the generalized Matsubara formalism and the compactified coordinates belong to the range $\tau \in \left[0,\beta\right]$ and $z \in \left[0,L_{z}\right]$~\cite{LivroAdolfo2009,EPJCEmerson2017,IJMPBEmerson2018,PhysicaAEmerson2021,PRDEmerson2022}.

From an analytical point of view, the application of generalized Matsubara formalism on the bosonic system at proper time representation can be done by 
%
%
\begin{eqnarray}
\int \frac{dp_{\tau}}{2\pi}\frac{dp_{z}}{2\pi} \, \exp\left[-\left(p_{\tau}^{2}+p_{z}^{2}\right)f(S)\right]&=& \frac{1}{4\pi \,f(S)} \, \theta_{3}\left\{\frac{i\mu \beta }{2} \,;\, \exp\left[-\frac{\beta^{2}}{4f(S)}\right]\right\} \nonumber \\
&\times& \theta_{3}\left\{\frac{i\mu_{z} L_{z} }{2} \,;\, \exp\left[-\frac{L_{z}^{2}}{4f(S)}\right]\right\},
\label{taca}
\end{eqnarray}
where $f(S)$ is some function dependent of proper time $S$ and the Jacobi theta function $\theta_{3}$ is defined by~\cite{Whittaker,Optics}
\begin{eqnarray}
\theta_{3}\,(u\,;q) &=&  \sum_{n=-\infty}^{+\infty}\, q^{n^2} \exp\left(2inu\right).
\end{eqnarray}

Using Eq.~(\ref{EeB1}) and the Matsubara generalized prescription written in Eq.~(\ref{Mat1}) and (\ref{Mat2}), we have
\begin{eqnarray}
G_{F}\left(T,\mu,L_{z},\mu_{z},{E},{B}\right) &=&\frac{1}{16\pi^{2}} \, \int_{0}^{+\infty} dS\,  \exp\left(-S \,m_{0}^{2}\right) \left[\frac{\omega_{E}}{\sin{(\omega_{E}S)}}\right]\left[\frac{\omega_{B}}{\sinh{(\omega_{B}S})}\right] \nonumber \\ 
&\times& \theta_{3}\left\{\frac{i\mu \beta} {2} \,;\, \exp\left[-\frac{  \beta^{2}\,\omega_{E}}{ 4\,\tan{\left(\omega_{E}S\right)}}\right]\right\}  \nonumber \\
 &\times&\theta_{3}\left\{\frac{i\mu_{z} L_{z}} {2} \,;\, \exp\left[-\frac{  L_{z}^{2}\,\omega_{E}}{ 4\,\tan{\left(\omega_{E}S\right)}}\right]\right\},
 \label{lastequation}
\end{eqnarray}
where we have calculated two Gaussian integrals and used Eq.~(\ref{taca}) for $f(S) = \tan{(\omega_{E}S)}/\omega_{E}$. Now we just plug Eq.~(\ref{lastequation}) into Eq.~(\ref{massacorrigida}).

In the limit of a pure electric or magnetic background, respectively, we obtain
\begin{eqnarray}
G_{F}\left(T,\mu,L_{z},\mu_{z},{E}\right) &=&\frac{1}{16\pi^{2}} \, \int_{0}^{+\infty} dS\,  \exp\left(-S \,m_{0}^{2}\right)  \left[\frac{\omega_{E}}{\sin{(\omega_{E}S)}}\right] \,\left(\frac{1}{S}\right)\nonumber \\ 
&\times& \theta_{3}\left\{\frac{i\mu \beta} {2} \,;\, \exp\left[-\frac{  \beta^{2}\,\omega_{E}}{ 4\,\tan{\left(\omega_{E}S\right)}}\right]\right\}  \nonumber \\
 &\times&\theta_{3}\left\{\frac{i\mu_{z} L_{z}} {2} \,;\, \exp\left[-\frac{  L_{z}^{2}\,\omega_{E}}{ 4\,\tan{\left(\omega_{E}S\right)}}\right]\right\},
 \label{lastequationE}
\end{eqnarray}
and
\begin{eqnarray}
G_{F}\left(T,\mu,L_{z},\mu_{z},{B}\right) &=&\frac{1}{16\pi^{2}} \, \int_{0}^{+\infty} dS\,  \exp\left(-S \,m_{0}^{2}\right)  \left[\frac{\omega_{B}}{\sinh{(\omega_{B}S)}}\right] \,\left(\frac{1}{S}\right)\nonumber \\ 
&\times& \theta_{3}\left[\frac{i\mu \beta} {2} \,;\, \exp\left(-\frac{  \beta^{2}}{ 4\,S}\right)\right]  \,\theta_{3}\left[\frac{i\mu_{z} L_{z}} {2} \,;\, \exp\left(-\frac{  L_{z}^{2}}{ 4\,S}\right)\right].
 \label{lastequationB}
\end{eqnarray}

The thermodynamic limit in these expressions can be computed by taking the limit $L_{z}\,\rightarrow \,\infty$ ($V \,\rightarrow\,\infty$, since $L_{x}$ and $L_{y}$ are unbounded) at fixed $\mu,T$, on the functions $\theta_{3}$ written in Eqs.~(\ref{lastequationE}) and (\ref{lastequationB}) (see e.g. Ref.~\cite{LimTermodinamico}):
\begin{eqnarray*}
\lim_{L_{z} \,\rightarrow \,\infty}\theta_{3}\left\{\frac{i\mu_{z} L_{z}} {2} \,;\, \exp\left[-\frac{  L_{z}^{2}}{ 4\,f(S)}\right]\right\} &=& \lim_{L_{z} \,\rightarrow \,\infty} \, \left\{1+2\,\sum_{n_{z}=1}^{+\infty}\,\left[\exp{\left(-\dfrac{n_{z}^{2}L_{z}^{2}}{4f(S)}\right)}\right] \right. \nonumber \\
&\times&\left.\frac{}{}\cos{(n_{z}\alpha_{z}\pi})\right\}, \nonumber \\
&\approx&1.
\end{eqnarray*}

In the next section, we shall perform the regularization of the model.
\section{Regularization of the model}\label{sec3}

\hspace{0.53cm}Since the model considered in this note presents divergencies in the ultraviolet sector of proper time integrals, we have to consider one of many methods to make the quantities finite. Here, we treat the proper time integrals within the MFIR (magnetic field independent regularization) procedure present on Ref.~\cite{Avancini-MFIR} adapted to the scalar field. Thanks to 
\begin{eqnarray*}
	\lim_{S \,\rightarrow \, 0^{+}} \,\theta_{3}\left\{v(S);\exp[-h(S)]\right\} = 1, \,\,\,\hbox{for}\,\,\,\,h(S\rightarrow 0^{+}) \approx +\infty,
\end{eqnarray*}
being $v(S)$ a well behaved function, and due to behaviour of $\left[\omega_{E}/S\sin{(\omega_{E}S)}\right]$ and $\left[\omega_{B}/S\sinh{(\omega_{B}S)}\right]$ for small arguments, Eqs.~(\ref{lastequationE}) and (\ref{lastequationB}) exhibits the singular term $1/S^2$ in the ultraviolet proper time sector. 

\subsection{Regularization under an electric background}

\hspace{0.53cm}Let us handle with Eq.~(\ref{lastequationE}) seeking to remove the singularity. To contour this issue, we subtract and add some terms in the propagator. For a pure electric external field, we have
\begin{eqnarray}
G_{F}\left(T,\mu,L_{z},\mu_{z},{E}\right) &=&\frac{1}{16\pi^{2}} \, \int_{0}^{+\infty} \,\frac{dS}{S^2}\,  \exp\left(-S \,m_{0}^{2}\right)  \left[\frac{(\omega_{E}S)}{\sin{(\omega_{E}S)}}-1-\frac{(\omega_{E}S)^{2}}{6}\right] \nonumber \\ 
&\times& \theta_{3}\left\{\frac{i\mu \beta} {2} \,;\, \exp\left[-\frac{  \beta^{2}\,\omega_{E}}{ 4\,\tan{\left(\omega_{E}S\right)}}\right]\right\} \,\theta_{3}\left\{\frac{i\mu_{z} L_{z}} {2} \,;\, \exp\left[-\frac{  L_{z}^{2}\,\omega_{E}}{ 4\,\tan{\left(\omega_{E}S\right)}}\right]\right\} \nonumber \\
 &+& \frac{\omega_{E}^{2}}{96\pi^{2}} \, \int_{0}^{+\infty} dS\,  \exp\left(-S \,m_{0}^{2}\right) \nonumber \\  &\times&\theta_{3}\left\{\frac{i\mu \beta} {2} \,;\, \exp\left[-\frac{  \beta^{2}\,\omega_{E}}{ 4\,\tan{\left(\omega_{E}S\right)}}\right]\right\}\,\theta_{3}\left\{\frac{i\mu_{z} L_{z}} {2} \,;\, \exp\left[-\frac{  L_{z}^{2}\,\omega_{E}}{ 4\,\tan{\left(\omega_{E}S\right)}}\right]\right\} \nonumber \\
 &+&\frac{1}{16\pi^{2}} \, \int_{0}^{+\infty} \,\frac{dS}{S^2}\,  \exp\left(-S \,m_{0}^{2}\right)  \nonumber \\
 &\times&\,\left\{\theta_{3}\left\{\frac{i\mu \beta} {2} \,;\, \exp\left[-\frac{  \beta^{2}\,\omega_{E}}{ 4\,\tan{\left(\omega_{E}S\right)}}\right]\right\} \, \theta_{3}\left\{\frac{i\mu_{z} L_{z}} {2} \,;\, \exp\left[-\frac{  L_{z}^{2}\,\omega_{E}}{ 4\,\tan{\left(\omega_{E}S\right)}}\right]\right\}-1\right\}\nonumber \\
 &+&\frac{1}{16\pi^{2}} \, \int_{0}^{+\infty} \,\frac{dS}{S^2}\,  \exp\left(-S \,m_{0}^{2}\right).
 \label{lastequationEN}
\end{eqnarray}
Notice the last term on the right side written in Eq.~(\ref{lastequationEN}). It does not depend on any external parameter like temperature or electric field intensity. This term is the vacuum contribution to the model. 
\subsection{Regularization under a magnetic background}

\hspace{0.53cm}In the same way, for a pure magnetic external field, we obtain after adding and subtracting appropriated quantities inside Eq.~(\ref{lastequationB}) the expression 
\begin{eqnarray}
G_{F}\left(T,\mu,L_{z},\mu_{z},{B}\right) &=&\frac{1}{16\pi^{2}} \, \int_{0}^{+\infty} \,\frac{dS}{S^2}\,  \exp\left(-S \,m_{0}^{2}\right)  \left[\frac{(\omega_{B}S)}{\sinh{(\omega_{B}S)}}-1+\frac{(\omega_{B}S)^{2}}{6}\right] \nonumber \\ 
&\times& \theta_{3}\left[\frac{i\mu \beta} {2} \,;\, \exp\left(-\frac{\beta^{2}}{4\,S}\right)\right]\,\theta_{3}\left[\frac{i\mu_{z} L_{z}}{2} \,;\, \exp\left(-\frac{L_{z}^{2}}{4\,S}\right)\right] \nonumber \\
 &-& \frac{\omega_{B}^{2}}{96\pi^{2}} \, \int_{0}^{+\infty} dS\,  \exp\left(-S \,m_{0}^{2}\right) \nonumber \\  &\times&\theta_{3}\left[\frac{i\mu \beta} {2} \,;\, \exp\left(-\frac{\beta^{2}}{4\,S}\right)\right]\,\theta_{3}\left[\frac{i\mu_{z} L_{z}}{2} \,;\, \exp\left(-\frac{L_{z}^{2}}{4\,S}\right)\right] \nonumber \\
 &+&\frac{1}{16\pi^{2}} \, \int_{0}^{+\infty} \,\frac{dS}{S^2}\,  \exp\left(-S \,m_{0}^{2}\right)  \nonumber \\
 &\times&\,\left\{\theta_{3}\left[\frac{i\mu \beta} {2} \,;\, \exp\left(-\frac{\beta^{2}}{4\,S}\right)\right]\,\theta_{3}\left[\frac{i\mu_{z} L_{z}}{2} \,;\, \exp\left(-\frac{L_{z}^{2}}{4\,S}\right)\right]-1\right\}\nonumber \\
 &+&\frac{1}{16\pi^{2}} \, \int_{0}^{+\infty} \,\frac{dS}{S^2}\,  \exp\left(-S \,m_{0}^{2}\right).
 \label{lastequationBN}
\end{eqnarray}

All integrals on Eqs.~(\ref{lastequationEN}) and (\ref{lastequationBN}) are finite, except the last ones. Indeed, as expressed above these last terms do not depend on external parameters $(T,\mu,L_{z},\mu_{z},E)$ or $(T,\mu,L_{z},\mu_{z},B)$ and are the vacuum contributions of electric and magnetic scenarios, respectively. As in all prescriptions of regularization, they present divergence as well. 

Since we parameterize the model by $m_{0}$, let us define the ultraviolet cutoff in terms of this parameter. Thus,
\begin{eqnarray}
\int_{0}^{+\infty} \,\frac{dS}{S^2}\,  \exp\left(-S \,m_{0}^{2}\right)
 \rightarrow \int_{1/m_{0}^2}^{+\infty}\,\frac{dS}{S^2}\,  \exp\left(-S \,m_{0}^{2}\right).
\label{cutoff}	
\end{eqnarray}

In the next section, we will investigate the phase structure of the scalar model in terms of the regularized expressions given in Eqs.~(\ref{lastequationEN}), (\ref{lastequationBN}) and (\ref{cutoff}).
%
\section{Phase structure}\label{sec4}

\hspace{0.53cm}In this section, we solved numerically Eq.~(\ref{massacorrigida}) taking into account the regularized propagators with ultraviolet cutoff and we found some critical points necessary for phase transition to occur. The system begins from the broken phase [ordered phase (in an analogy with a ferromagnetic material, at $T<T_{\mathrm{Curie}}$)] and we will investigate the behavior of the system when the external parameters are changed, i.e., let us study if and how the self-interacting bosonic system undergoes phase transition passing to the restored phase [disordered phase (again in analogy ferromagnetic material case, at $T>T_{\mathrm{Curie}}$)]. Also, we fix as input configuration of the model, the effective coupling constant and the mass parameter at zero temperature as $\lambda_{eff} \equiv (\lambda_{0}/32 \pi^{2}) \approx 0.9$ and $m_{0}\equiv 0.500~\mathrm{GeV}$ Ref.~\cite{MesonSigma}.

\subsection{Mass parameter of the system}

\hspace{0.53cm}The results for the corrected mass parameter are shown in Fig.~$2$ under an external electric field at zero $\mathrm{B}$ for film thickness drop values. At the top, we have the PBC case and we notice a continuous decrease in mass until the value zero at the critical temperature $T_{c}$. At this point in the topological space, the phase transition occurs. The system passes from the broken phase to the restored phase. The effect caused by the finite size is to decrease the critical temperature as $L_{z}$ decreases. For temperatures higher than $T_{c}$, the system mass parameter starts to increase and the system is placed in the restored phase. Still in Fig.~$2$ (middle panel), we observe the effects of QBC on the system. In this case, the system showed little dependence while the thickness $L_{z}$ decreased. However, now the quasiperiodic boundary conditions induce slightly higher critical temperatures, considering the same values for finite size $L_{z}$. The ABC case is discussed in the lower part of Fig.~$2$. The finite size effects this time induce higher critical temperatures on the system. Thus, periodic and antiperiodic boundary conditions exhibit opposite behaviors.

On the left side of Fig.~$2$, we have chemical equilibrium $(\mu =0)$, and on the right side, we consider finite chemical potential $(\mu=0.100~\mathrm{GeV})$. In PBC, QBC, and ABC cases, finite chemical potential makes little augmentation to the $T_{c}$ values concerning the zero chemical potential. 
\begin{figure}
\centering
\includegraphics[{width=6.49cm}]{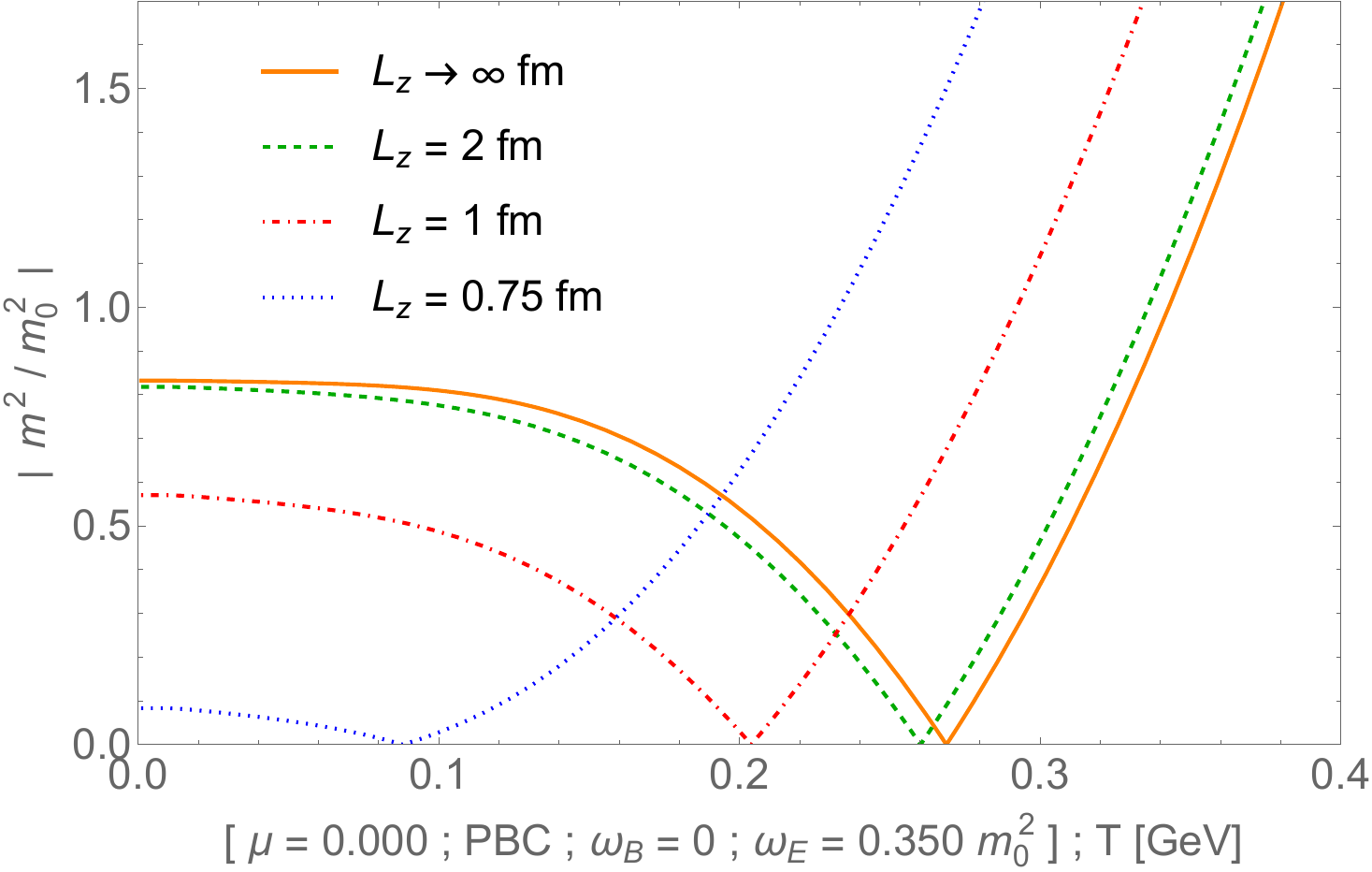}
\includegraphics[{width=6.49cm}]{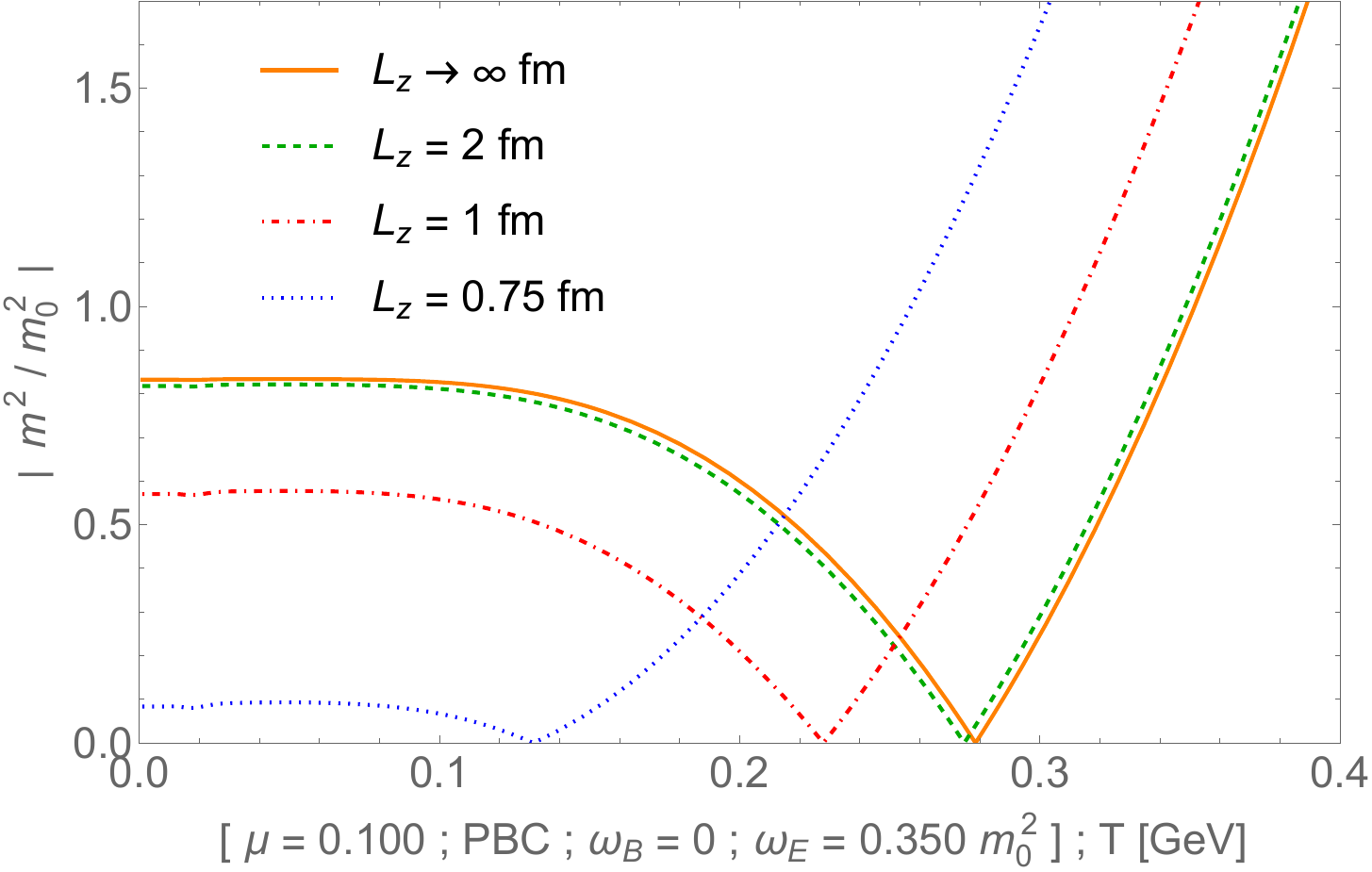} \\
\includegraphics[{width=6.49cm}]{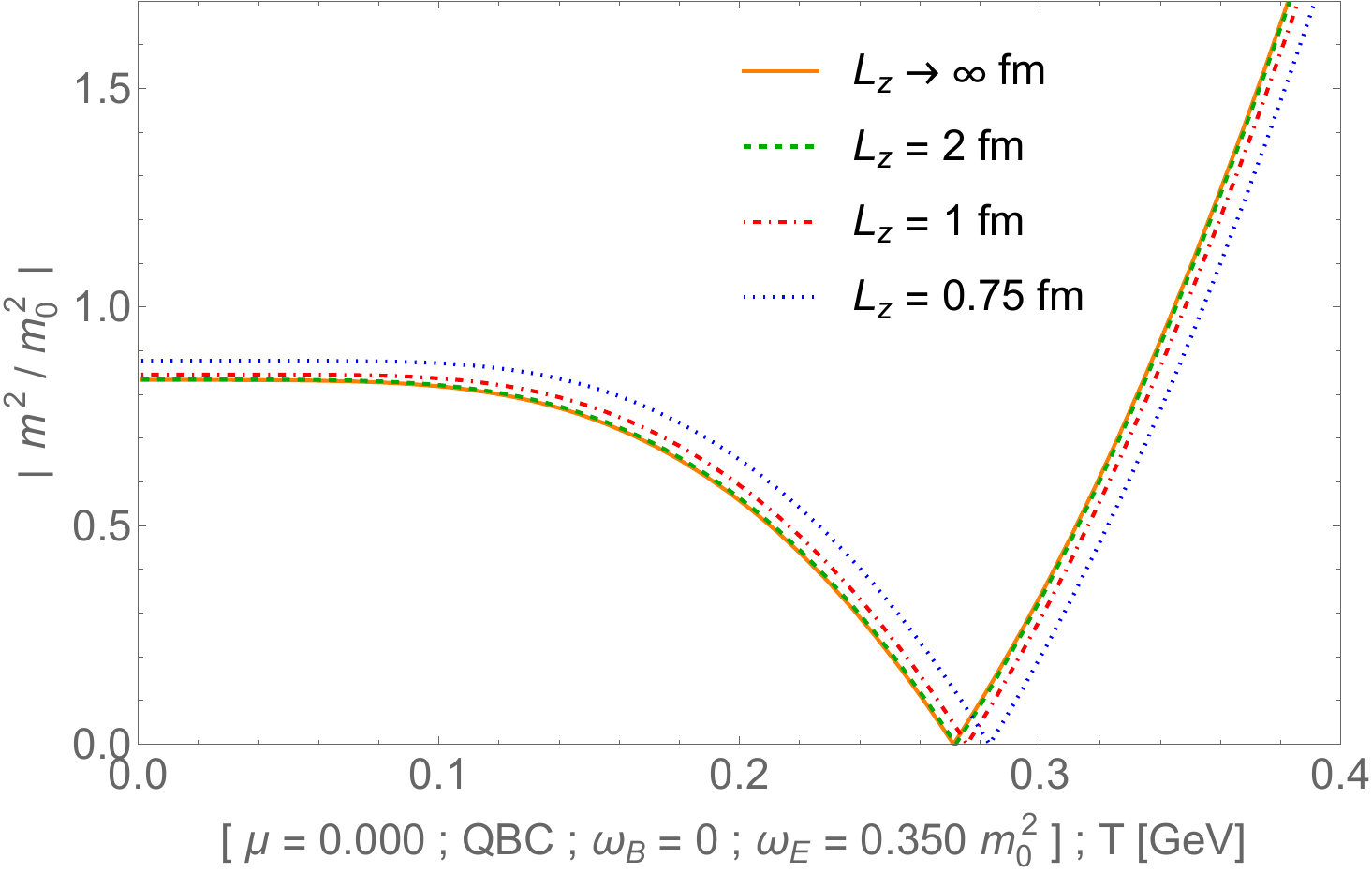} 
\includegraphics[{width=6.49cm}]{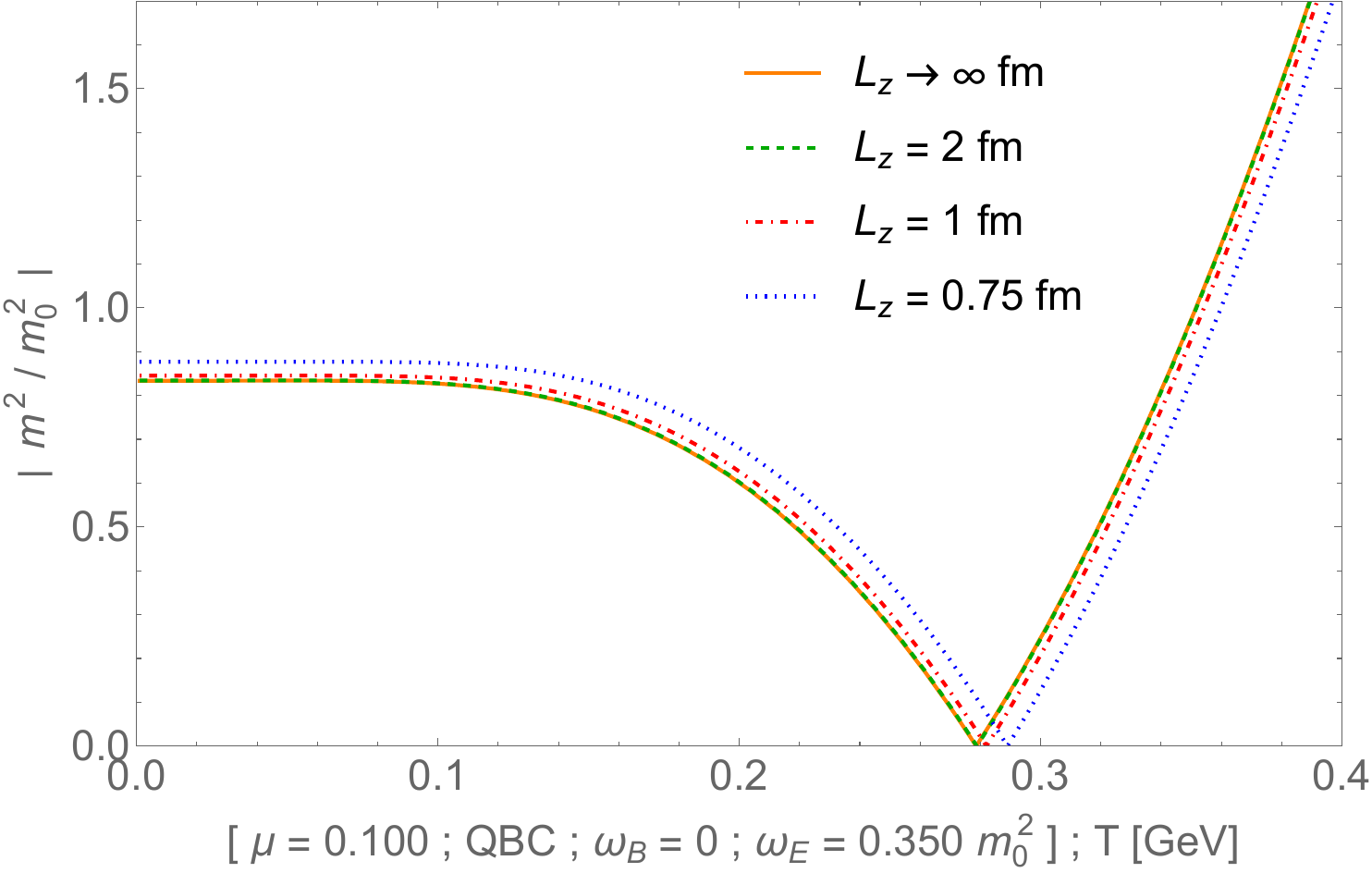} \\
\includegraphics[{width=6.49cm}]{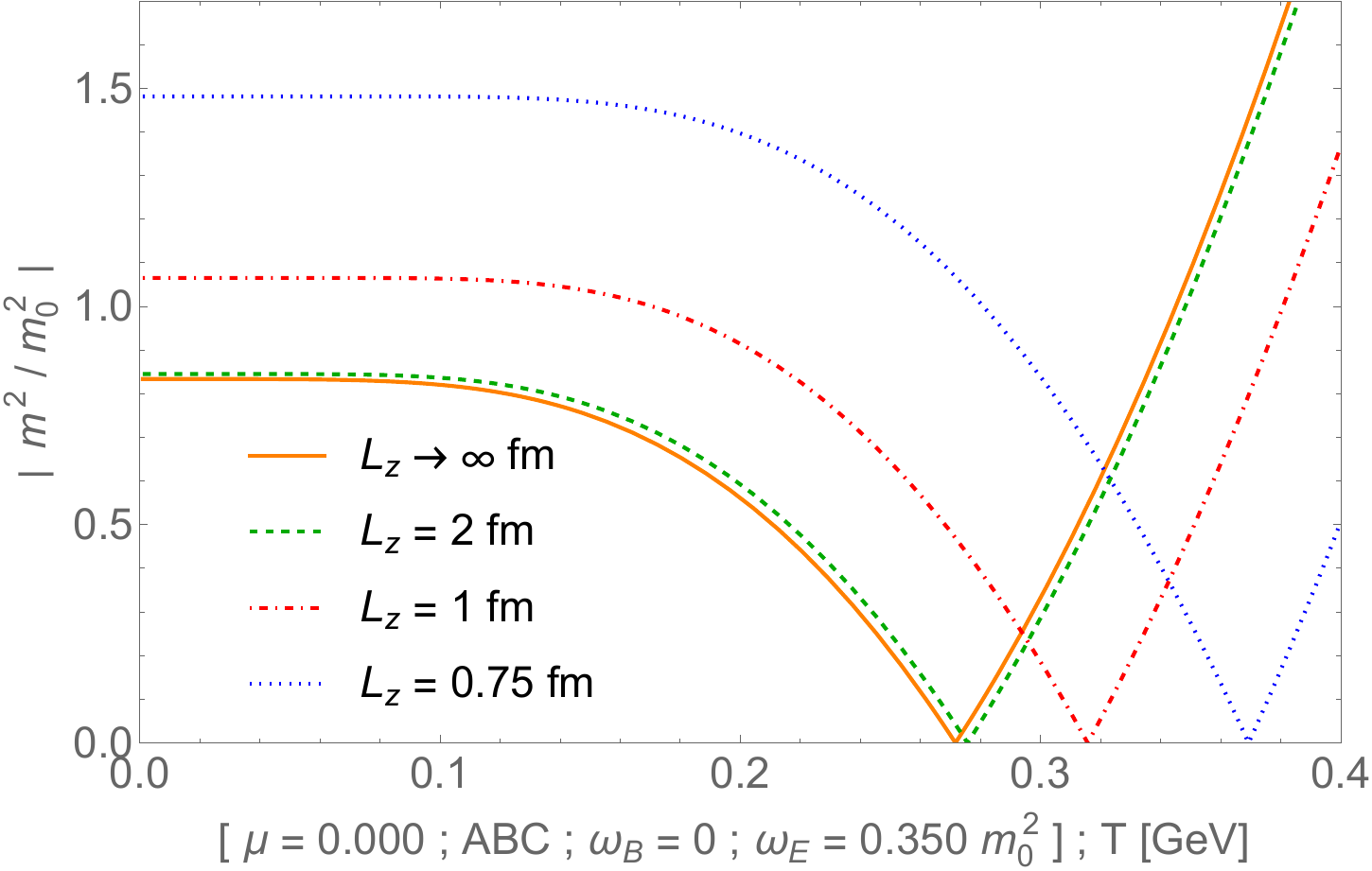}
\includegraphics[{width=6.49cm}]{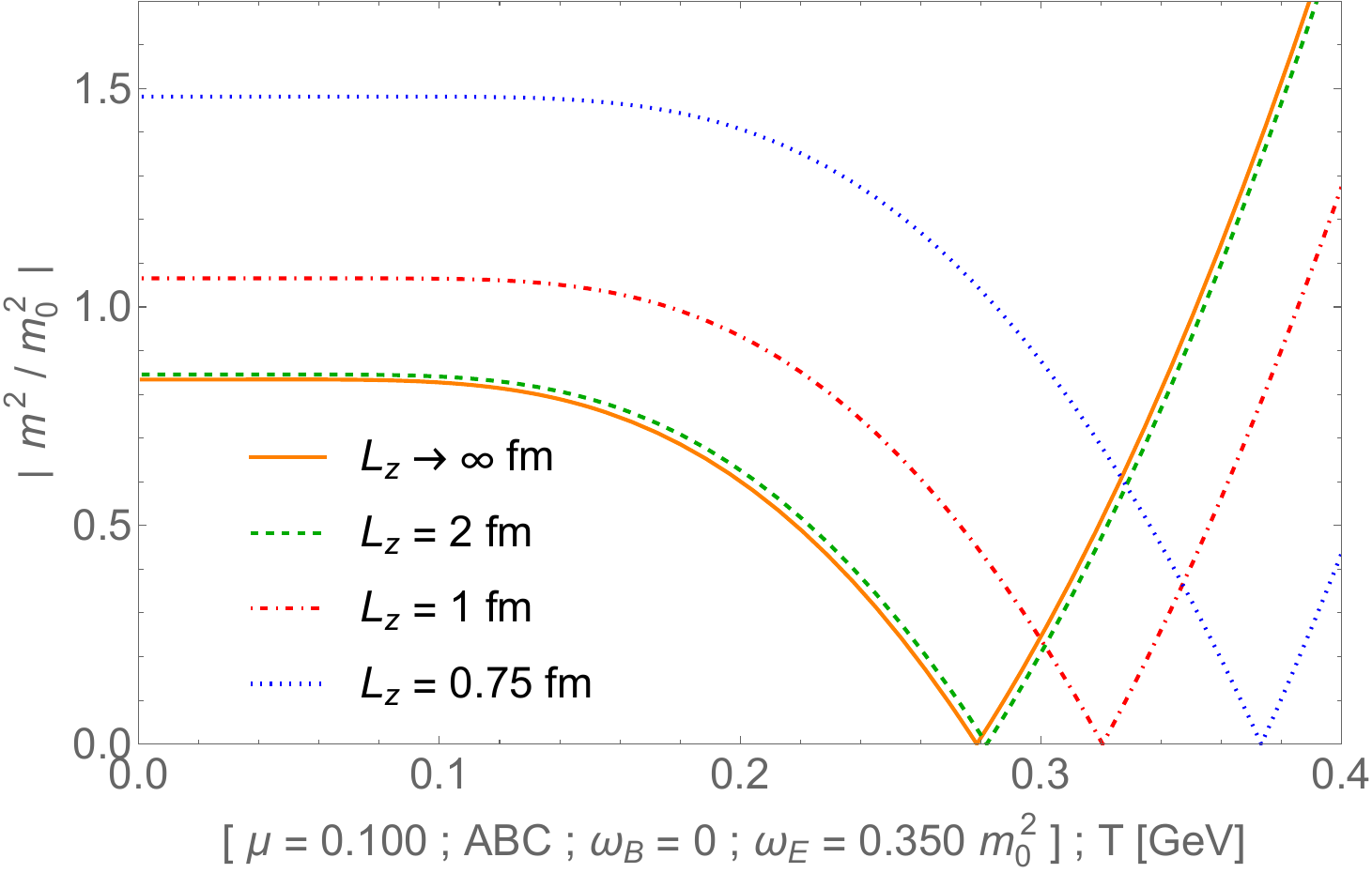} \\
\caption{~Corrected mass as a function of temperature for several values of thickness of the film (null chemical potential at the left and finite chemical potential on the right).  We have PBC (top panel), QBC (middle panel), and ABC (bottom panel) on $z$ coordinate.}
\end{figure}

To better describe the system under the electric backgrounds, in Fig.~$3$ we plot the corrected mass for different strengths of $\mathbf{E}$ upon the same separation between the planes on the $z$-direction. In the PBC case (top panels), the picture displays one not trivially evolution on the system critical temperature under increasing in electric field intensity.  For weak values of $\mathbf{E}$ the critical temperature gets lower values under an increase in the external electric field. Contrarily, for strong values of $\mathbf{E}$, we have higher critical temperatures for higher strengths electric field. Therefore, for PBC choice, the effect of the electric field on the system is to disfavor the phase transition for lower electric fields and cooperate with it for higher ones.  In other words, we found inverse electric catalysis (IEC) for weak values of $\mathbf{E}$ and electric catalysis (EC) for strong values of $\mathbf{E}$. These phenomena due to the application of periodic boundary conditions are more pronounced for small lengths. 

Under QBC and ABC, the system shows just the IEC phenomenon (middle and bottom panels of Fig.~$3$, respectively). For small lengths (numerically below $0.95$~$\mathrm{fm}$) and antiperiodic boundary conditions on the compacted spatial coordinate, the system seems not to suffer IEC. 
\begin{figure}
\centering
\includegraphics[{width=6.49cm}]{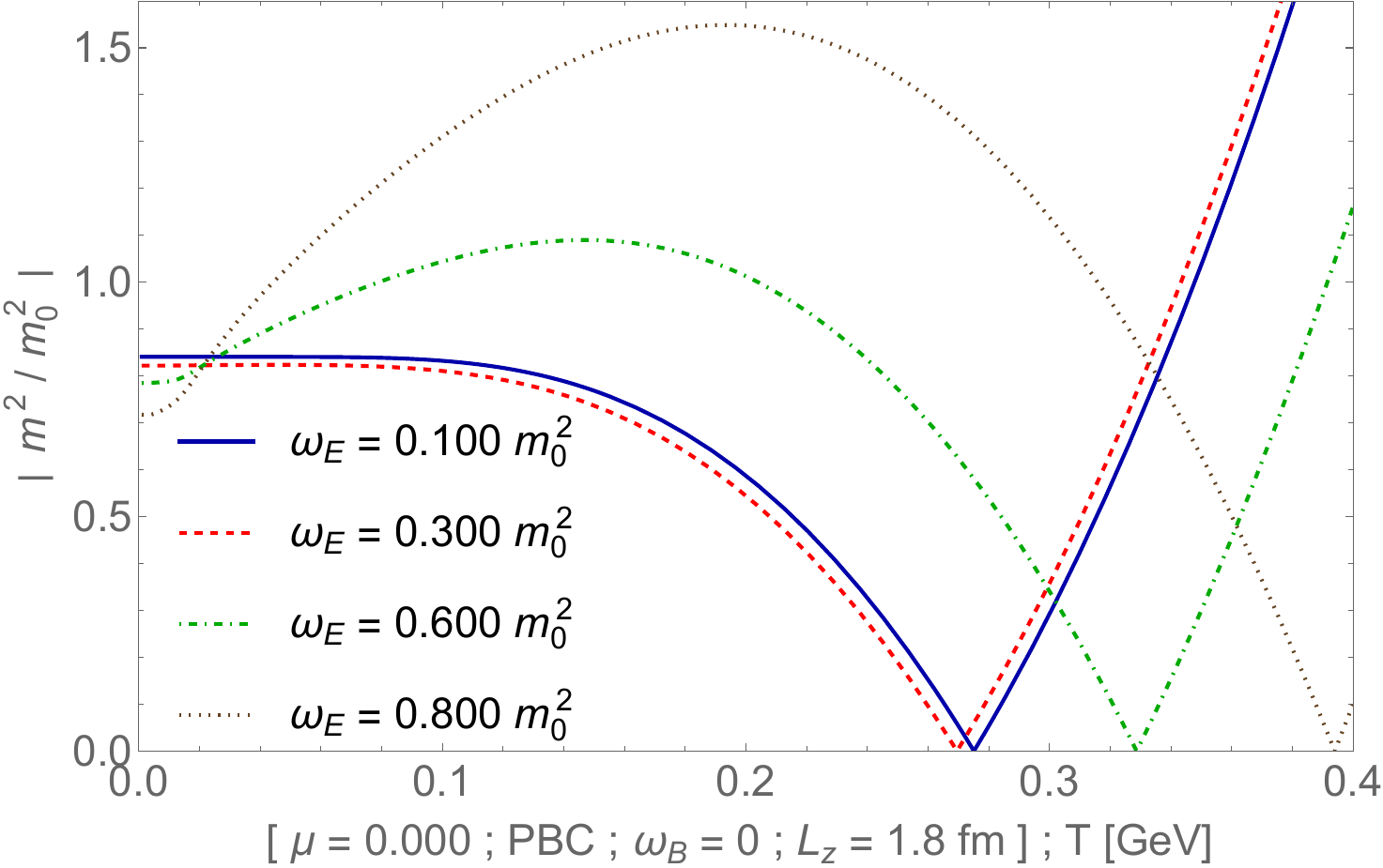}
\includegraphics[{width=6.49cm}]{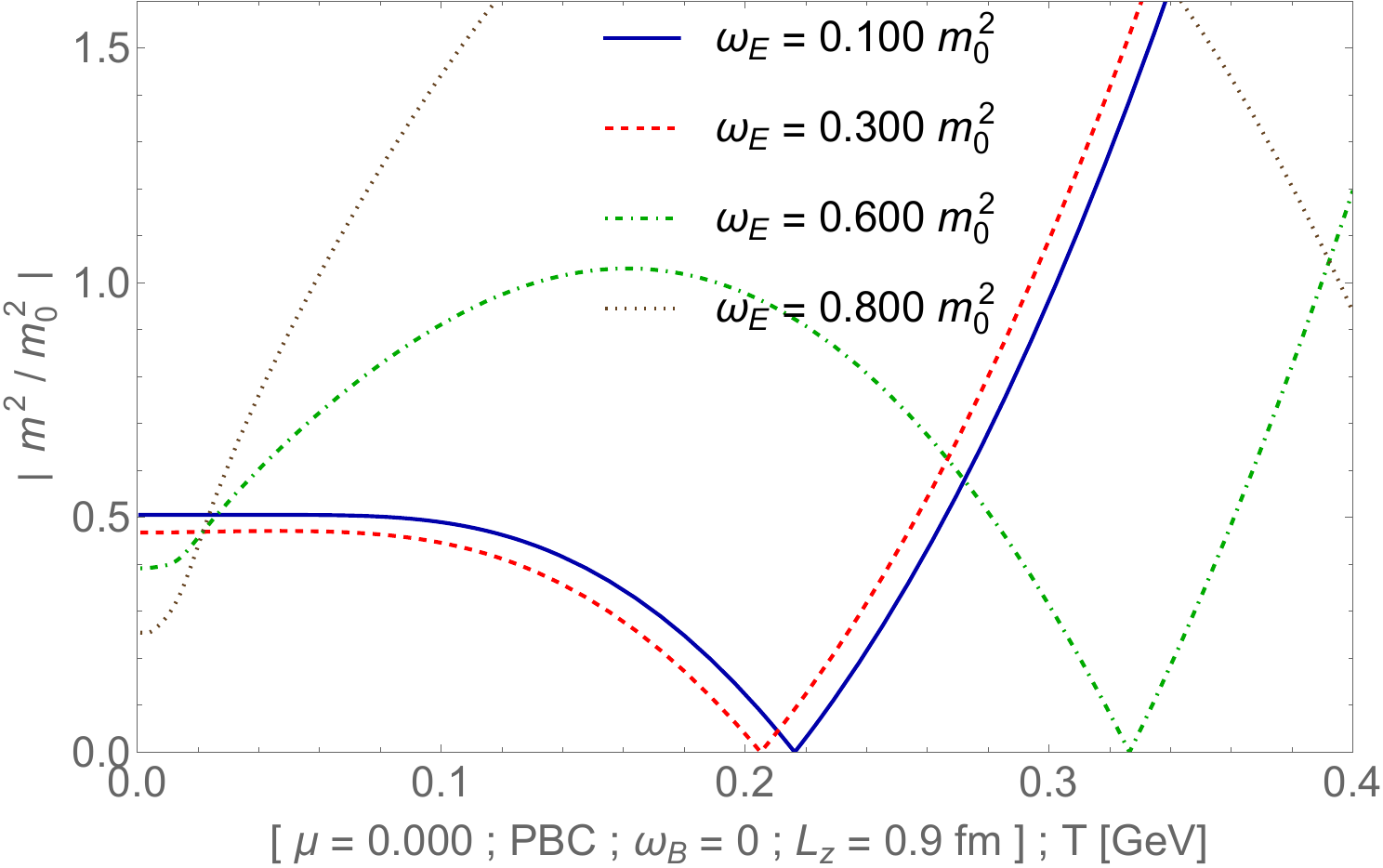} \\
\includegraphics[{width=6.49cm}]{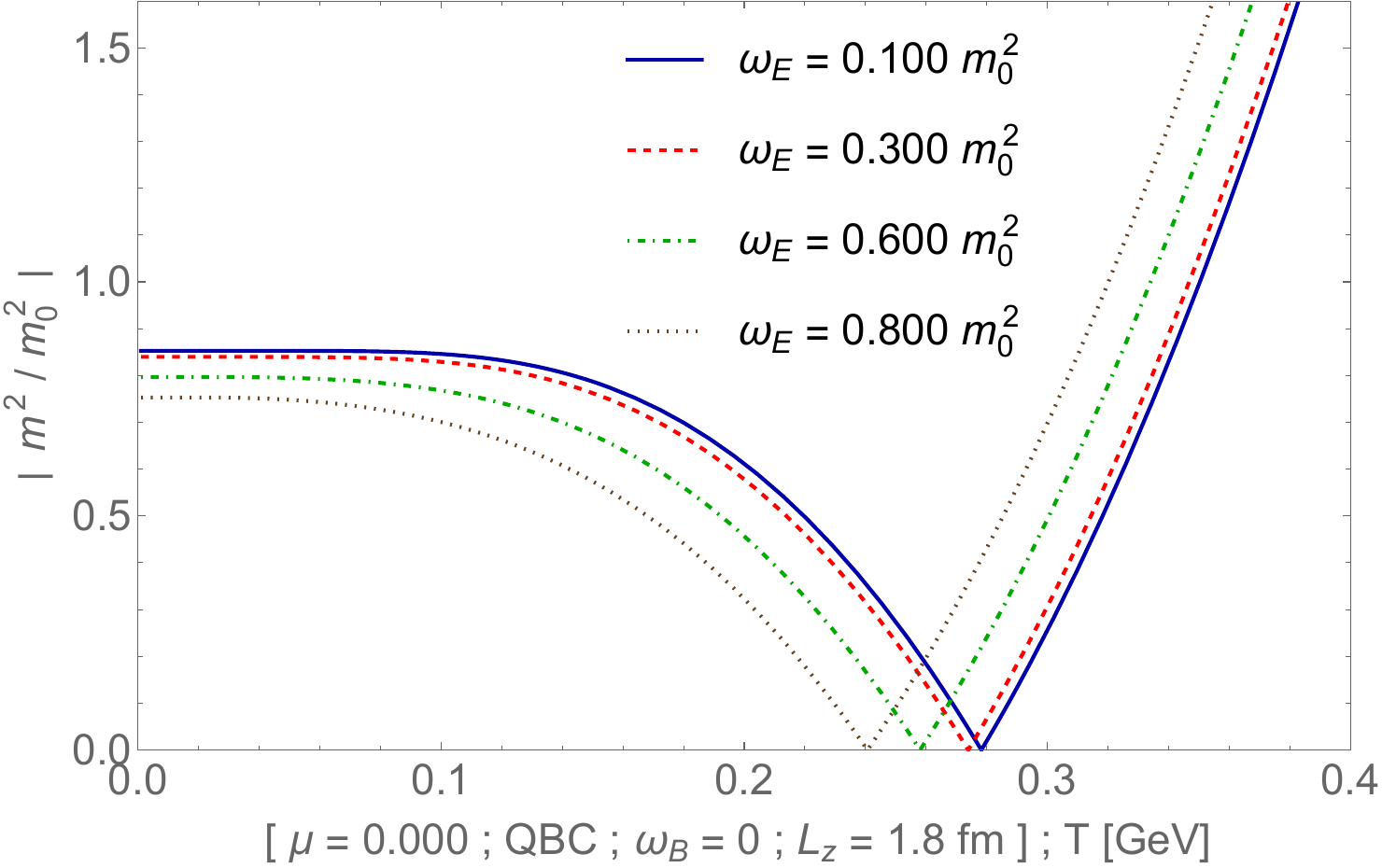}
\includegraphics[{width=6.49cm}]{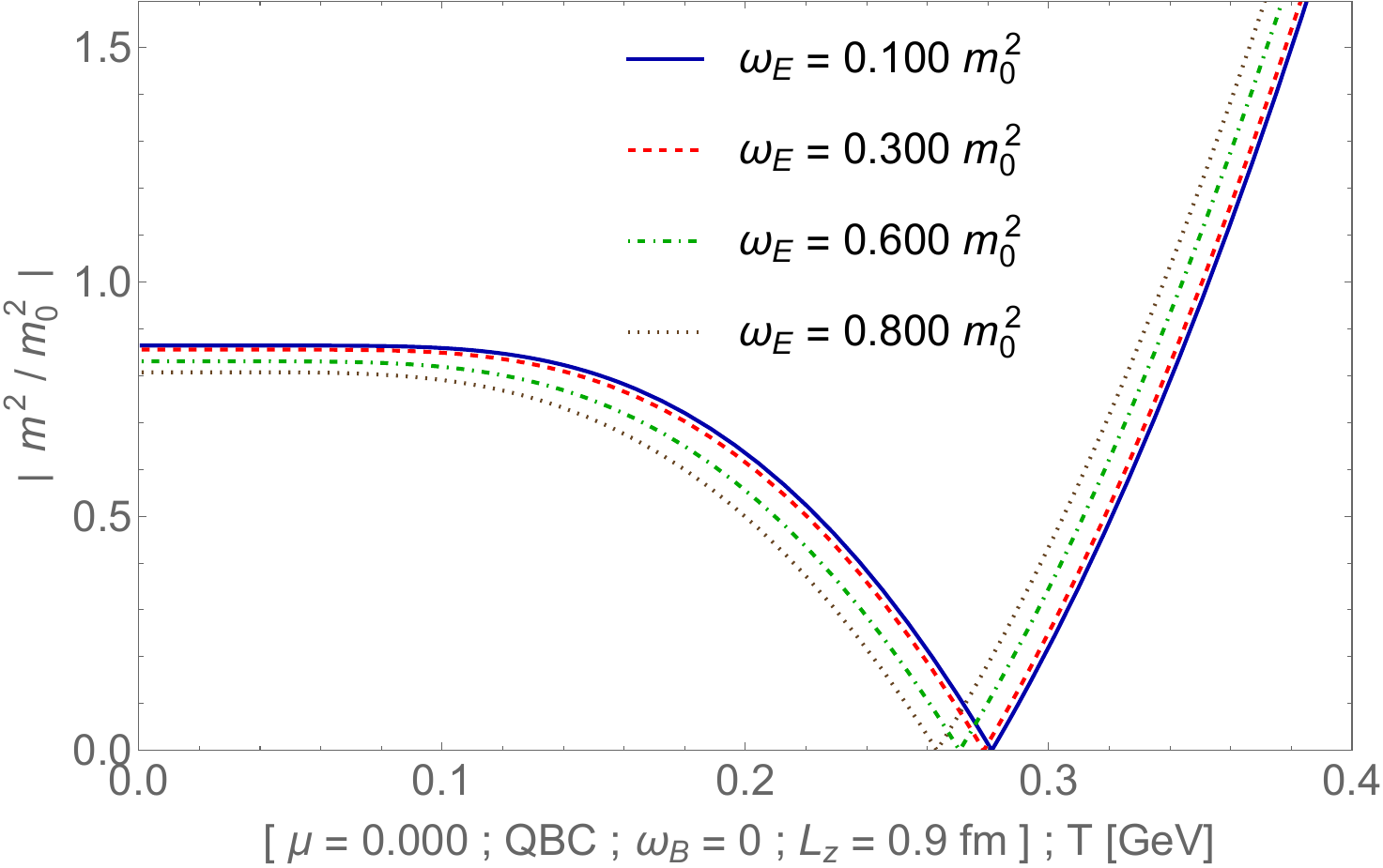} \\
\includegraphics[{width=6.49cm}]{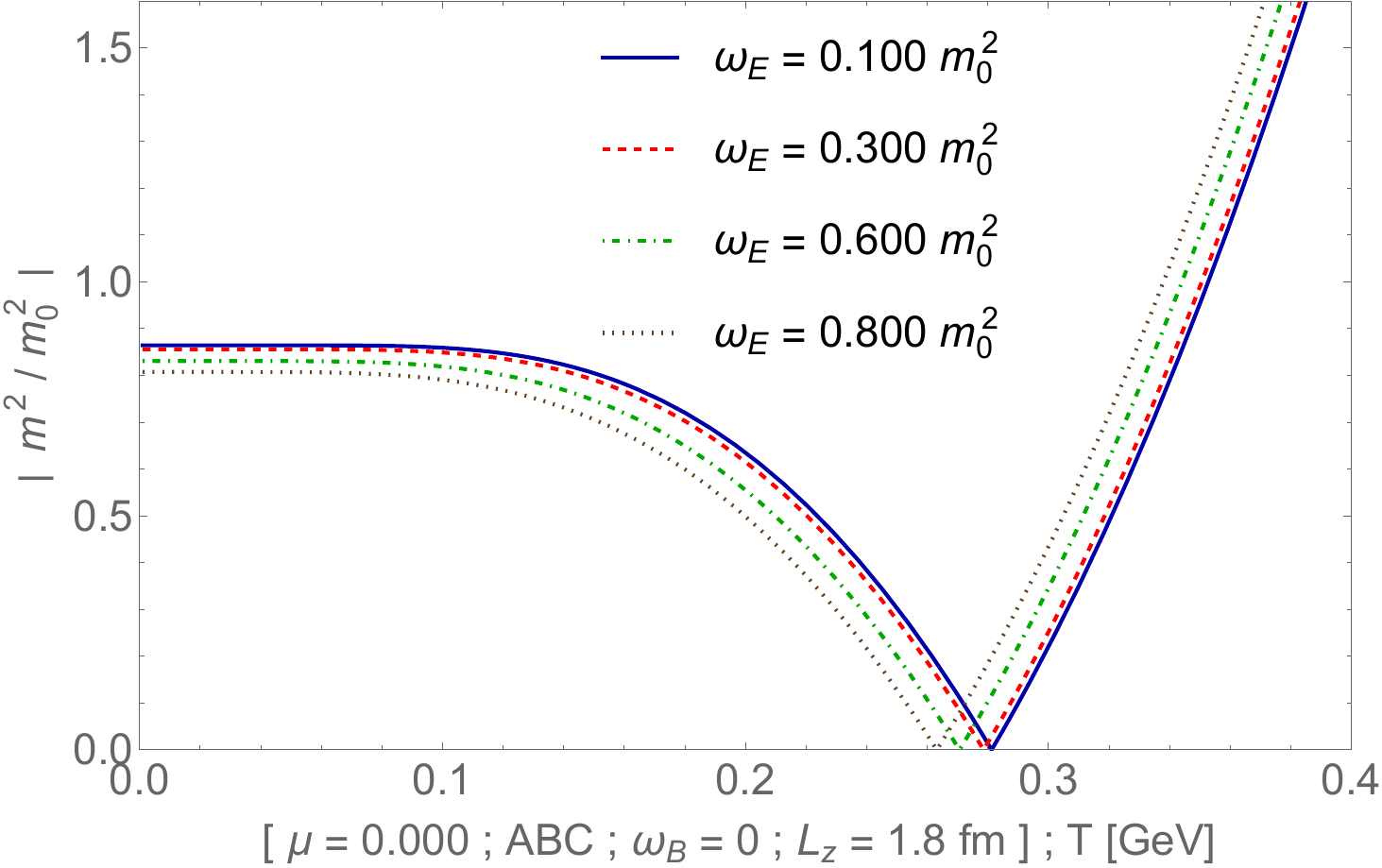}
\includegraphics[{width=6.49cm}]{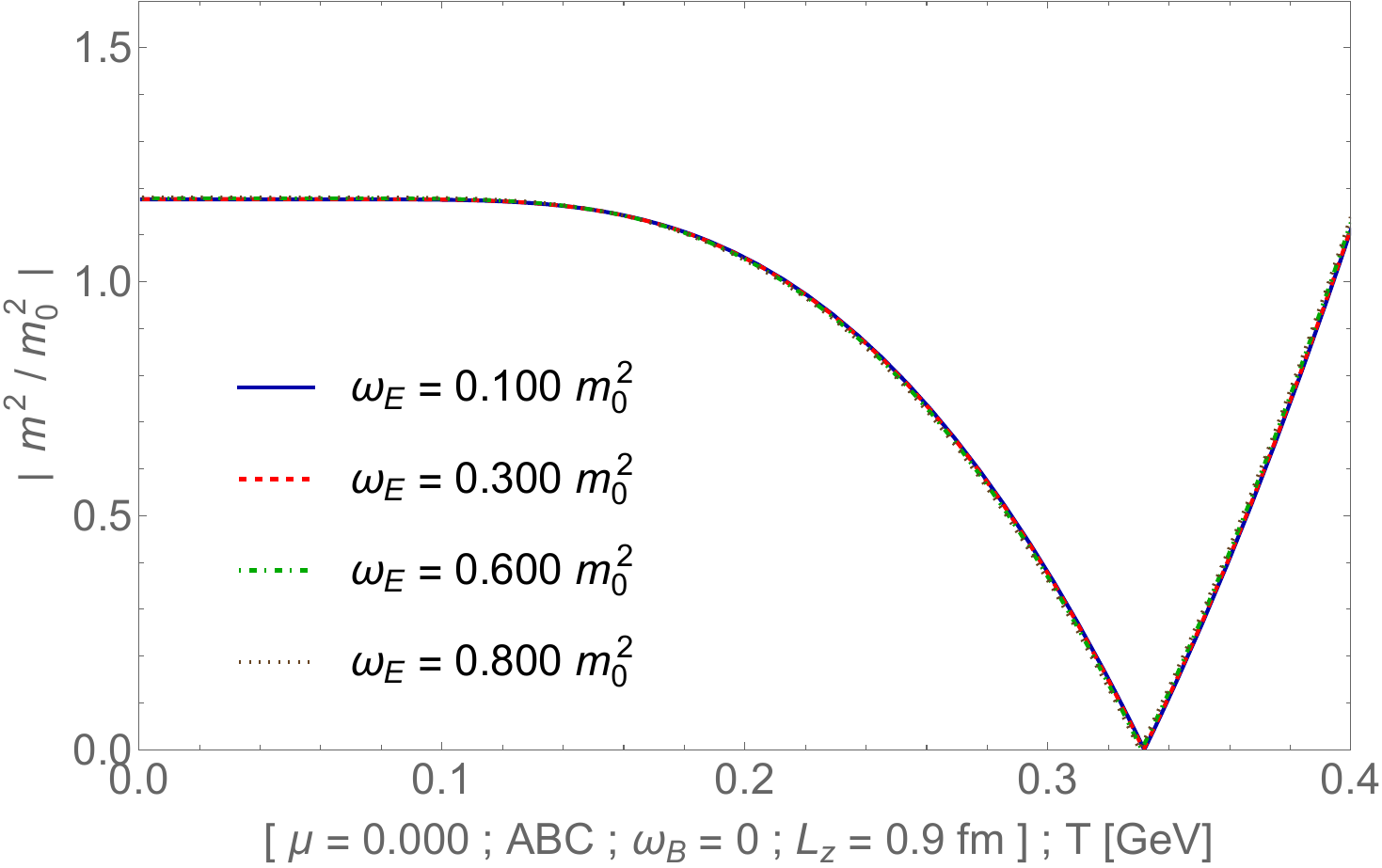} \\
\caption{~Corrected mass as a function of temperature for several strength values of the external electric field. At the left panels, we fixed the thickness of the film at $1.8~\mathrm{fm}$. In the same way, at the right panels, we fixed $L_{z}$ at $0.9~\mathrm{fm}$.}
\label{Fig3}
\end{figure}

Let us focus on the magnetic effects on the bosonic system. In Fig.~$4$, we see the same effects induced by the finite length of the system on $z$-axes that have occurred in the electric background for $L_{z}$ decreasing. Namely, at the fixed magnetic background and considering PBC, QBC, and ABC, we obtain $T_{c}$ smallest, $T_{c}$ almost constant, and $T_{c}$ highest, respectively.
\begin{figure}
\centering
\includegraphics[{width=6.49cm}]{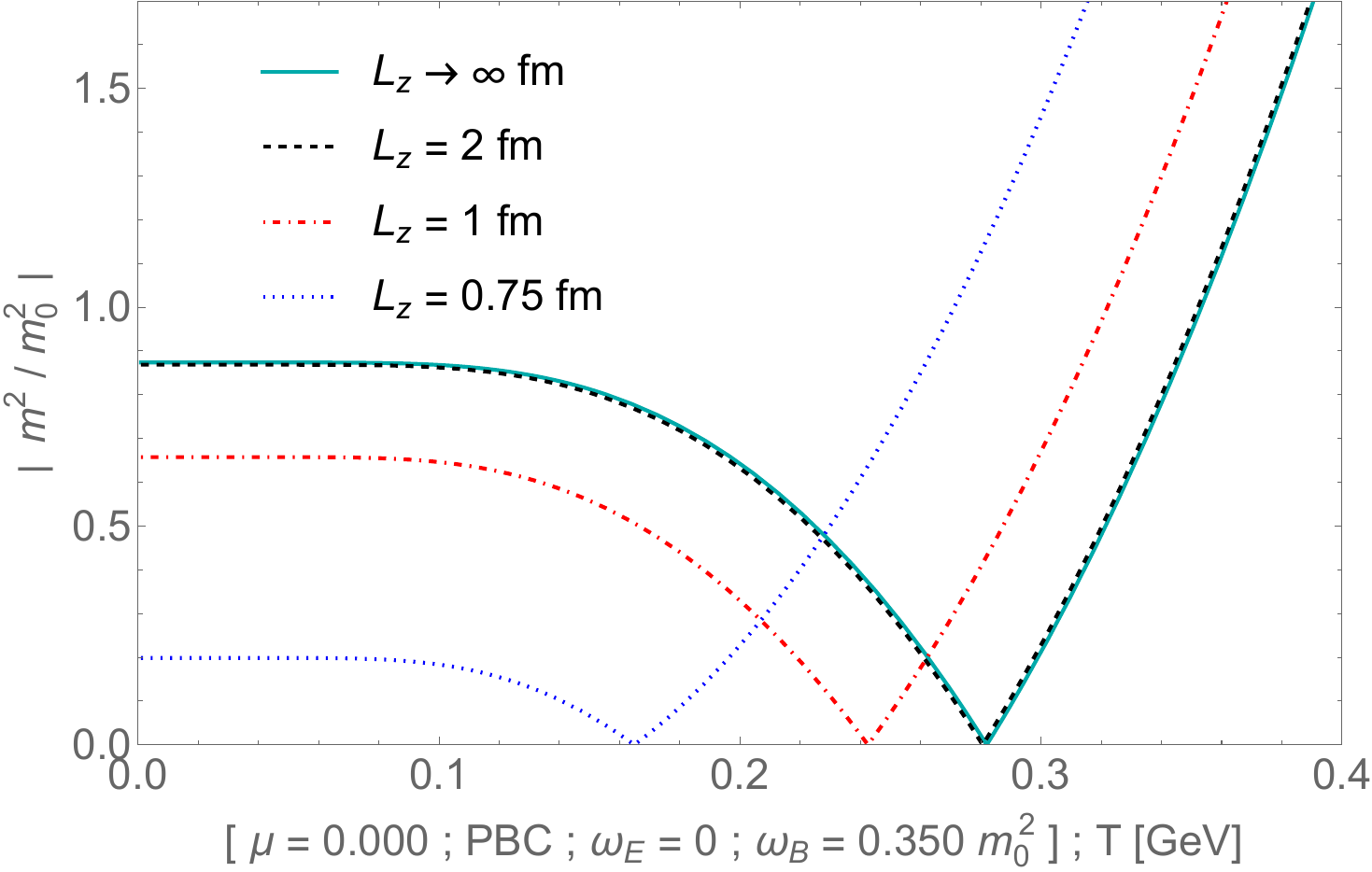}
\includegraphics[{width=6.49cm}]{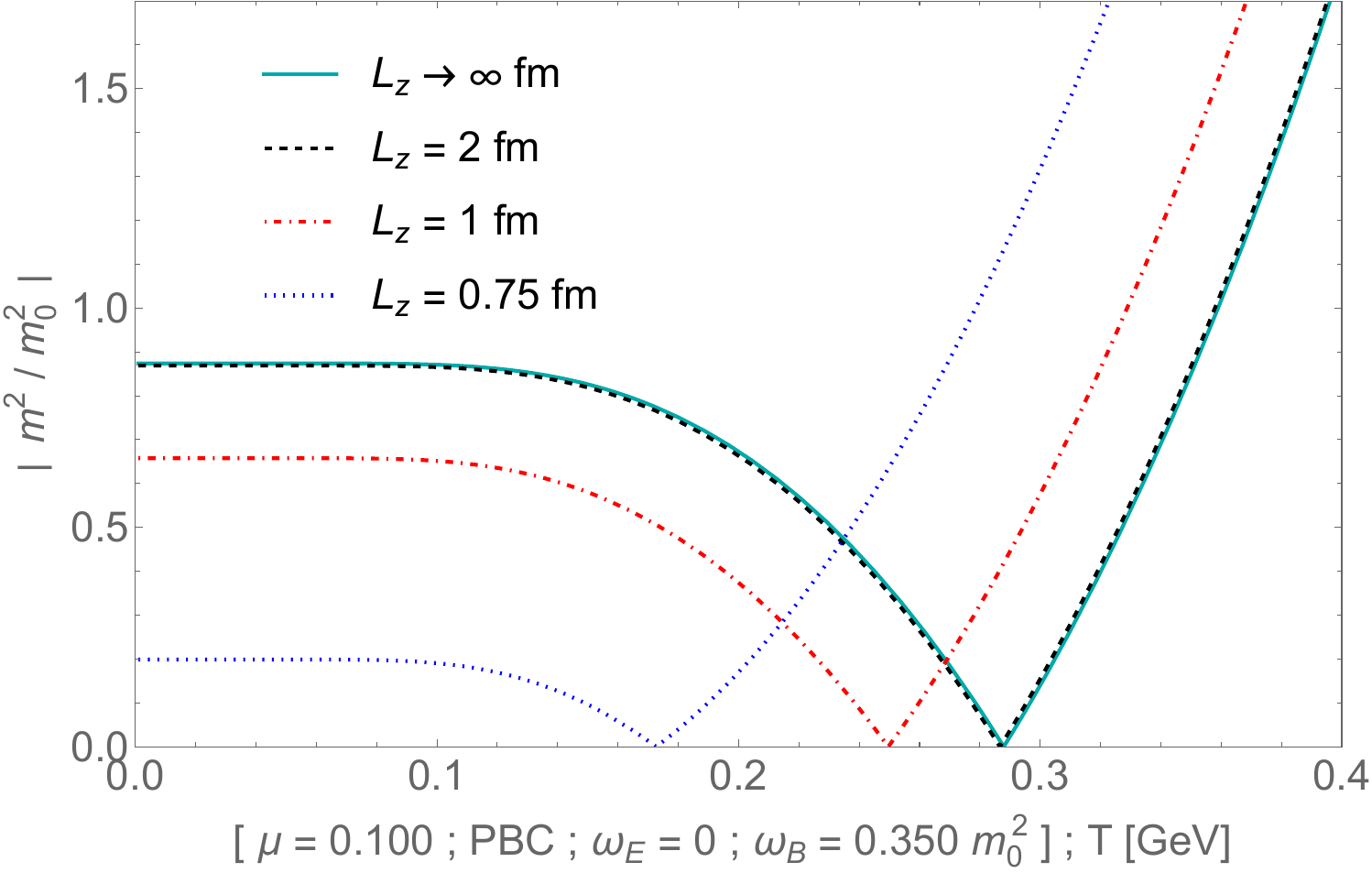} \\
\includegraphics[{width=6.49cm}]{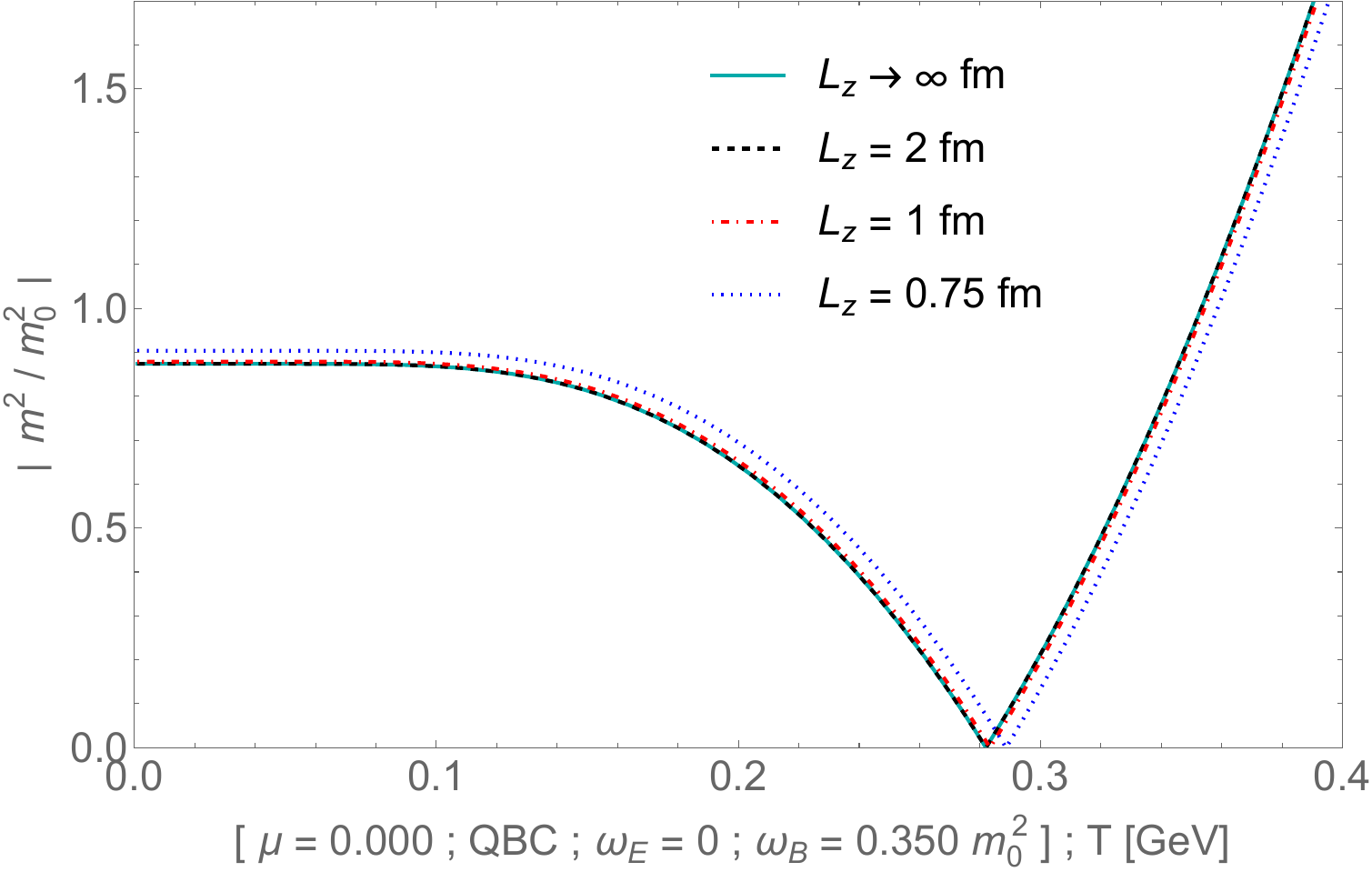} 
\includegraphics[{width=6.49cm}]{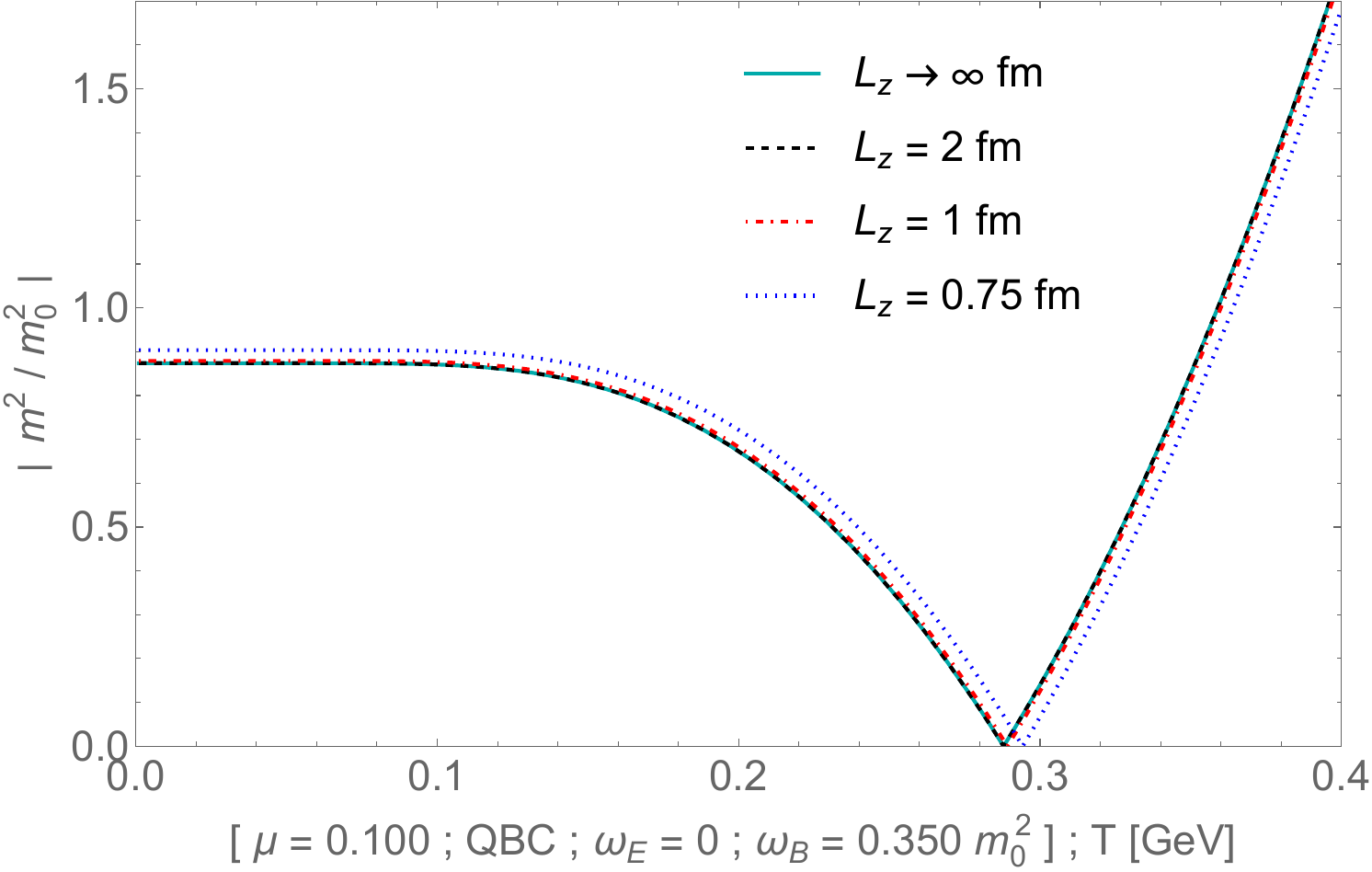} \\
\includegraphics[{width=6.49cm}]{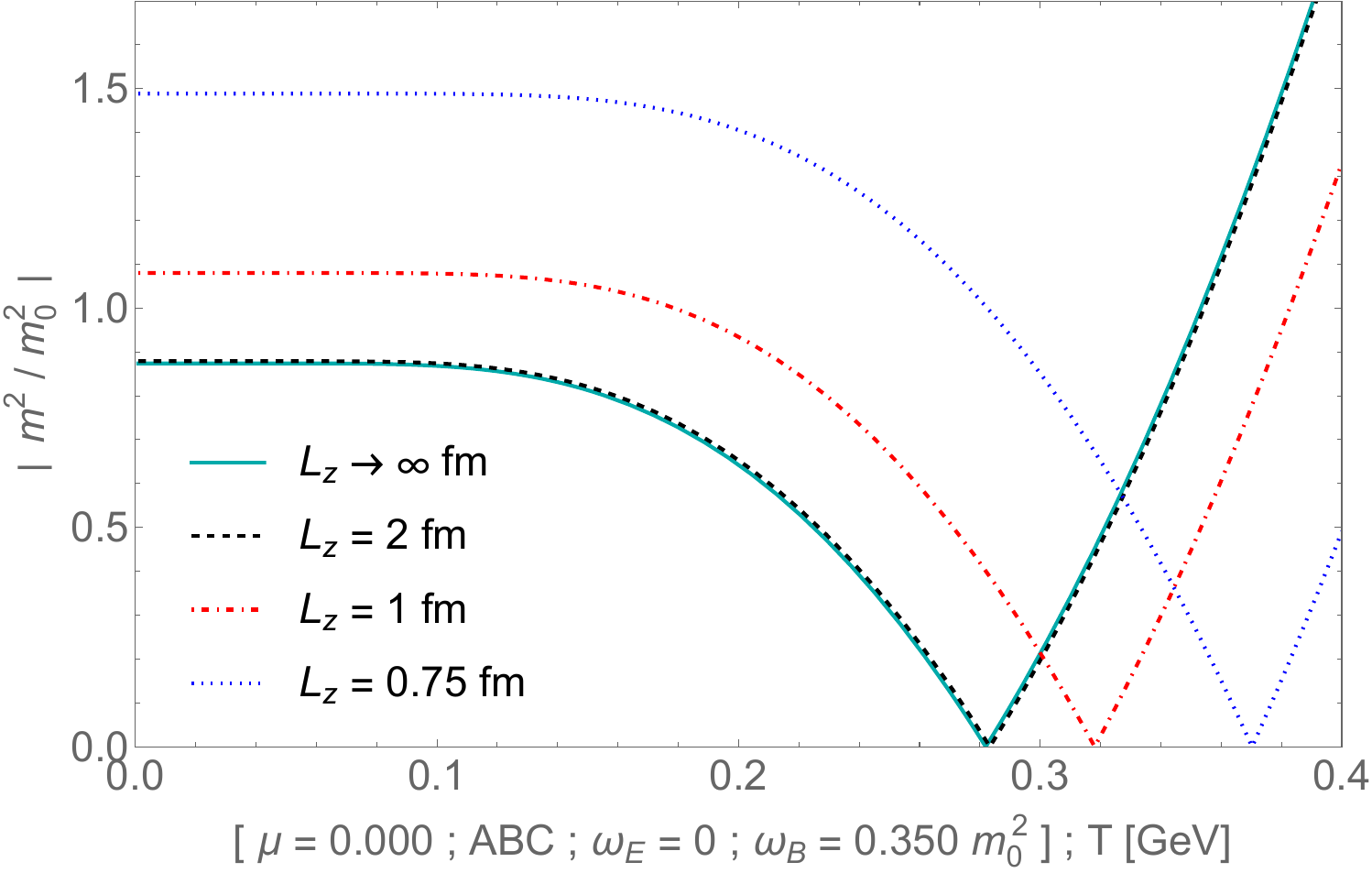}
\includegraphics[{width=6.49cm}]{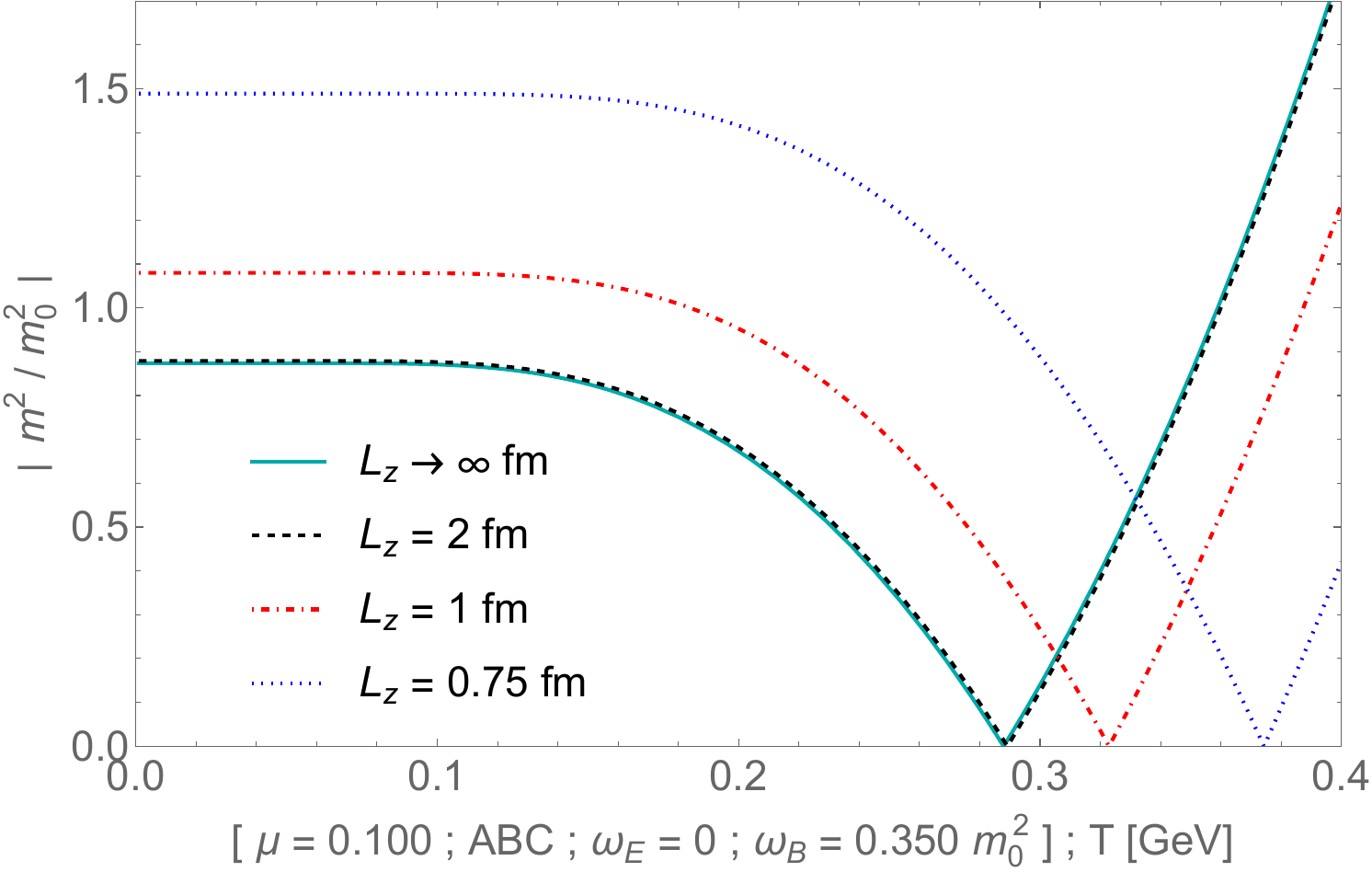} \\
\caption{~Corrected mass as a function of temperature for several values of thickness of the film (zero chemical potential at the left and finite chemical potential on the right).  We have PBC (top panel), QBC (middle panel), and ABC (bottom panel) on $z$-coordinate.}
\label{Fig4}
\end{figure}

Nevertheless, at fixed finite sizes and different magnetic field strengths, Fig.~$5$ depicts the magnetic catalysis phenomenon. The plots indicate higher critical temperature values at raised external magnetic field intensities. Then, the magnetic field increases the broken phase region, resulting in higher transition temperature values, independent of the type boundary condition used.
\begin{figure}
\centering
\includegraphics[{width=6.49cm}]{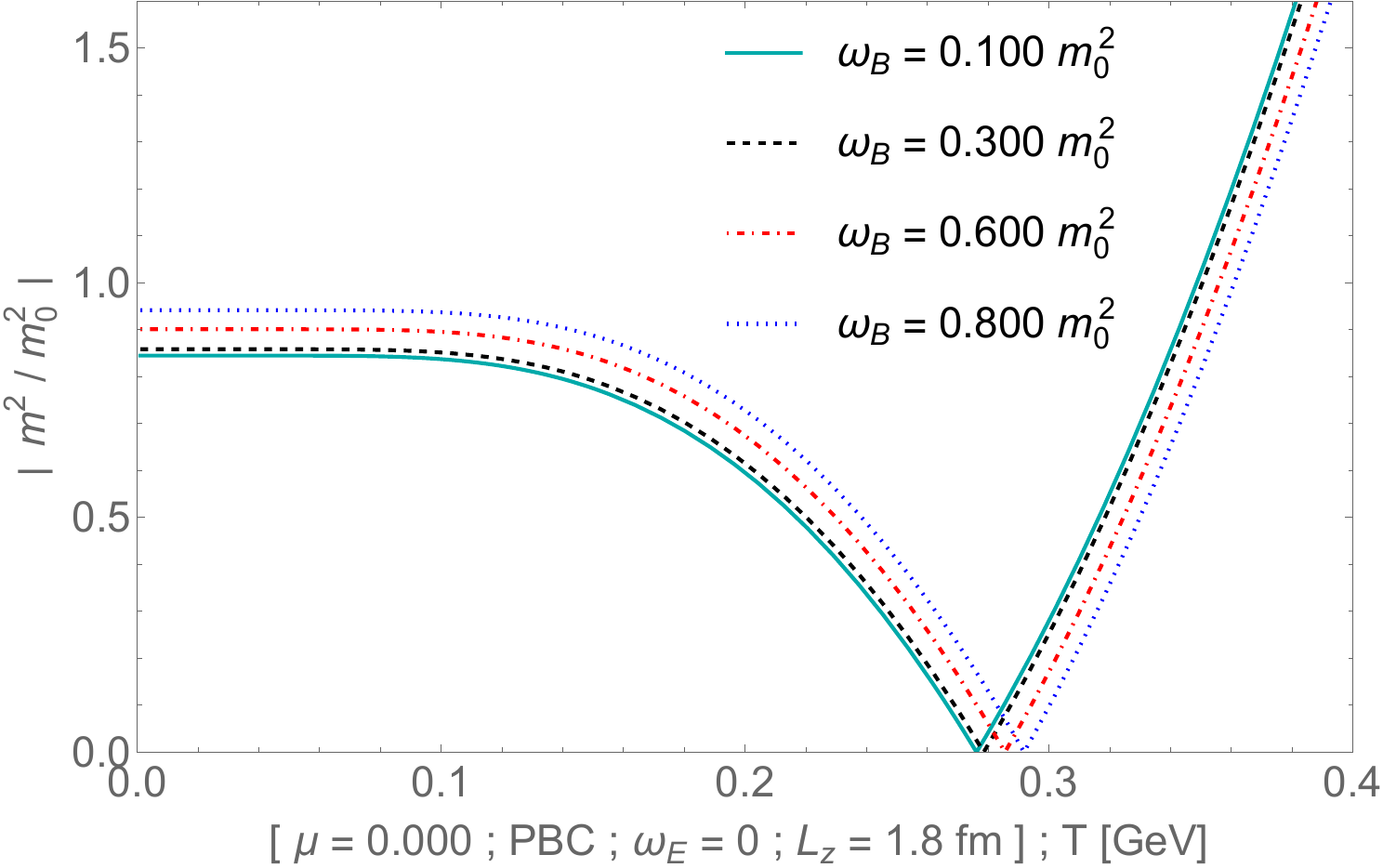}
\includegraphics[{width=6.49cm}]{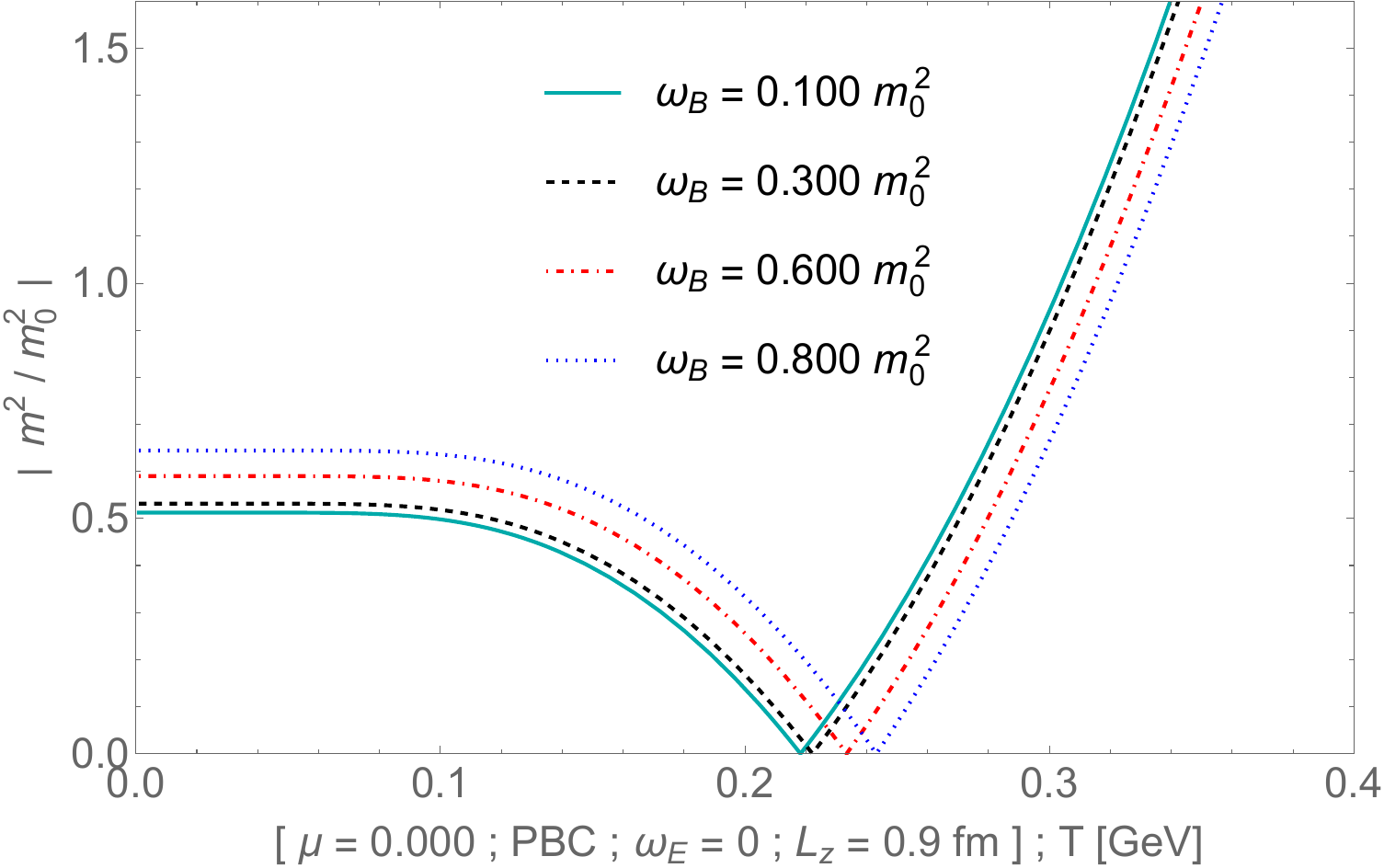} \\
\includegraphics[{width=6.49cm}]{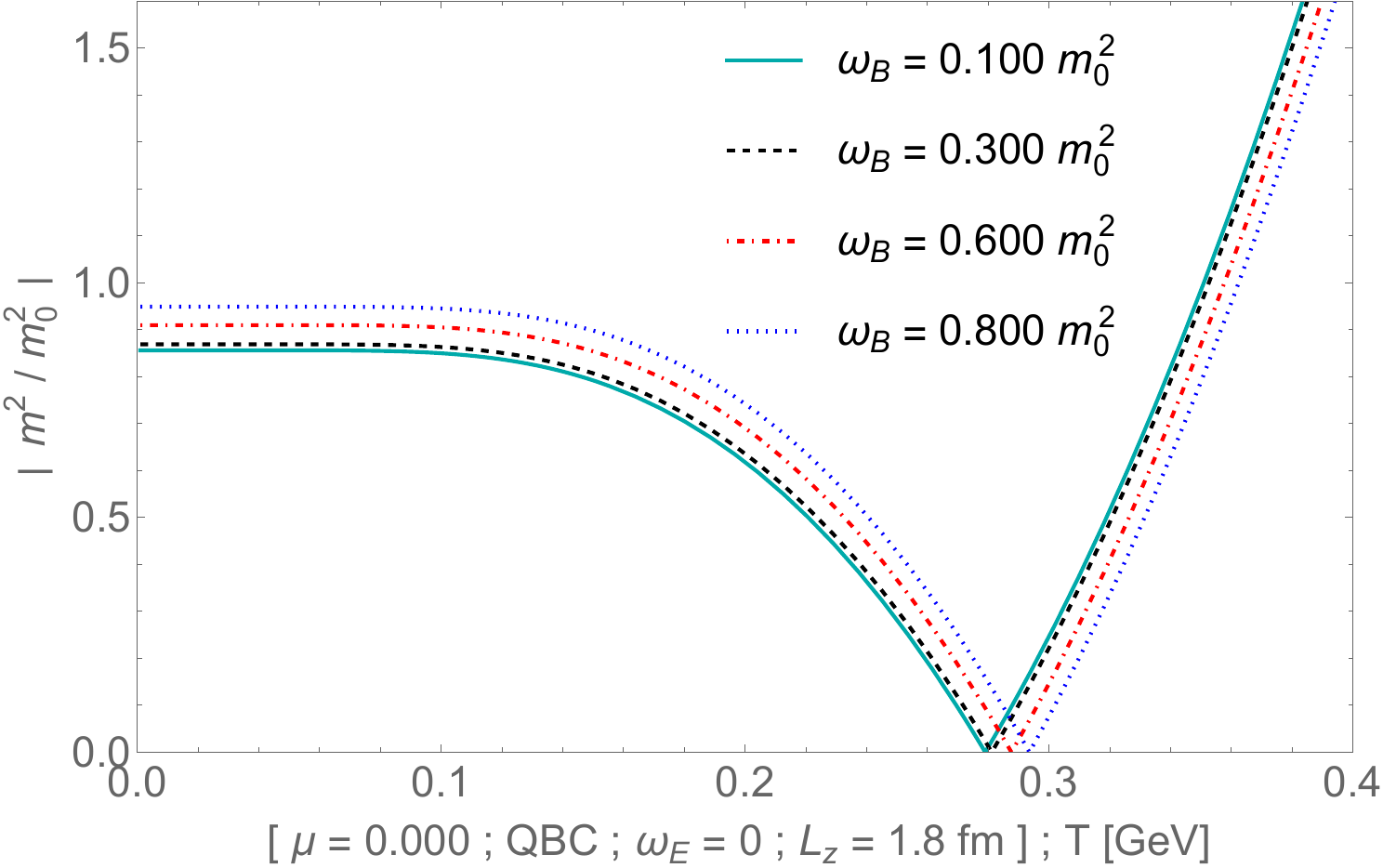}
\includegraphics[{width=6.49cm}]{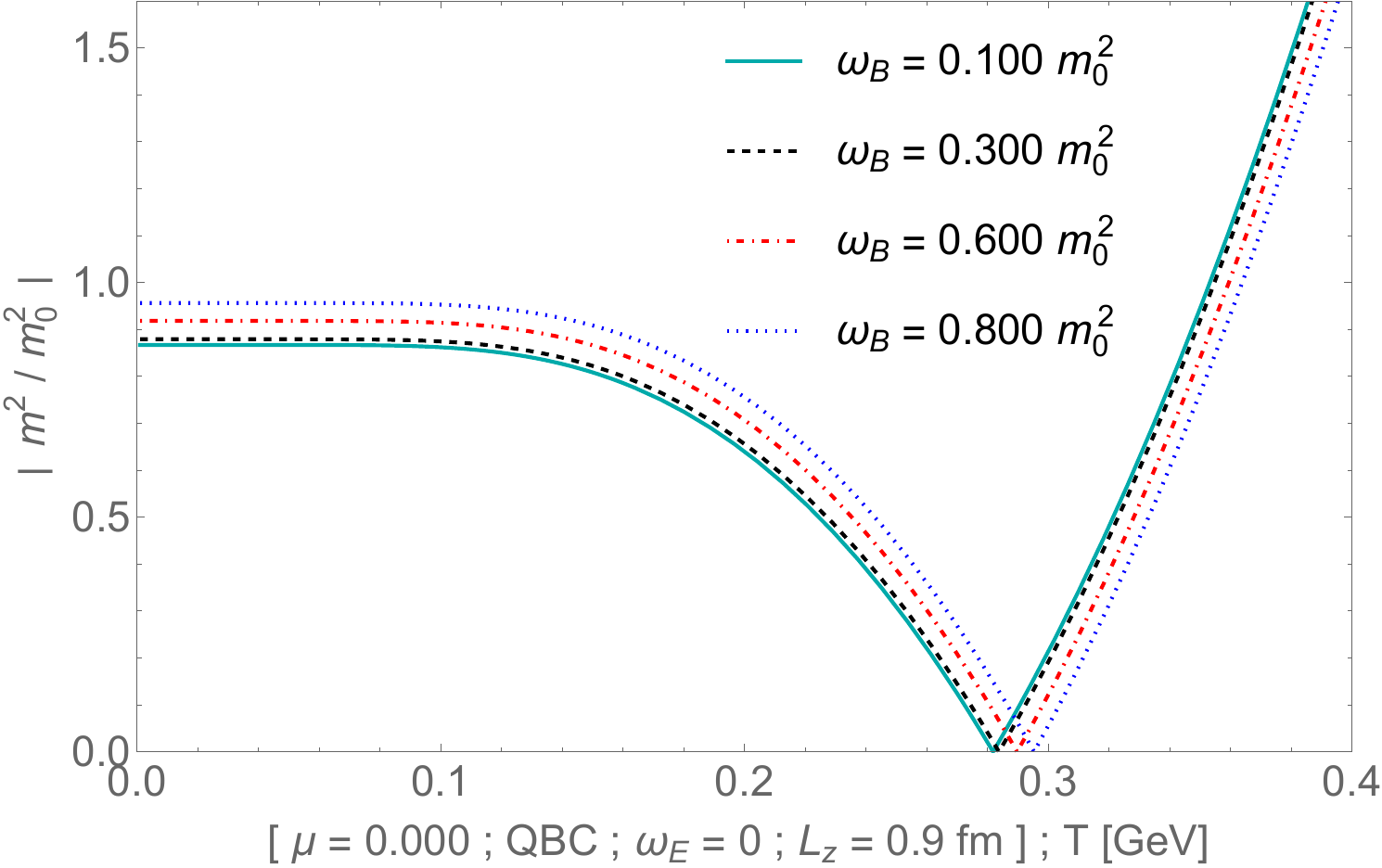} \\
\includegraphics[{width=6.49cm}]{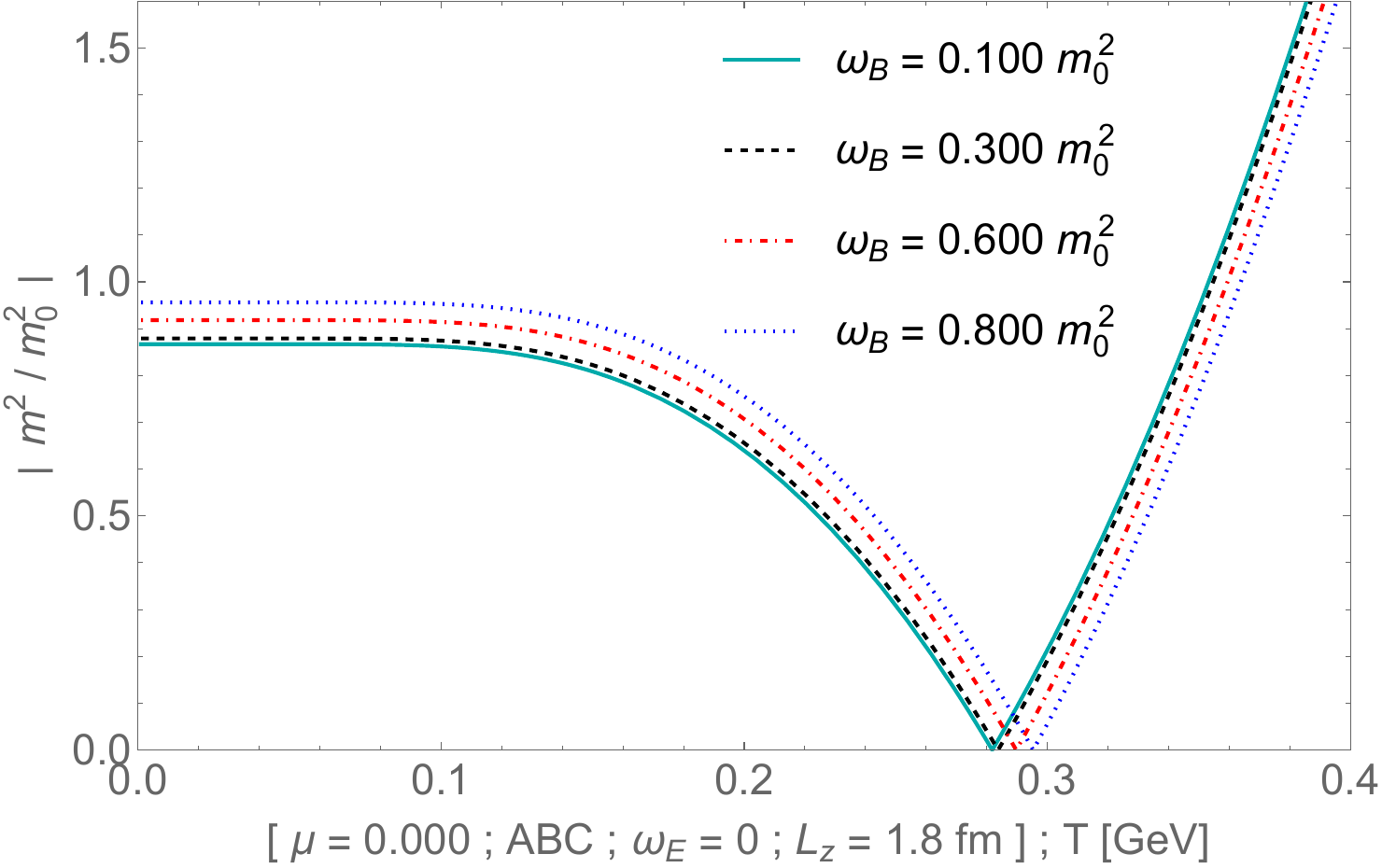}
\includegraphics[{width=6.49cm}]{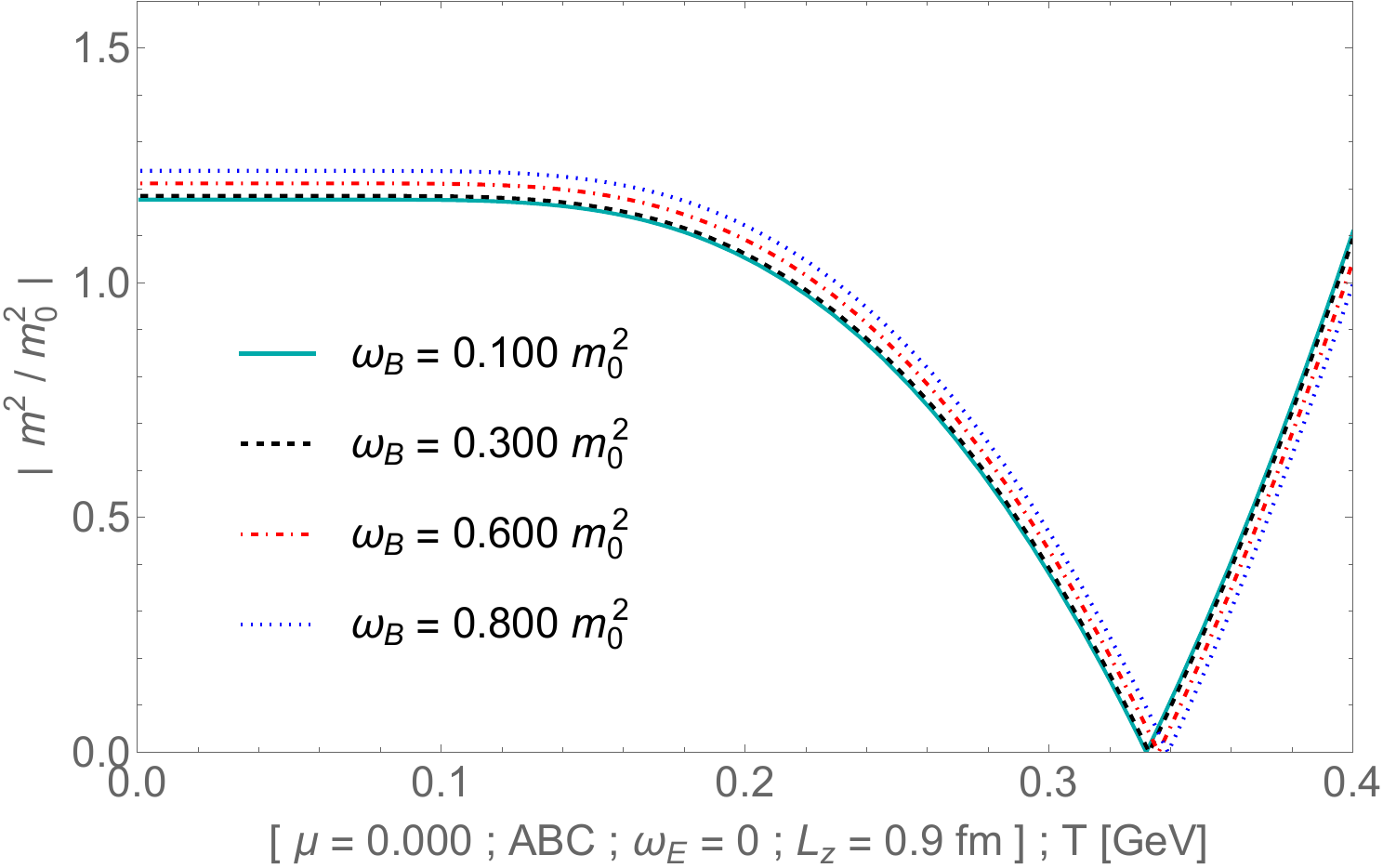} \\
\caption{~Corrected mass as a function of temperature for several values of strength of the external magnetic field. At the left panels, we fixed the thickness of the film at $1.8~\mathrm{fm}$. In the same way, at the right panels, we fixed $L_{z}$ at $0.9~\mathrm{fm}$.}
\label{Fig5}
\end{figure}

In Figs.~$6,7,8$ and $9$, we present a continuous interval of separation between the planes defined on region $0.5\,\mathrm{fm}\leq L_{z} \leq 5\,\mathrm{fm}$. As before, the graphs were made with PBC (top panels), QBC (middle panels), and ABC (bottom panels) in the finite size range declared above. In Fig.~$6$, we set four different temperature values. For the chosen temperatures, a phase transition upon an electric external field takes place just for periodic boundary conditions. In this case, we have found the values of the critical lengths that ensure it, $L_{c}$. Notice that the smallest critical lengths occur for lower temperatures. The effect due to $\mu$ is to decline $L_{c}$ (see right side of Fig.~$6$ on top).
\begin{figure}
\centering
\includegraphics[{width=6.49cm}]{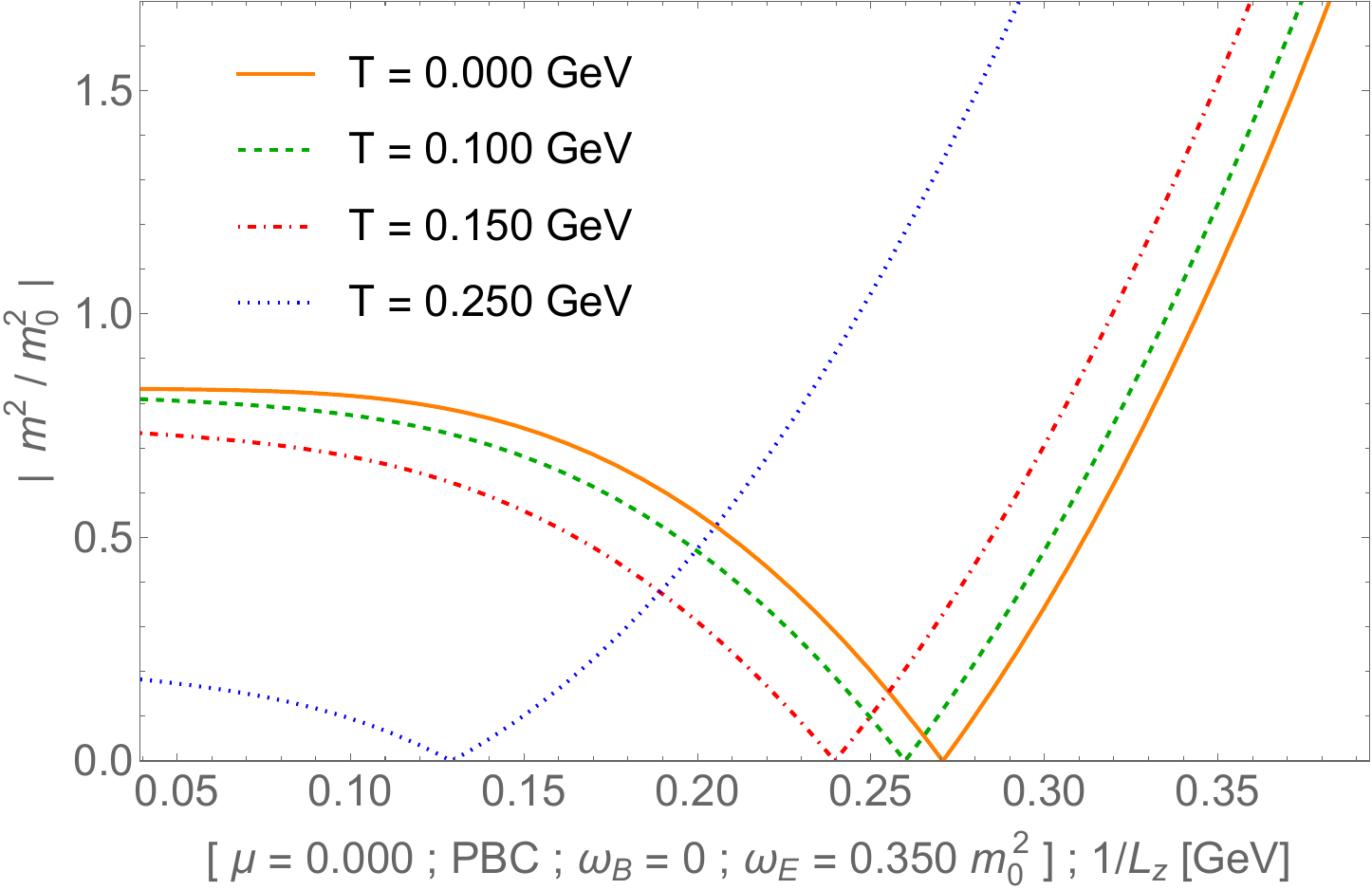}
\includegraphics[{width=6.49cm}]{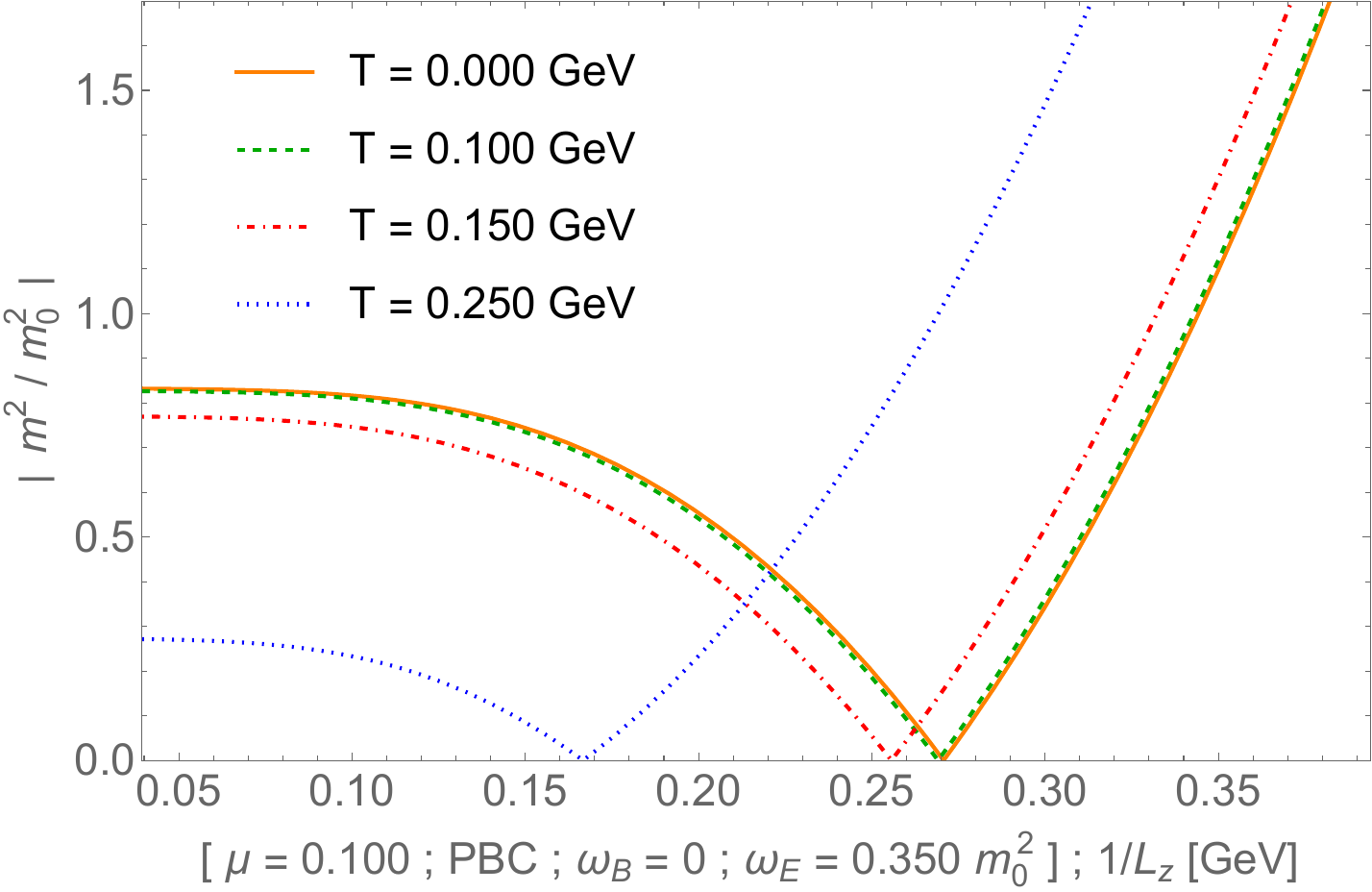} \\
\includegraphics[{width=6.49cm}]{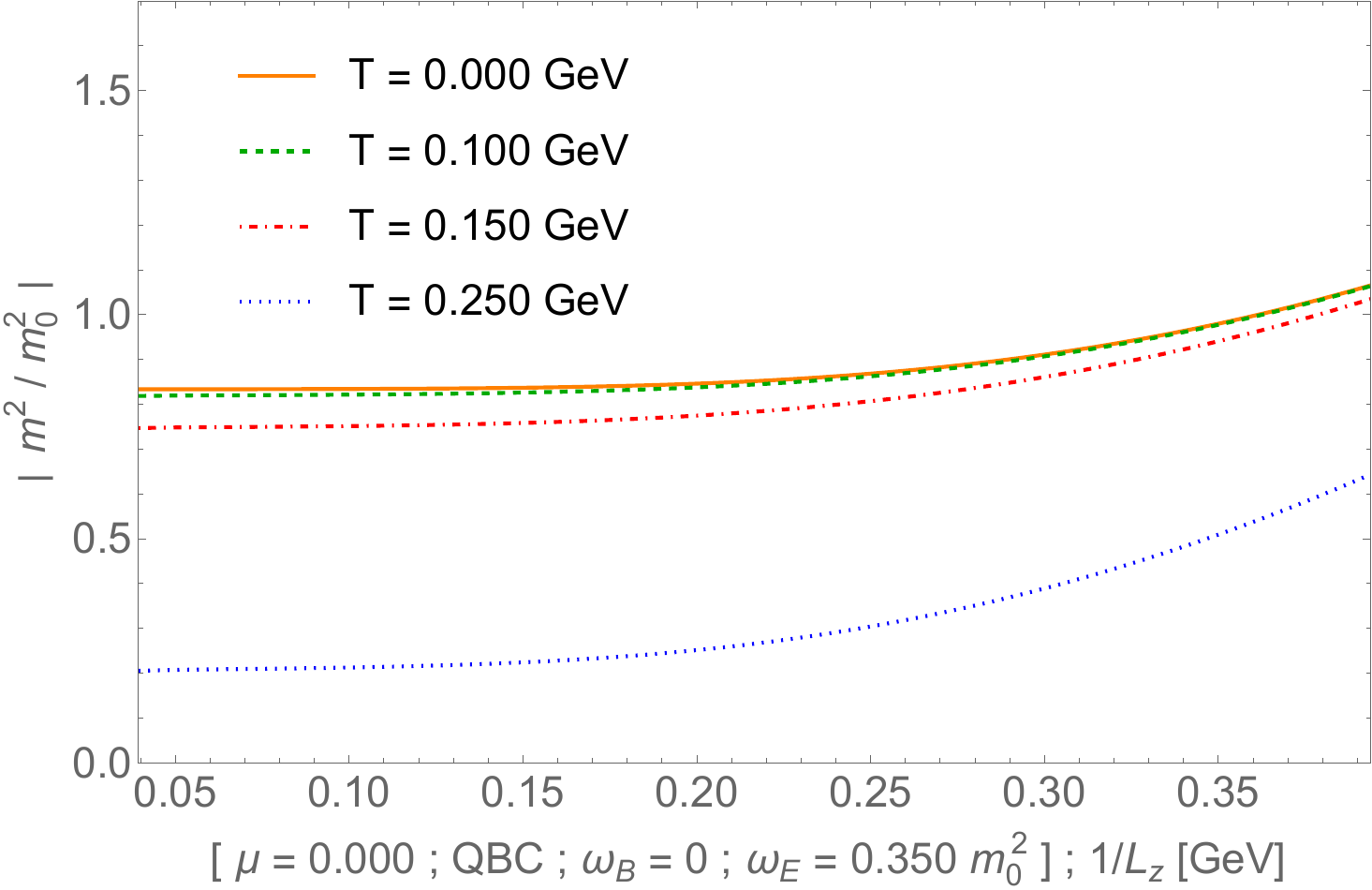} 
\includegraphics[{width=6.49cm}]{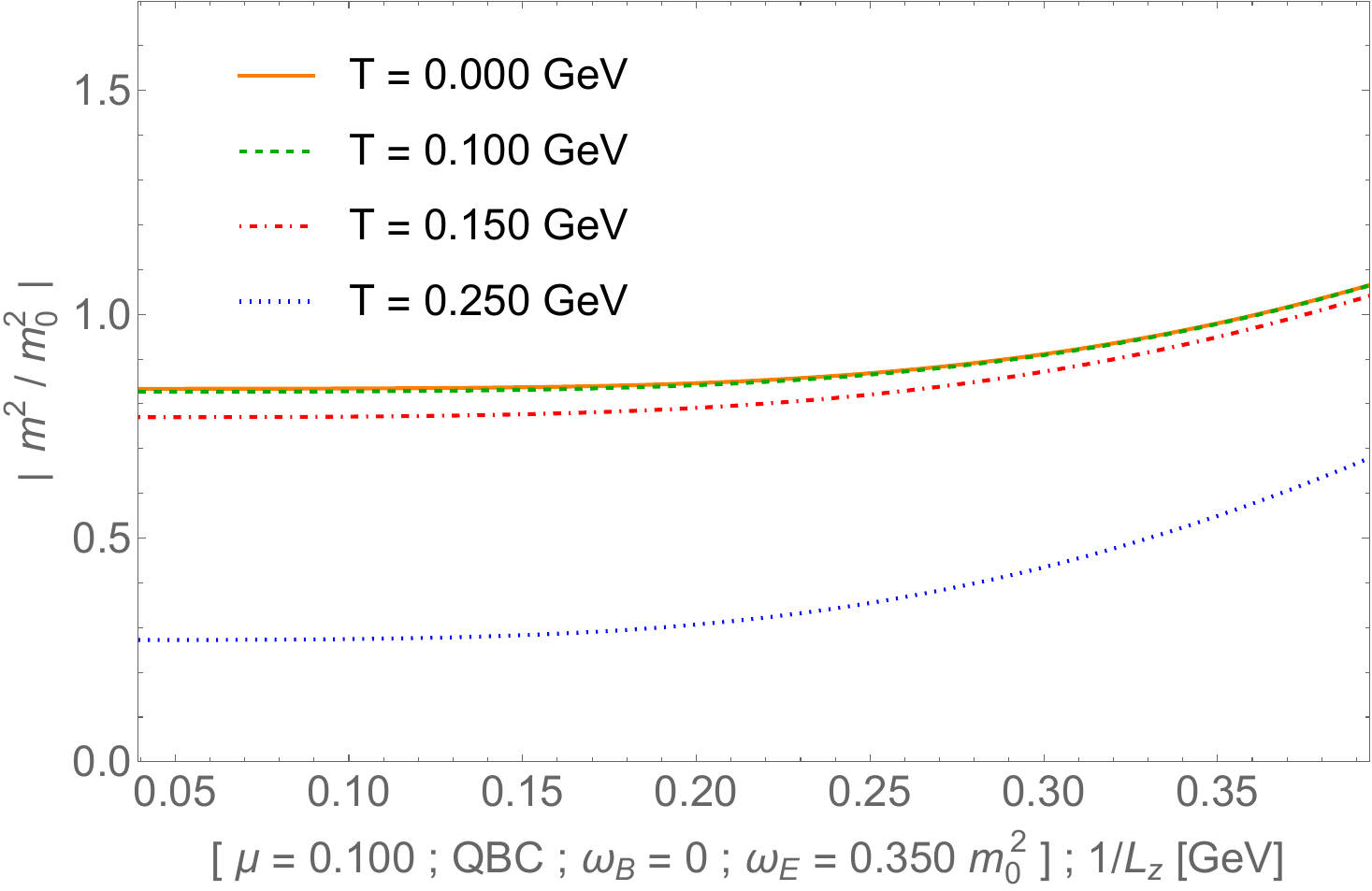} \\
\includegraphics[{width=6.49cm}]{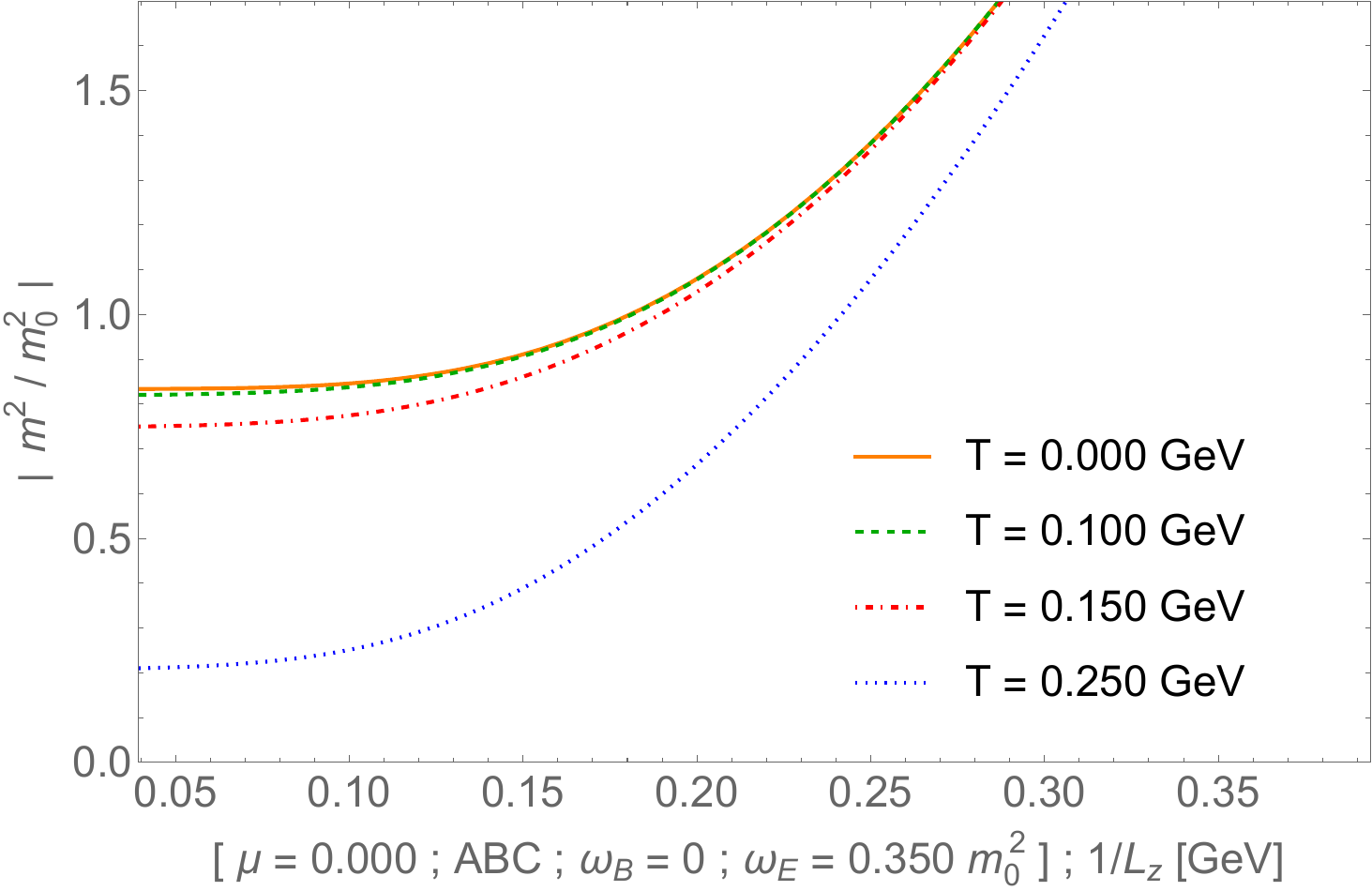}
\includegraphics[{width=6.49cm}]{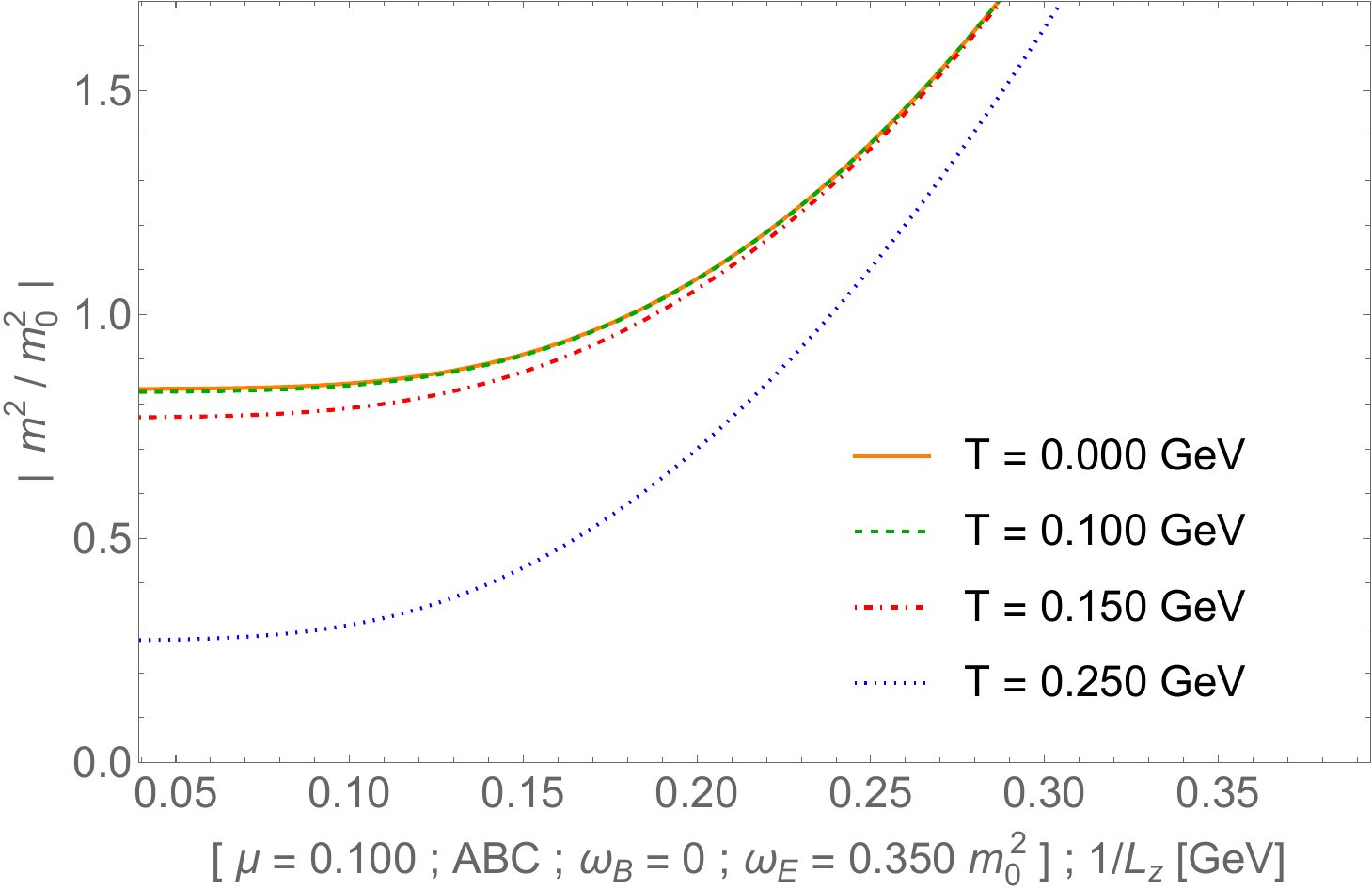} \\
\caption{~Corrected mass as a function of the inverse of length for several values of temperatures of the film (null chemical potential at the left and finite chemical potential on the right).  We have PBC (top panel), QBC (middle panel), and ABC (bottom panel) boundary conditions on $z$-coordinate.}
\label{Fig6}
\end{figure}

In Fig.~$7$, we observe the topological analog of the IEC and EC phenomena (only for the PBC case), that is, for weaker intensities of $\mathbf{E}$, the system passes from the broken phase to the restored phase, in smaller inverses of length. Conversely, for greater external electric field intensities, the interacting bosonic system undergoes a phase transition at increased inverse lengths. This topological effect can be more easily observed for higher temperatures, as shown on the right side of this figure. No phase transition was observed for the QBC and ABC cases, taking into account the temperature values fixed in the graphs. However, for large values of the separation $L_{z}$ $( 1/L_{z} \, \rightarrow \, 0) $, the influence of the external electric field on the corrected mass parameter is mixed for PBC (depending on the values of $\textbf{E}$), but always makes it smaller for QBC and ABC.
\begin{figure}
\centering
\includegraphics[{width=6.49cm}]{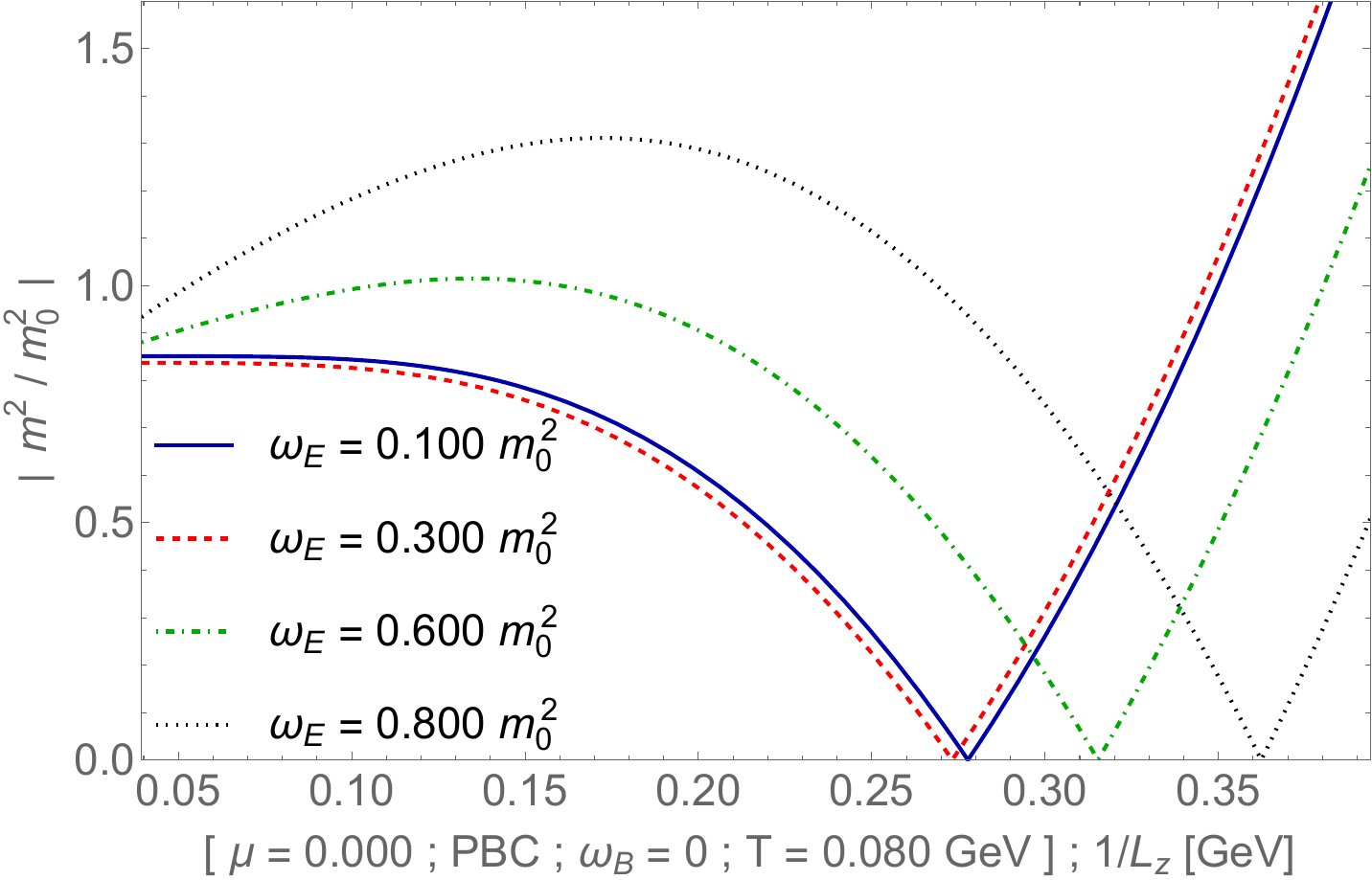}
\includegraphics[{width=6.49cm}]{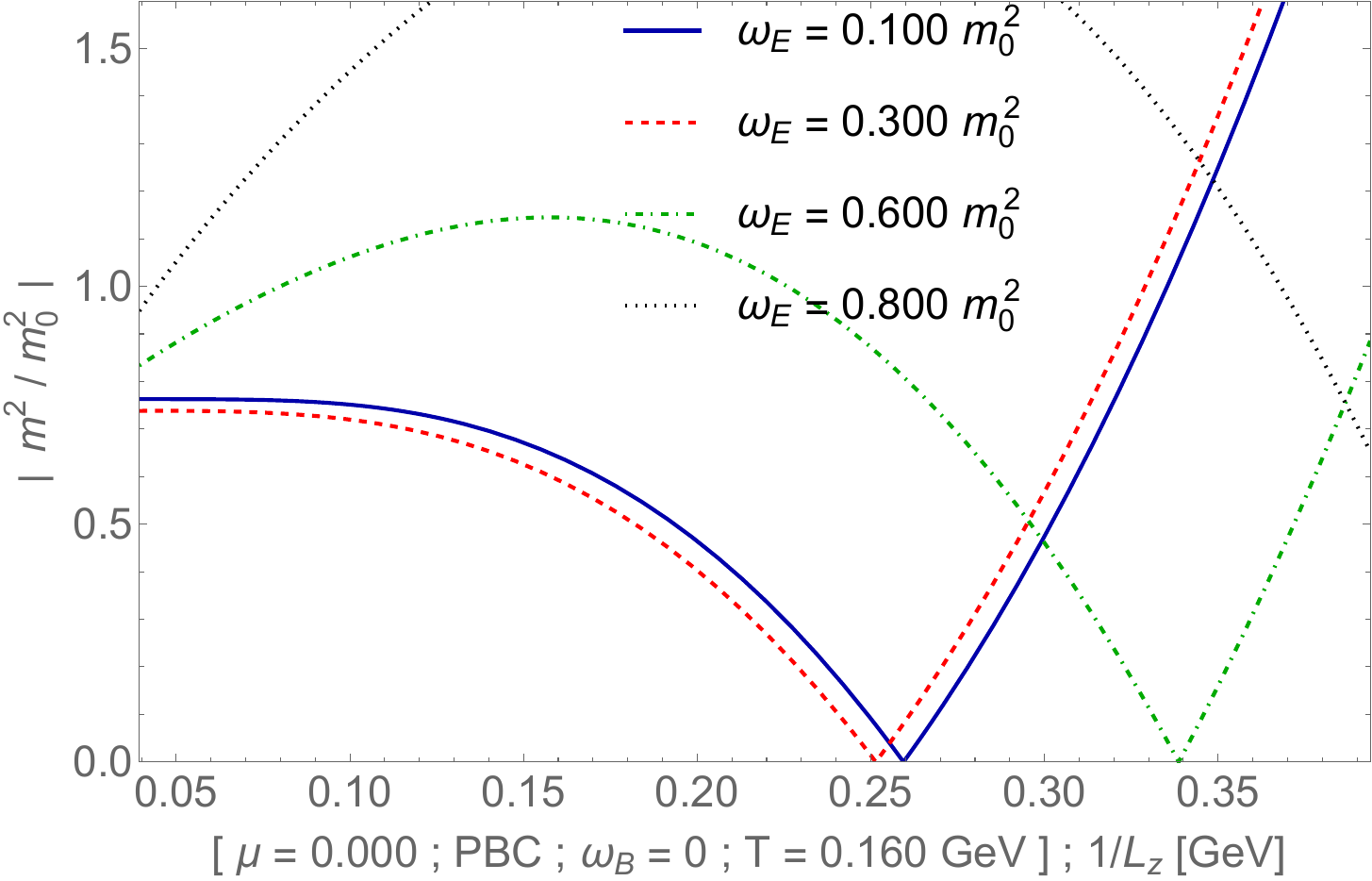} \\
\includegraphics[{width=6.49cm}]{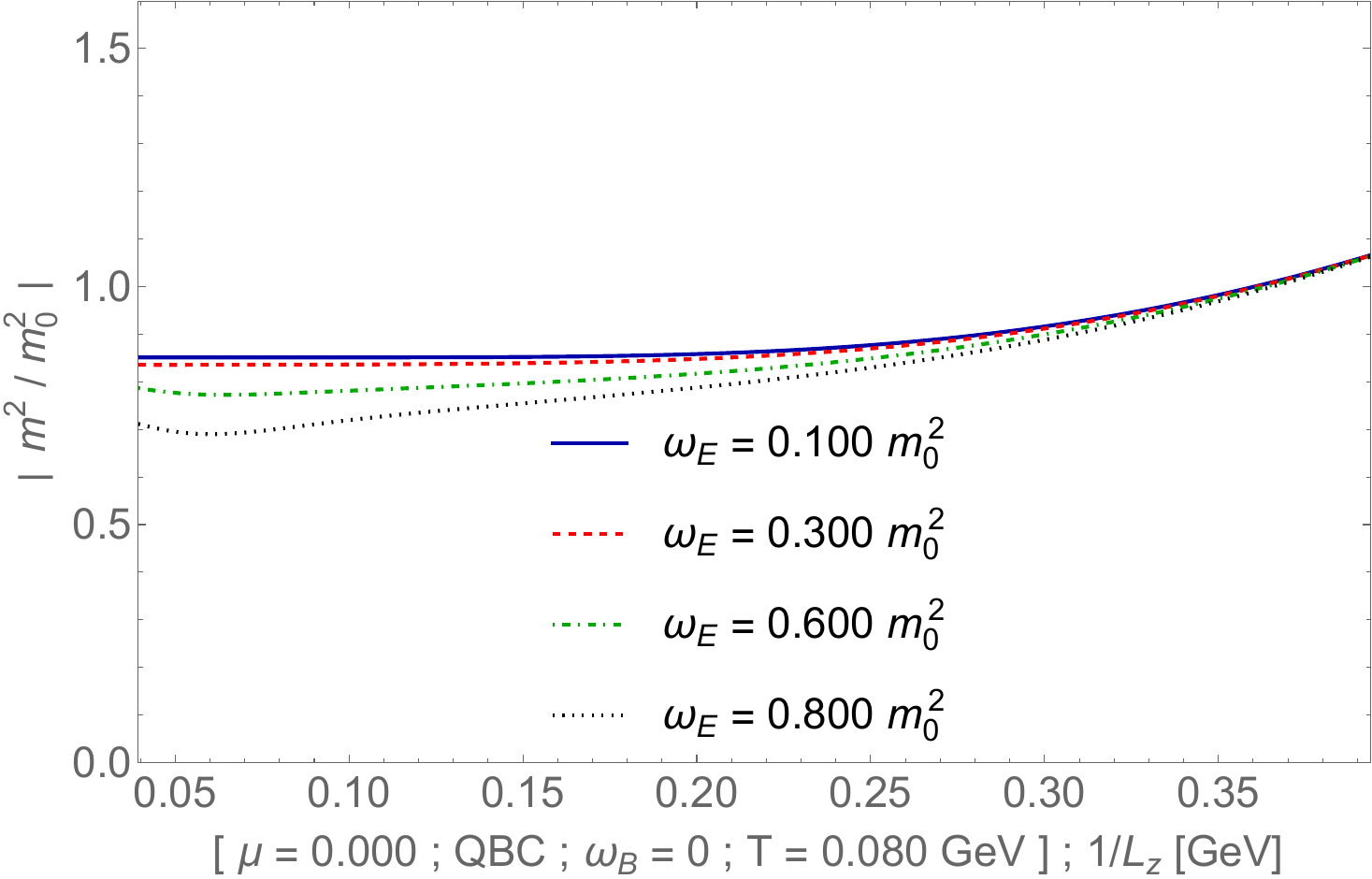}
\includegraphics[{width=6.49cm}]{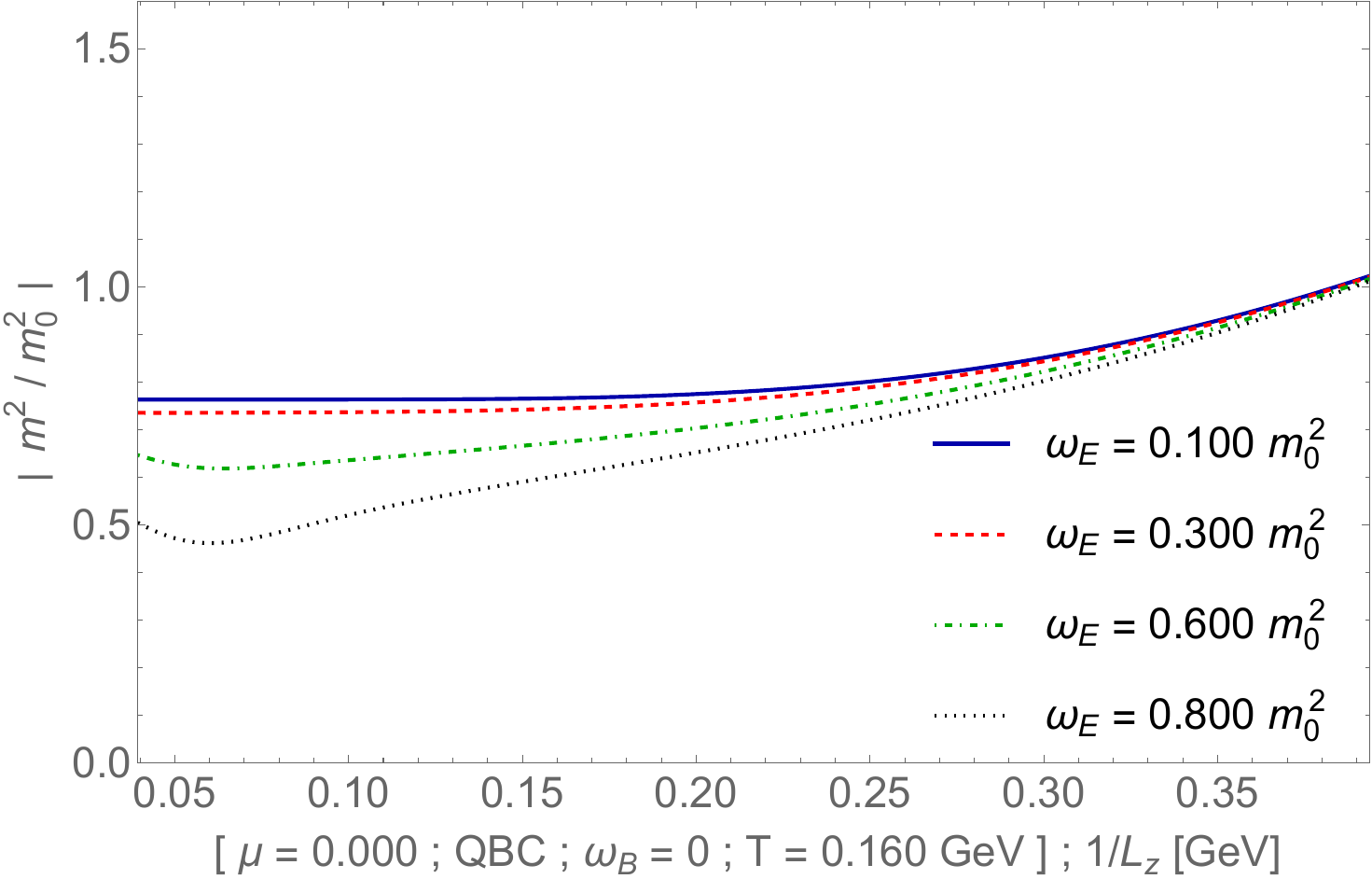} \\
\includegraphics[{width=6.49cm}]{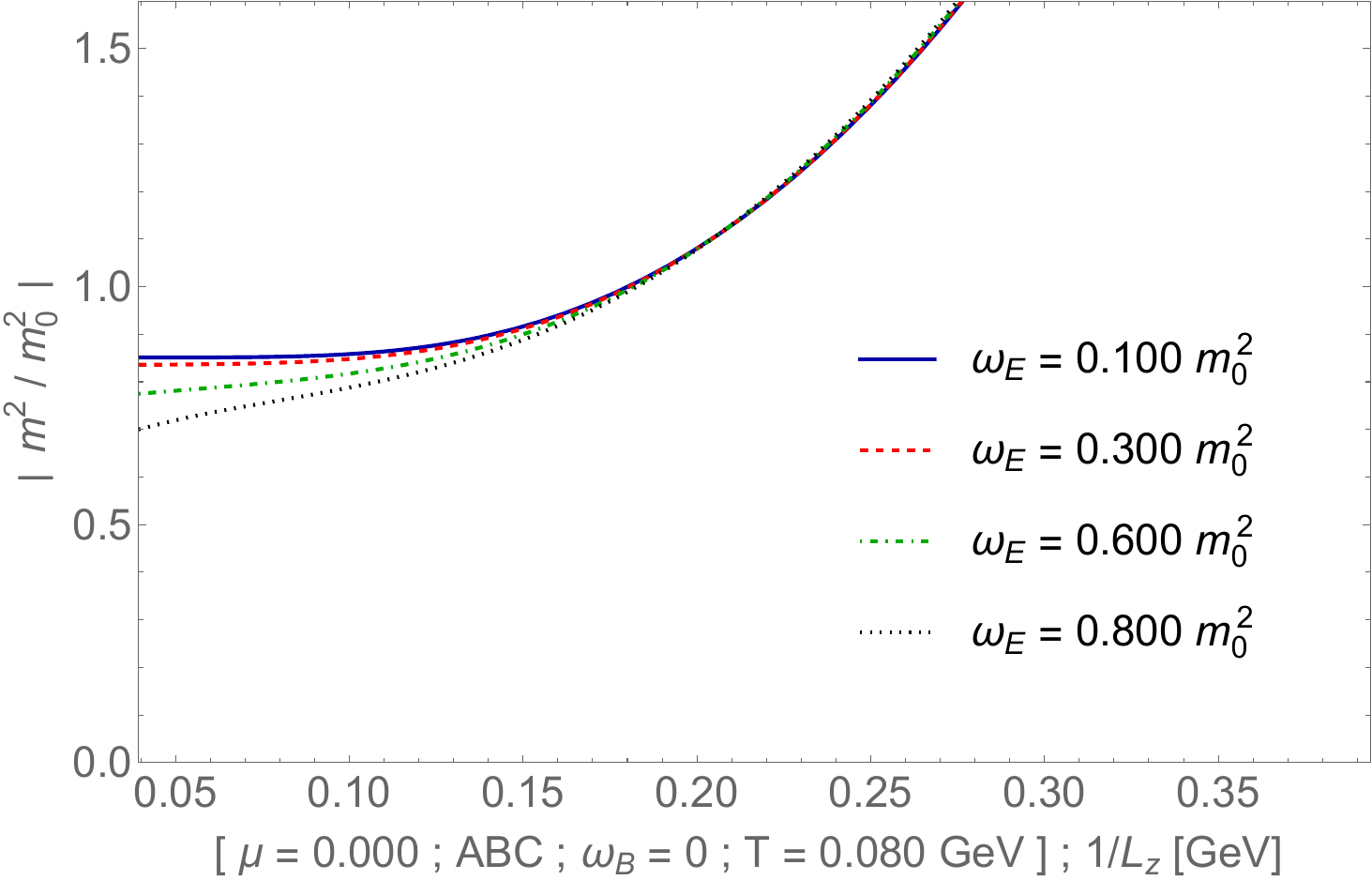}
\includegraphics[{width=6.49cm}]{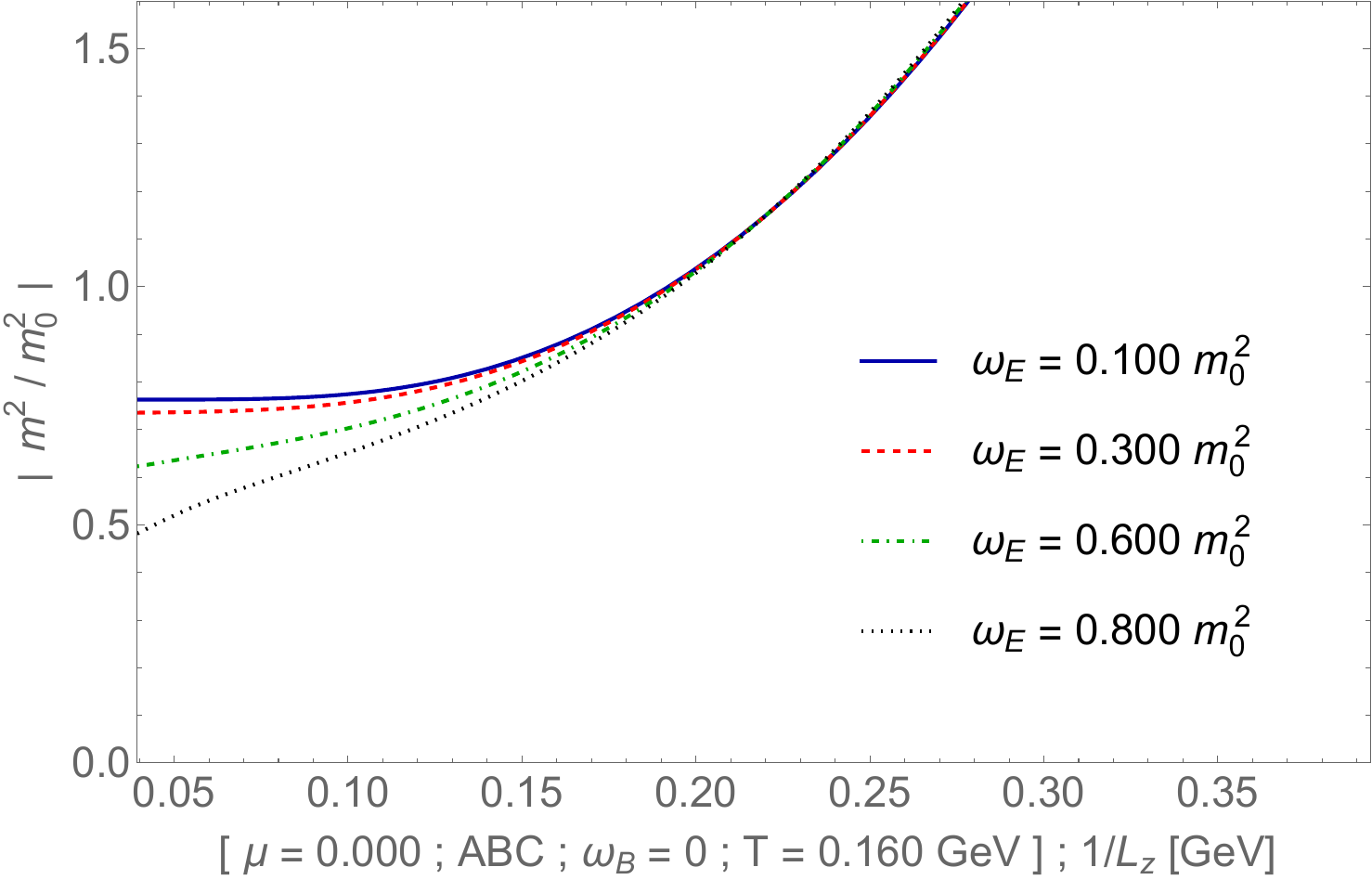} \\
\caption{~Corrected mass as a function of the inverse of length for several values of strength of the external electric field. At the left panels, we fixed the temperature of the film at $0.080~\mathrm{GeV}$. In the same way, at the right panels, we fixed $T$ at $0.160~\mathrm{GeV}$.}
\label{Fig7}
\end{figure}

Fig.~$8$ is plotted corrected mass as a function of the continuous range of finite size under a magnetic external field. For fixed temperatures, we observe qualitatively the same effects induced by finite chemical potential on the system when considering it in an electric background, namely: the finite $\mu$ diminishes the critical length of the system.
\begin{figure}
\centering
\includegraphics[{width=6.49cm}]{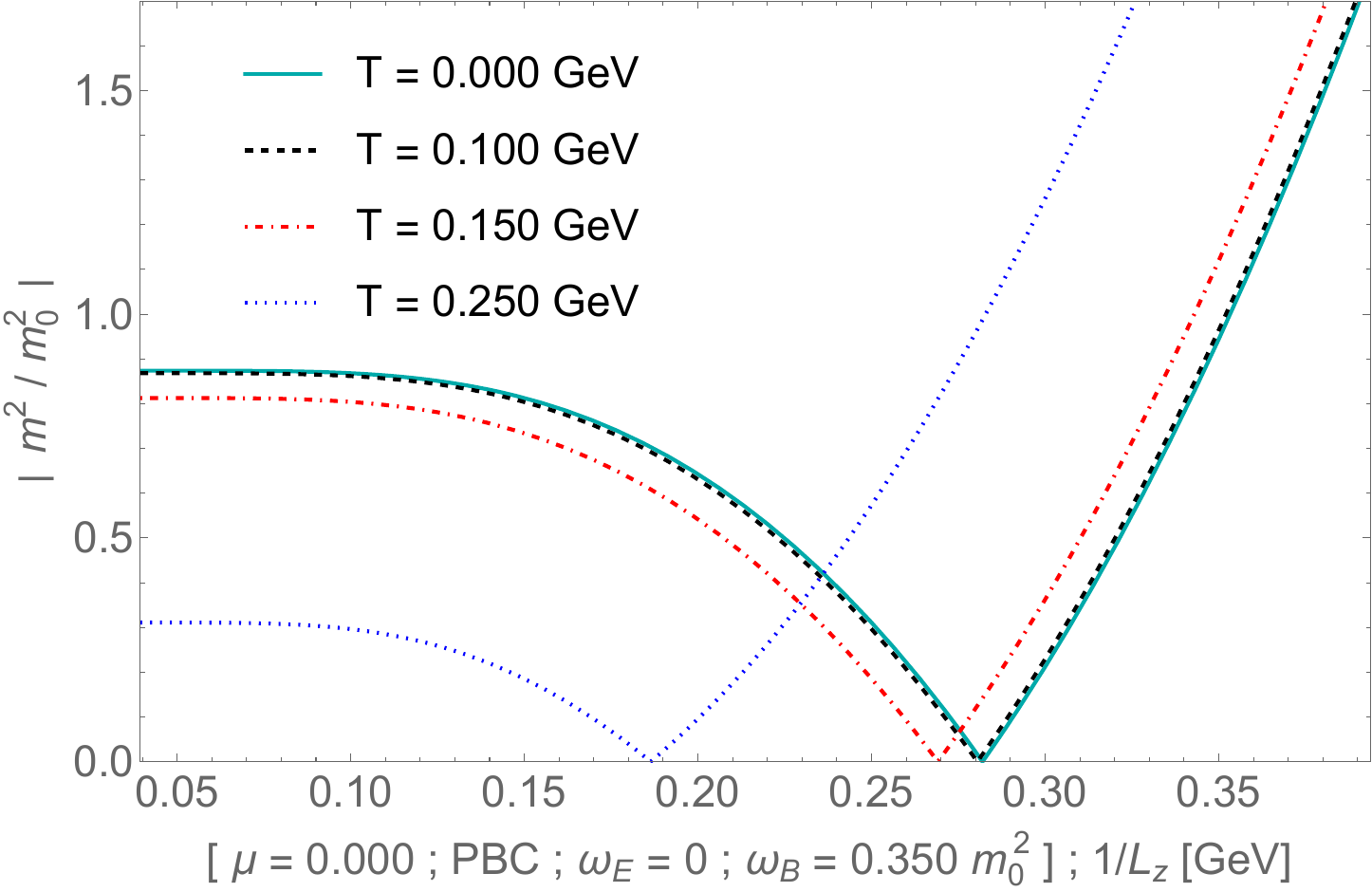}
\includegraphics[{width=6.49cm}]{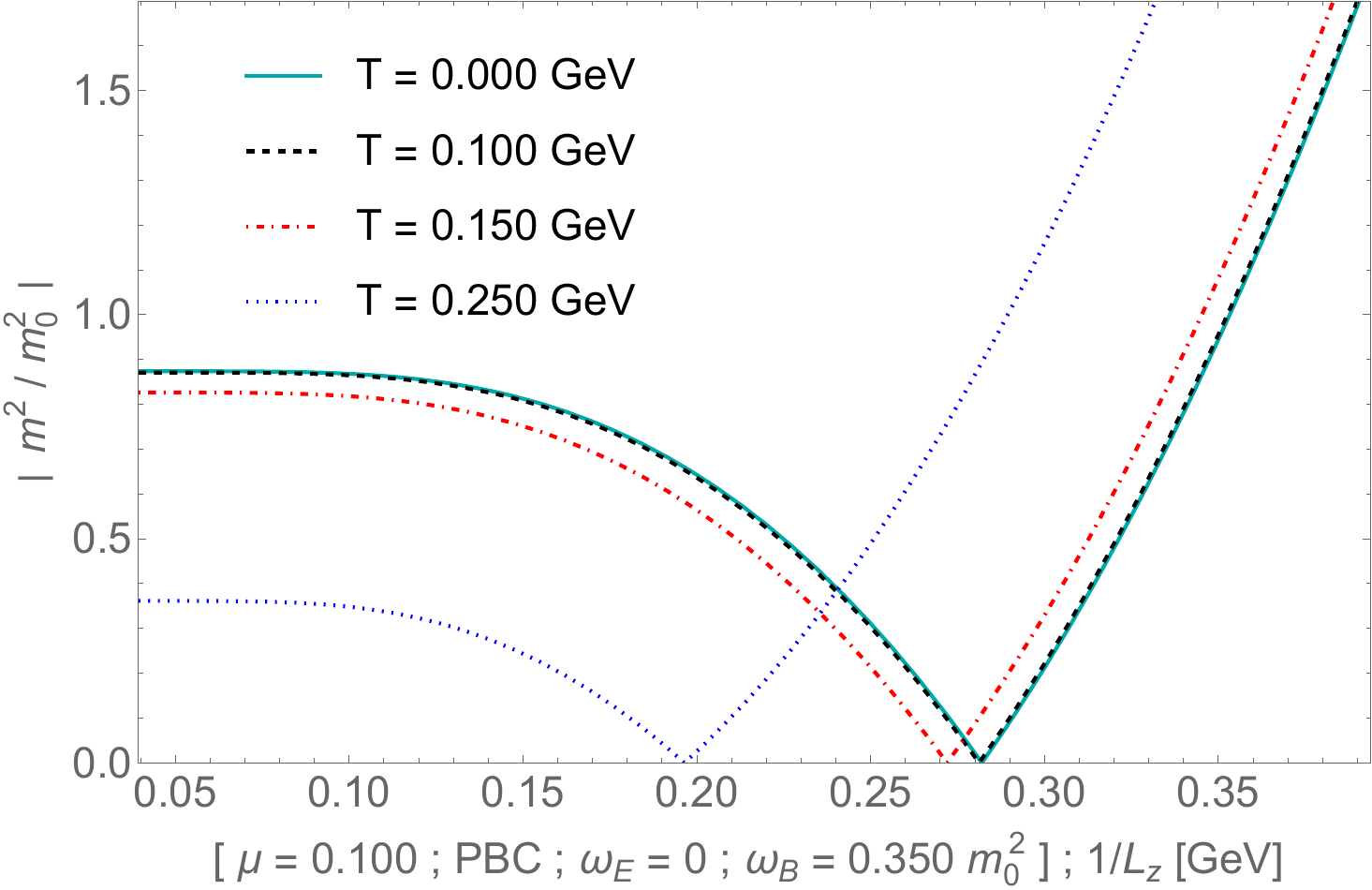} \\
\includegraphics[{width=6.49cm}]{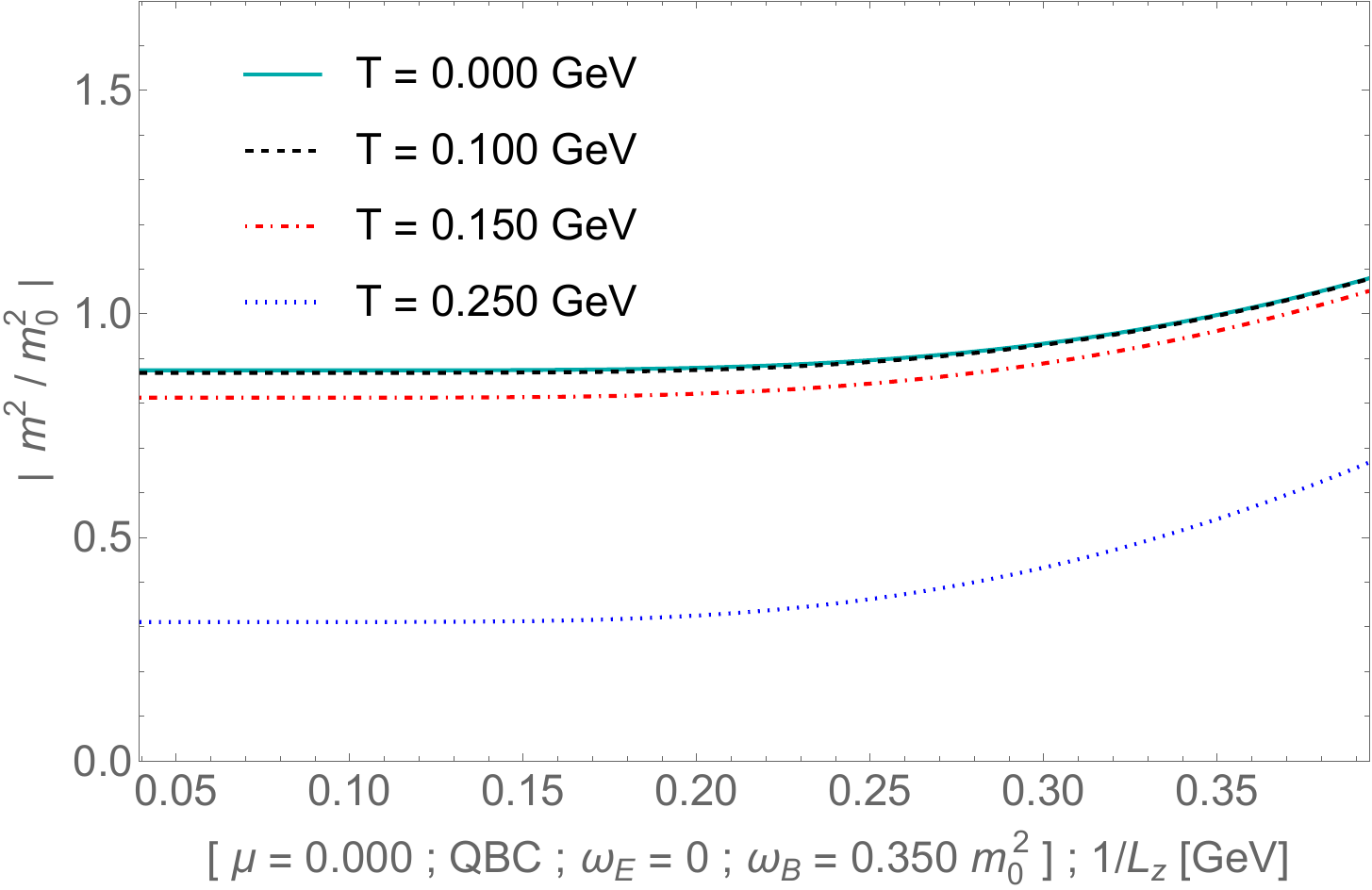} 
\includegraphics[{width=6.49cm}]{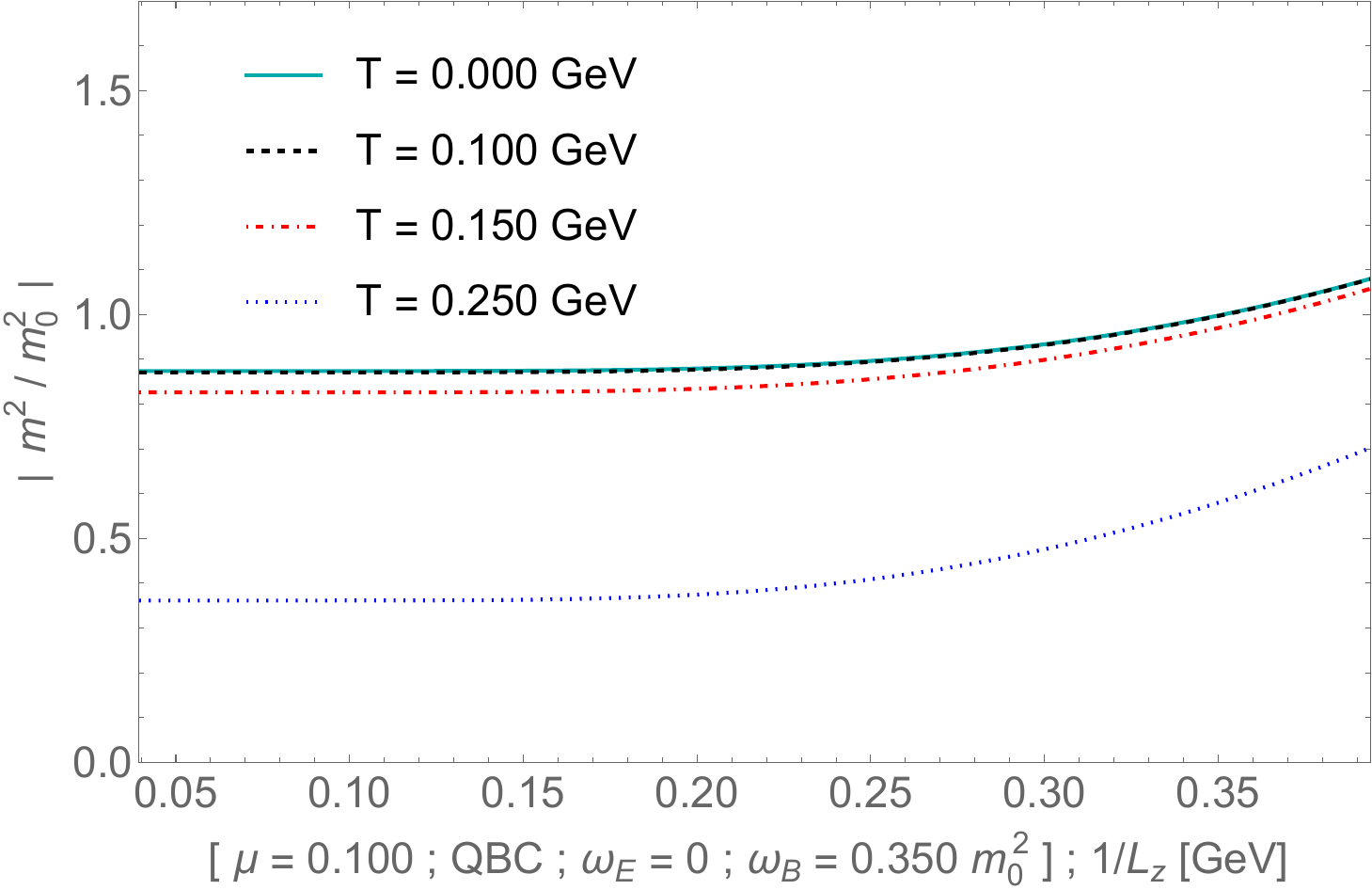} \\
\includegraphics[{width=6.49cm}]{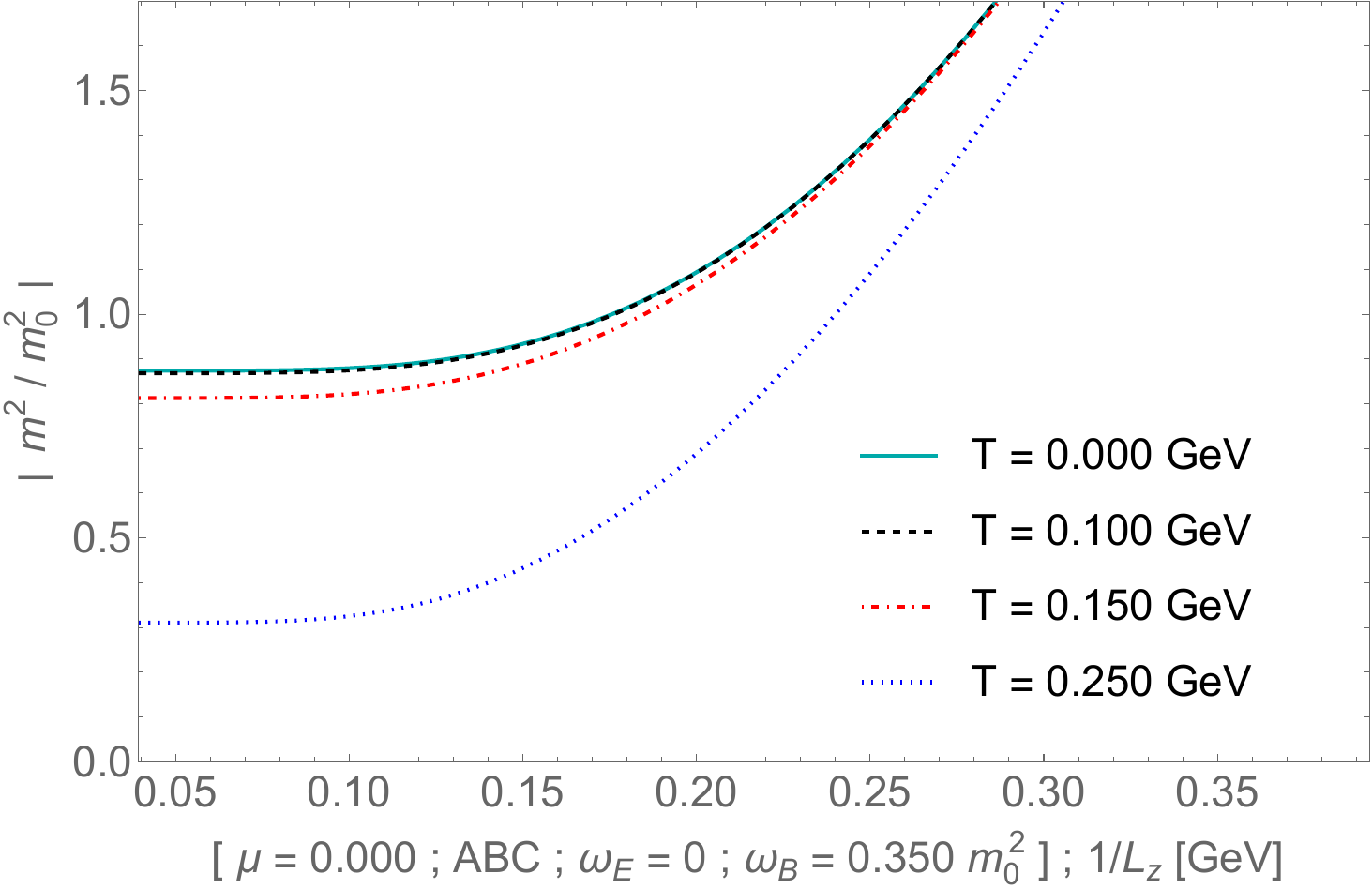}
\includegraphics[{width=6.49cm}]{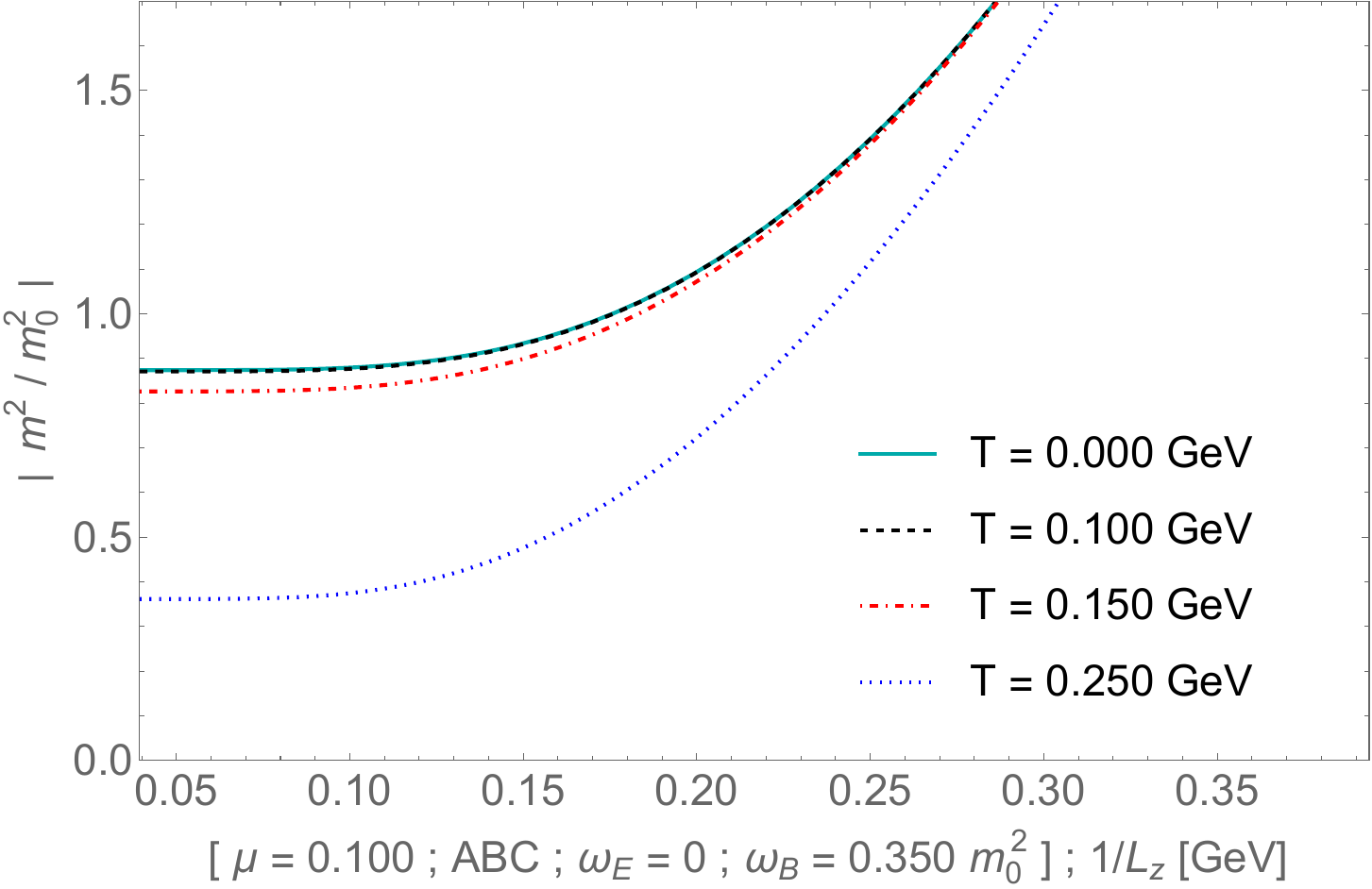} \\
\caption{~Corrected mass as a function of the inverse of length for several values of temperatures of the film (zero chemical potential at the left and finite chemical potential on the right).  We have PBC (top panel), QBC (middle panel), and ABC (bottom panel) boundary conditions on $z$-coordinate.}
\label{Fig8}
\end{figure}

The topological analogous of the MC phenomenon is depicted in Fig.~$9$. For fixed temperatures, we notice that higher intensities of magnetic external field make the system suffer phase transition at the highest $1/L_{c}$, for the PBC case. The cases QBC and ABC have not induced phase transition on the model for the temperatures fixed.
\begin{figure}
\centering
\includegraphics[{width=6.49cm}]{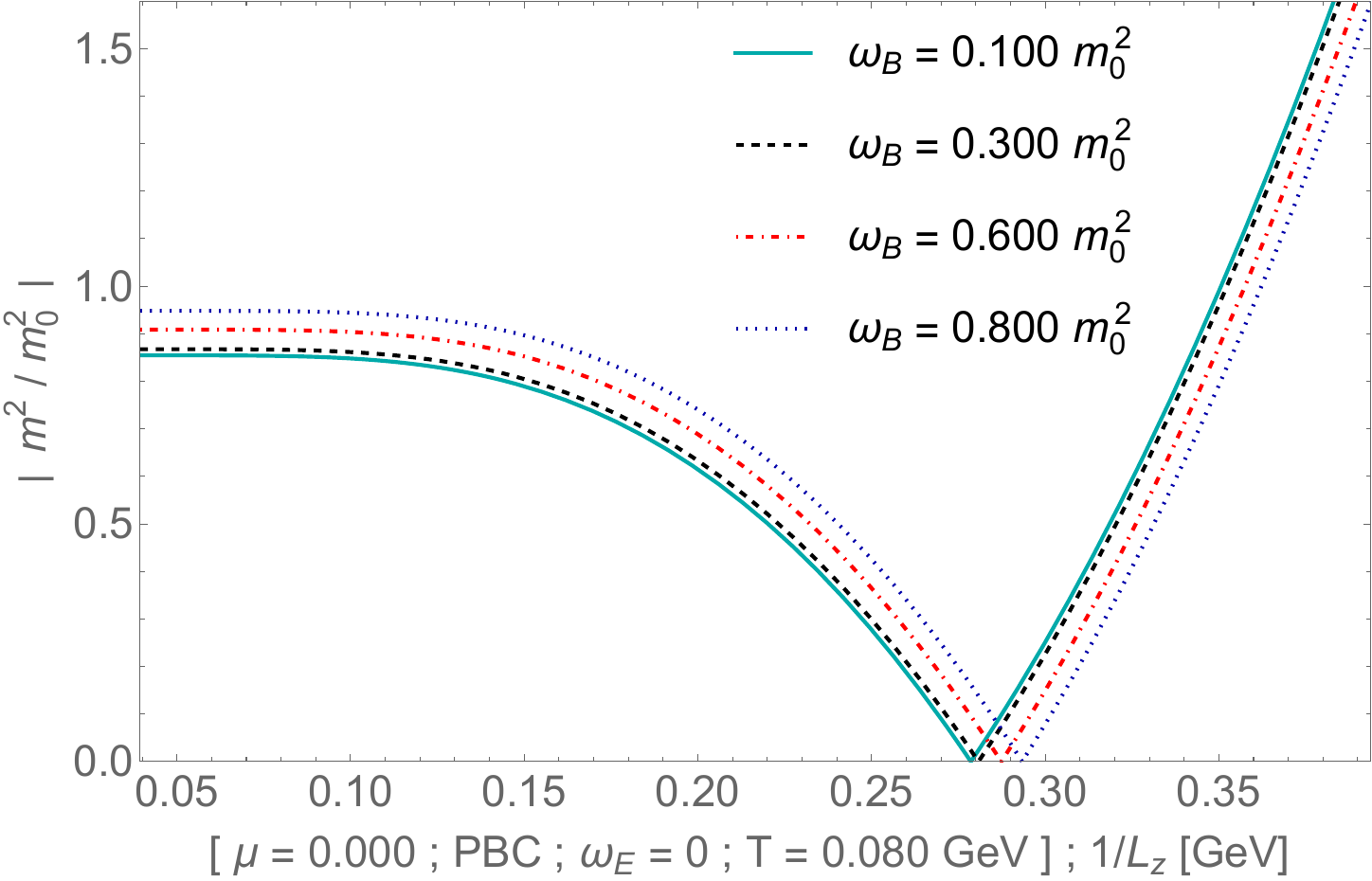}
\includegraphics[{width=6.49cm}]{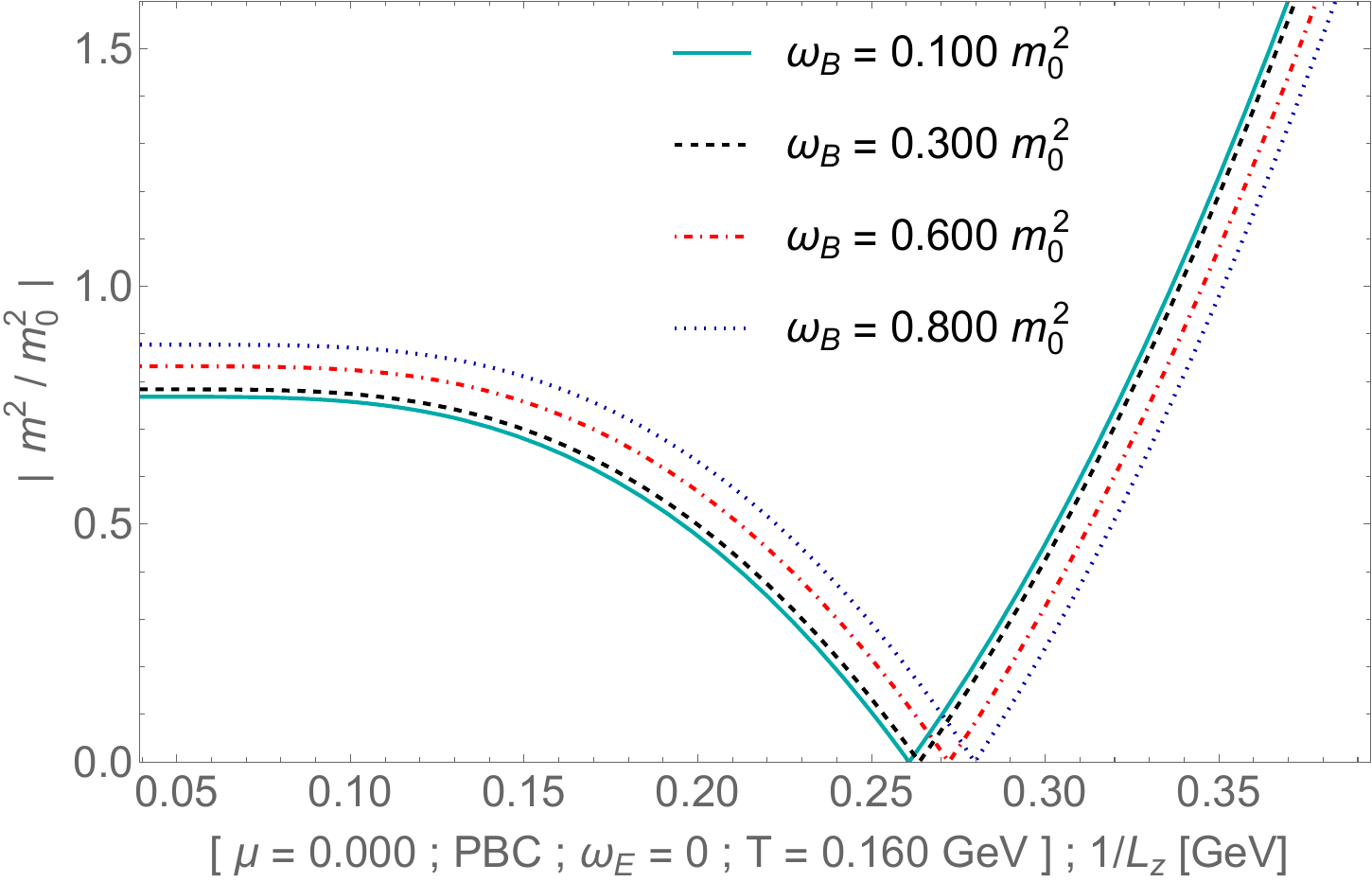} \\
\includegraphics[{width=6.49cm}]{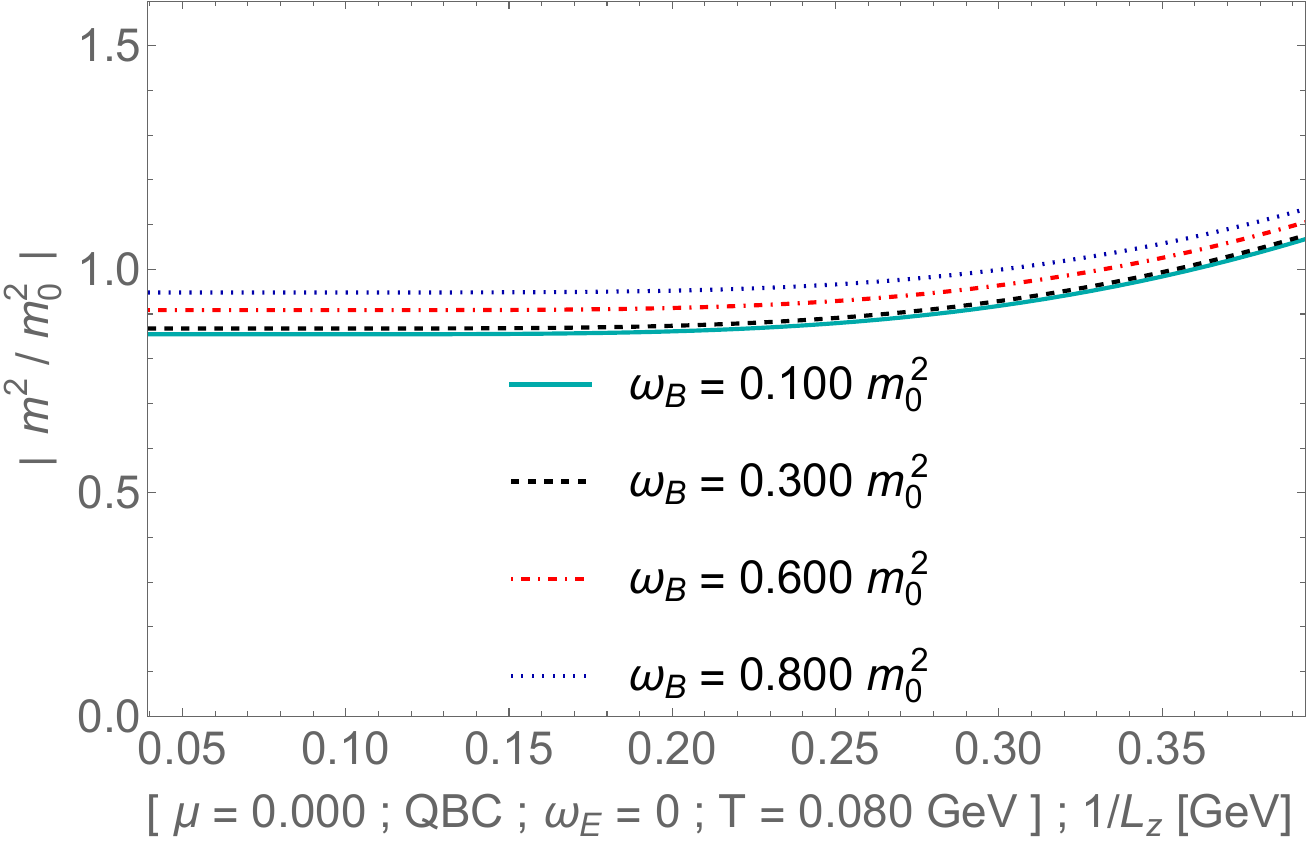}
\includegraphics[{width=6.49cm}]{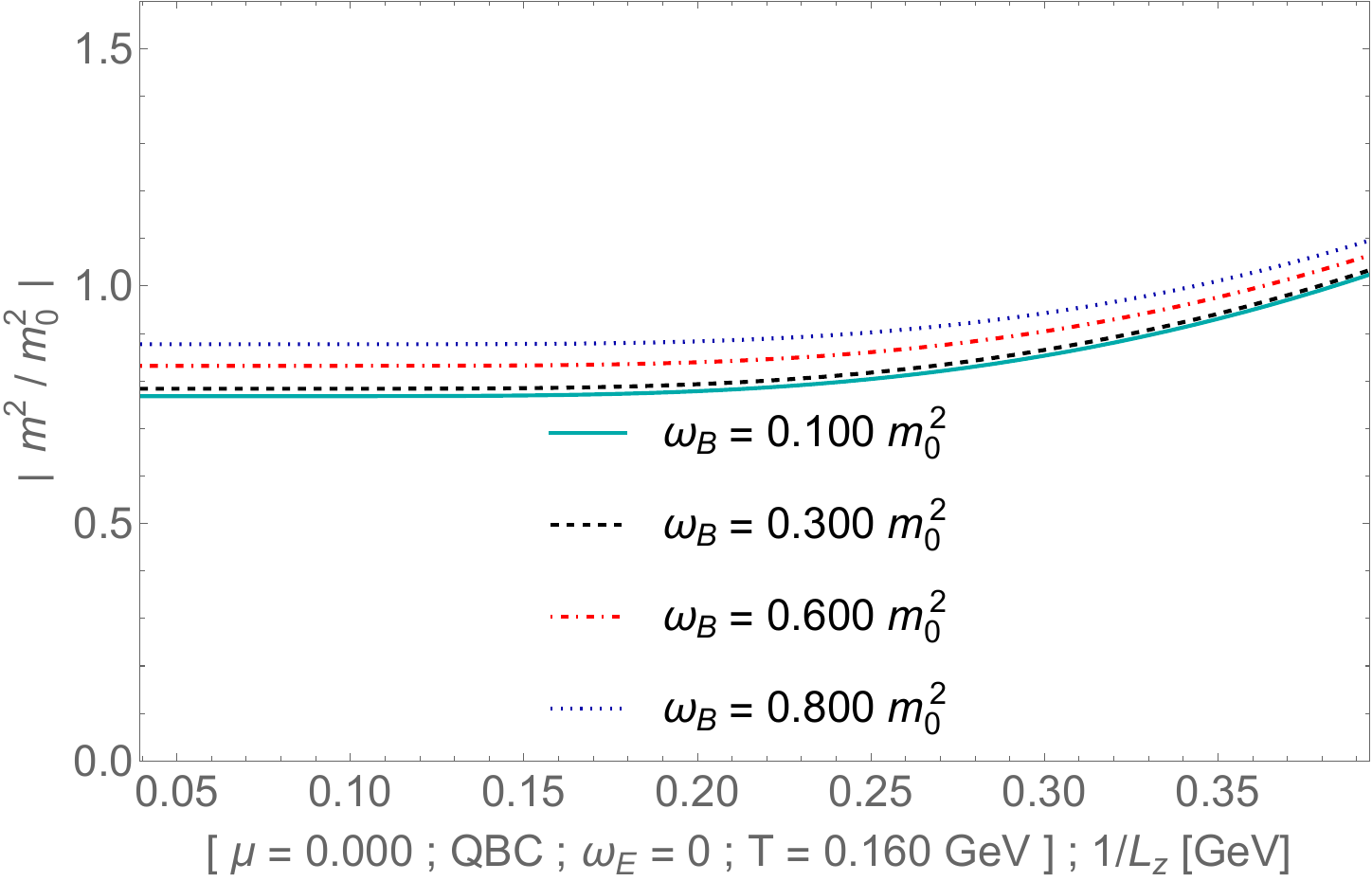} \\
\includegraphics[{width=6.49cm}]{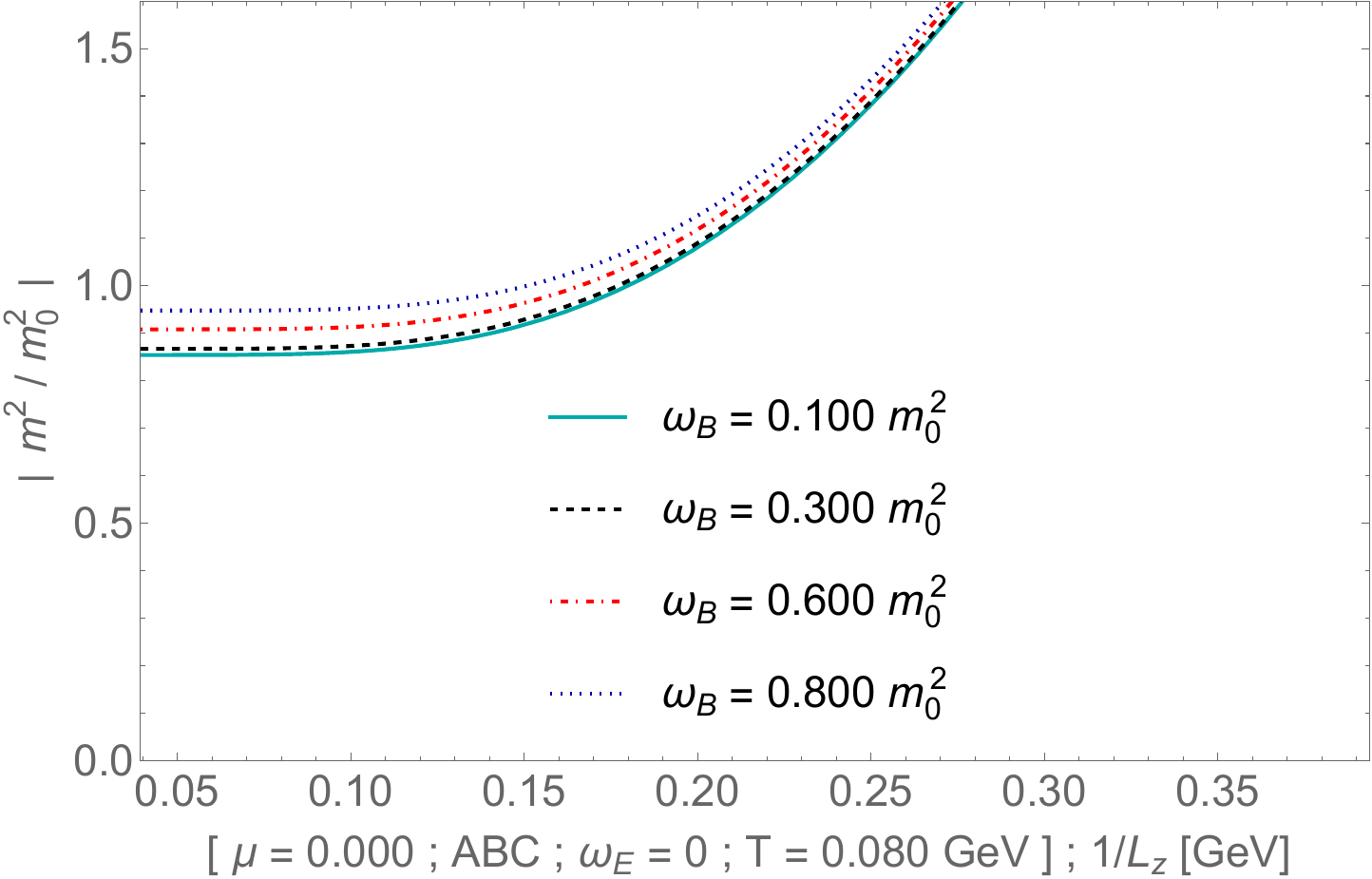}
\includegraphics[{width=6.49cm}]{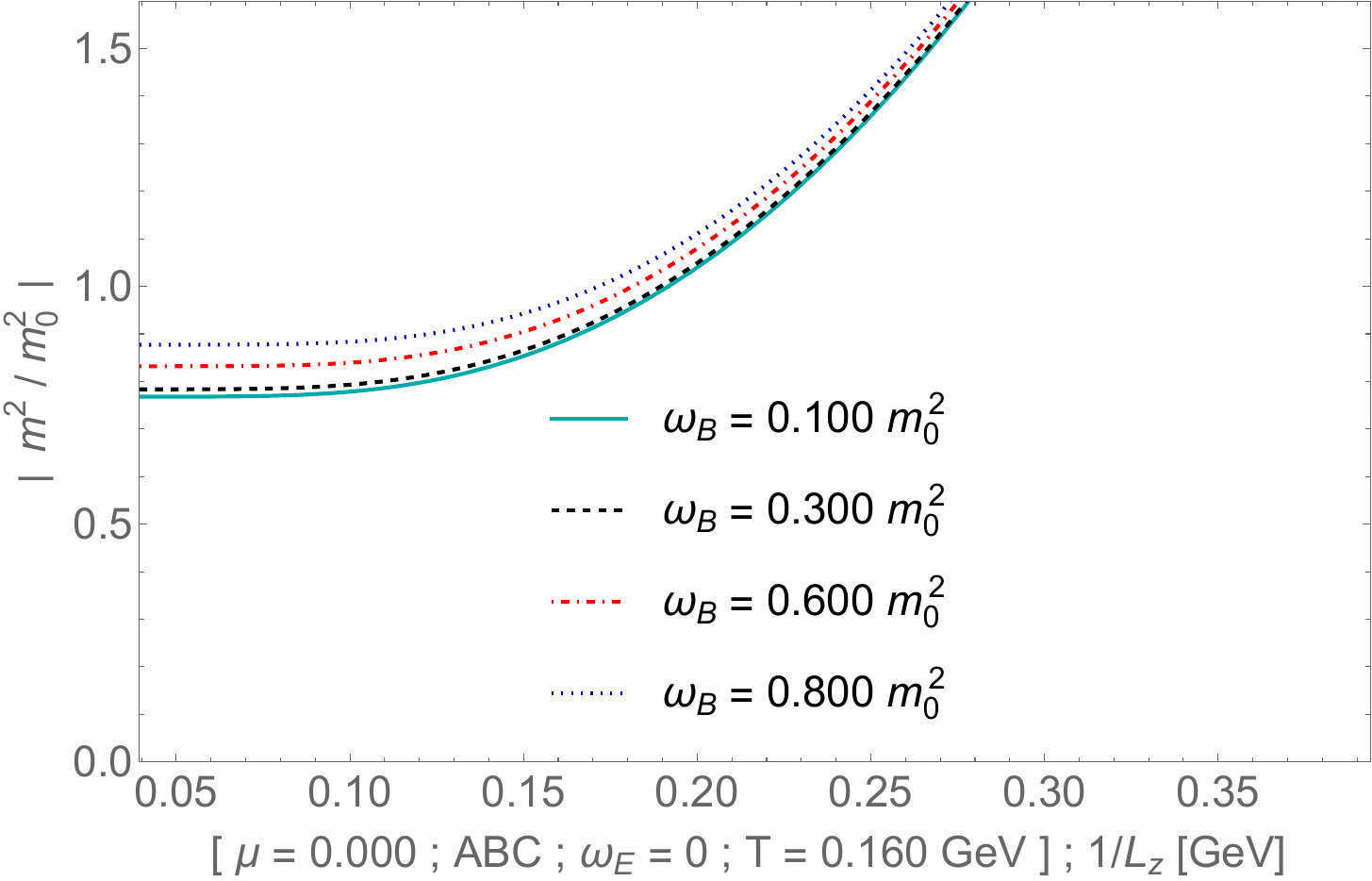} \\
\caption{~Corrected mass as a function of the inverse of length for several values of strength of the external magnetic field. At the left panels, we fixed the temperature of the film at $0.080~\mathrm{GeV}$. In the same way, at the right panels, we fixed $T$ at $0.160~\mathrm{GeV}$.}
\label{Fig9}
\end{figure}
\subsection{Free energy density of the system}

\hspace{0.53cm}Now we shall study the free energy density of the system \textit{à la Ginzburg-Landau model}. The analogous of Ginzburg--Landau free energy density is given by
\begin{eqnarray}
\mathcal{F} \equiv m^{2} \,\varphi_{c}^{2} + \lambda_{eff} \,\varphi_{c}^{4},
\end{eqnarray}
where the vacuum expectation value, as declared, is defined by $ \varphi_{c} \equiv\langle 0\,|\sqrt{\Phi^{*}\Phi} \,|0\rangle$. Thus, the classical field plays the role of the order parameter of theory and the phase transition takes place in phase space where the mass parameter vanishes~\cite{Landau,LeBellac}. Therefore, the critical temperature is such that $m(T_{c},\mu,L_{z},\mu_{z}.E,B)=0$.

In Figs.~$10$ and $12$, we display the dimensionless free energy density behavior under several finite sizes of the system as a function of the rescaled classical field $\varphi_{c} \,\rightarrow\, \varphi_{c}/m_{0}$. Again, the effect caused by periodic boundary conditions over the system frontier is a decrease in the critical temperature, when the finite size $L_z$ diminishes, at fixed external fields. However, the quasiperiodic and antiperiodic boundary conditions have the opposite effect, i.e., QBC and ABC induced higher critical temperatures in the phase structure of the system, while its thickness declined, keeping the external fields fixed.

Figs.~$11$ and $13$ suggest one more time the distinct behaviors that electric and magnetic backgrounds have on the bosonic thermal system.  Strong augmentation in electric fields tends to increase the critical temperature, considering the PBC on the $z$-direction (electric catalysis). On the other hand, these strong electric fields contribute to the decline of critical temperature in QBC and ABC cases. In contrast, greater intensities in the external magnetic field always promote an increase in the system's critical temperature.
\begin{figure}
\centering
\includegraphics[{width=6.49cm}]{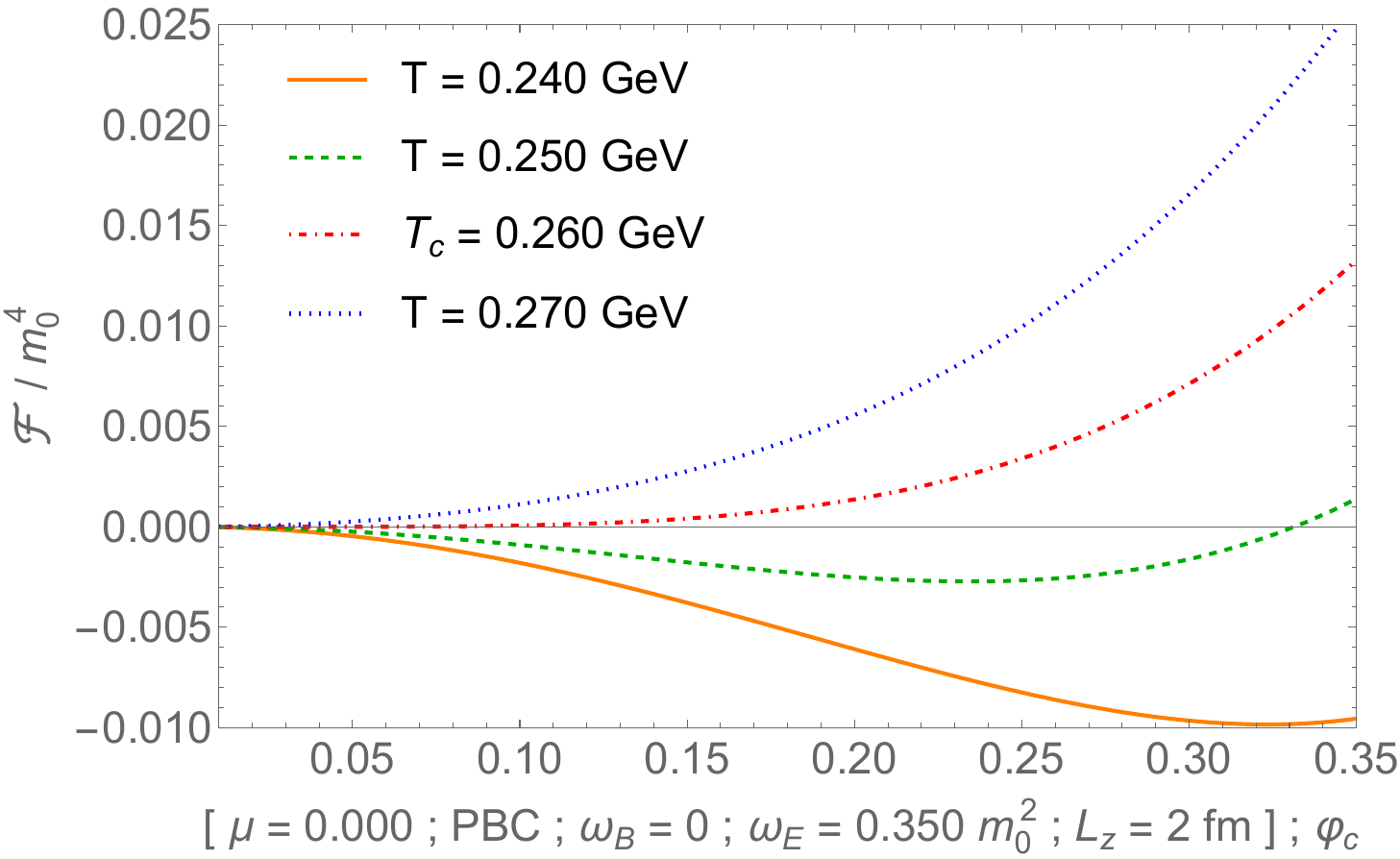}
\includegraphics[{width=6.49cm}]{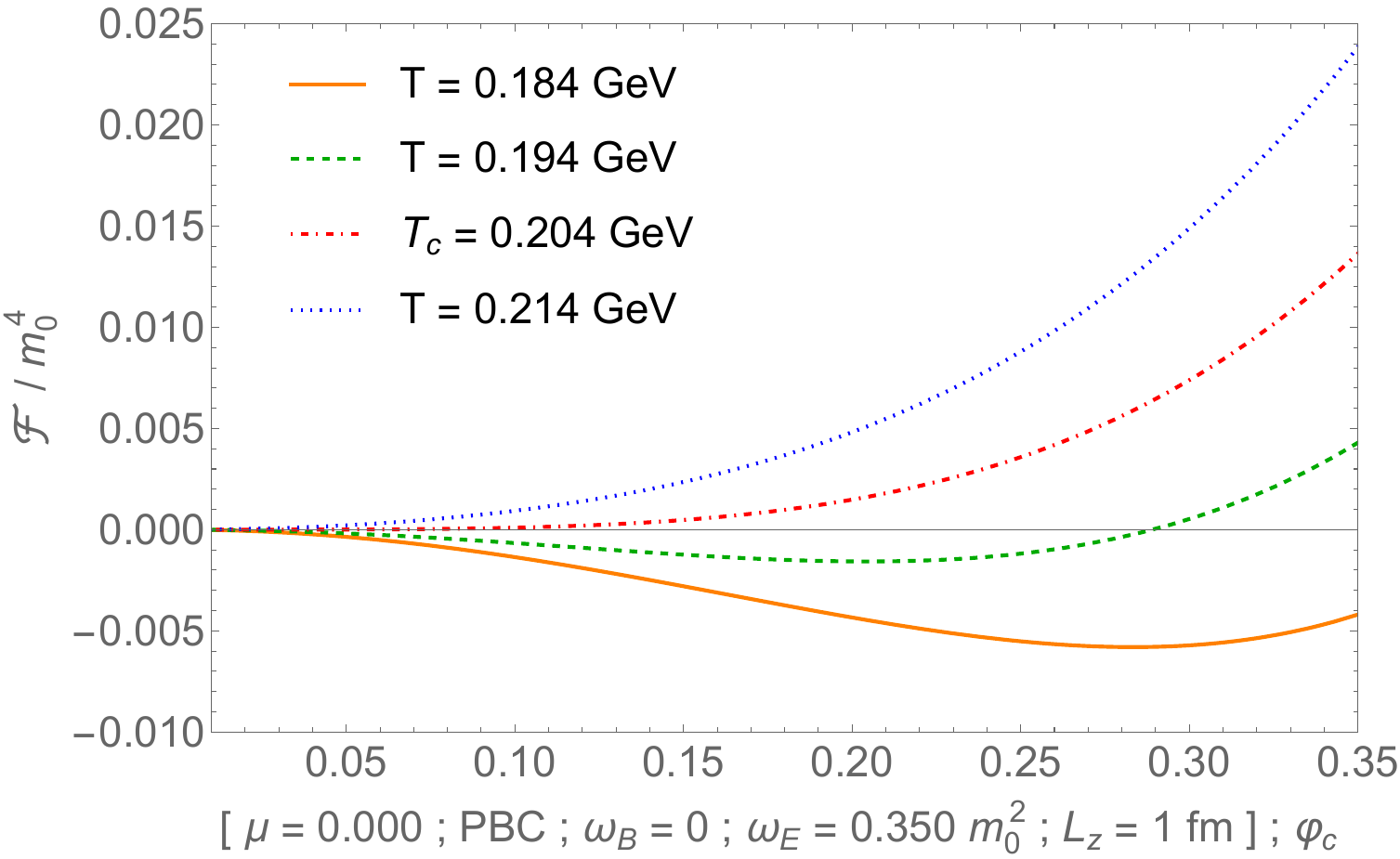} \\
\includegraphics[{width=6.49cm}]{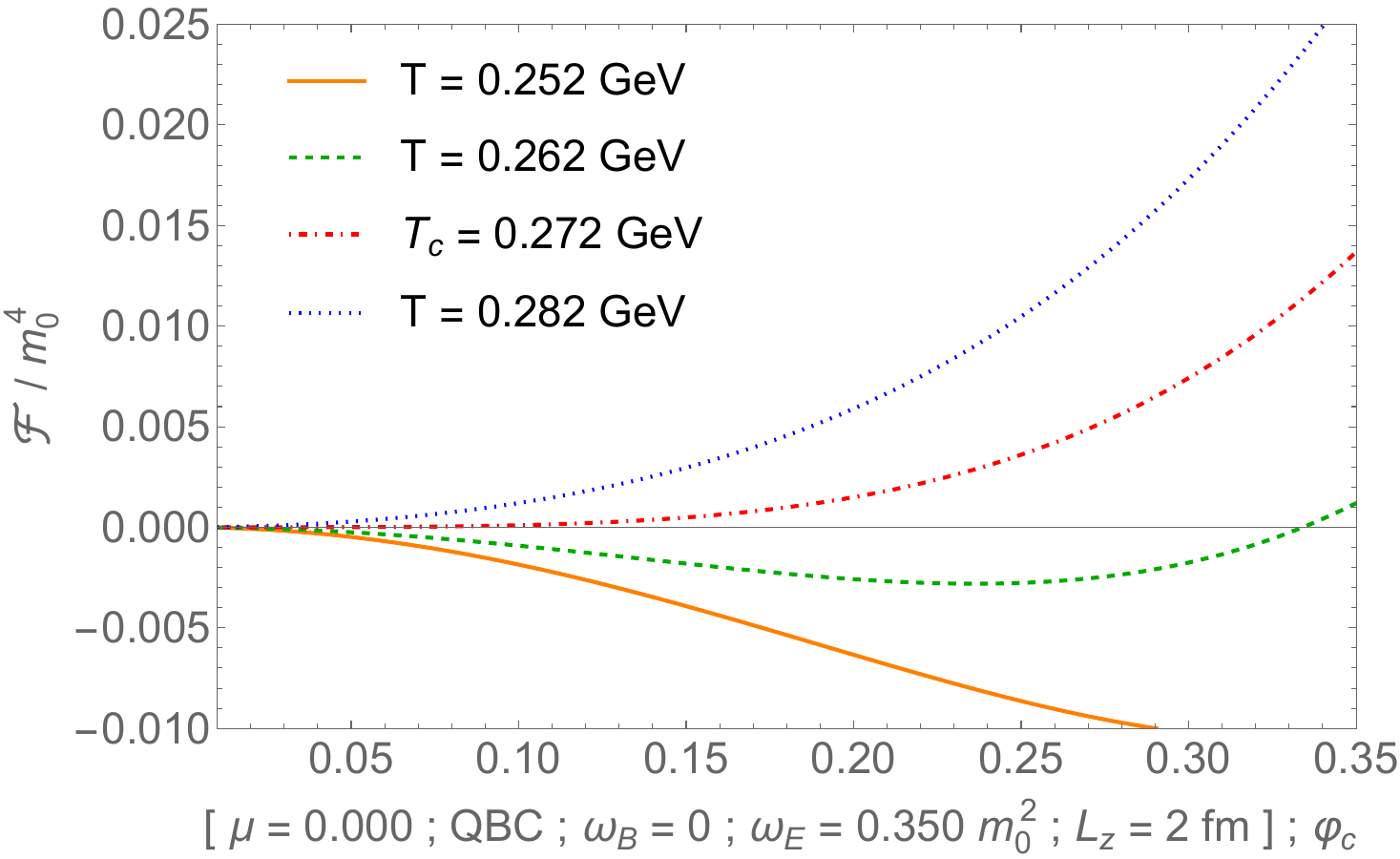}
\includegraphics[{width=6.49cm}]{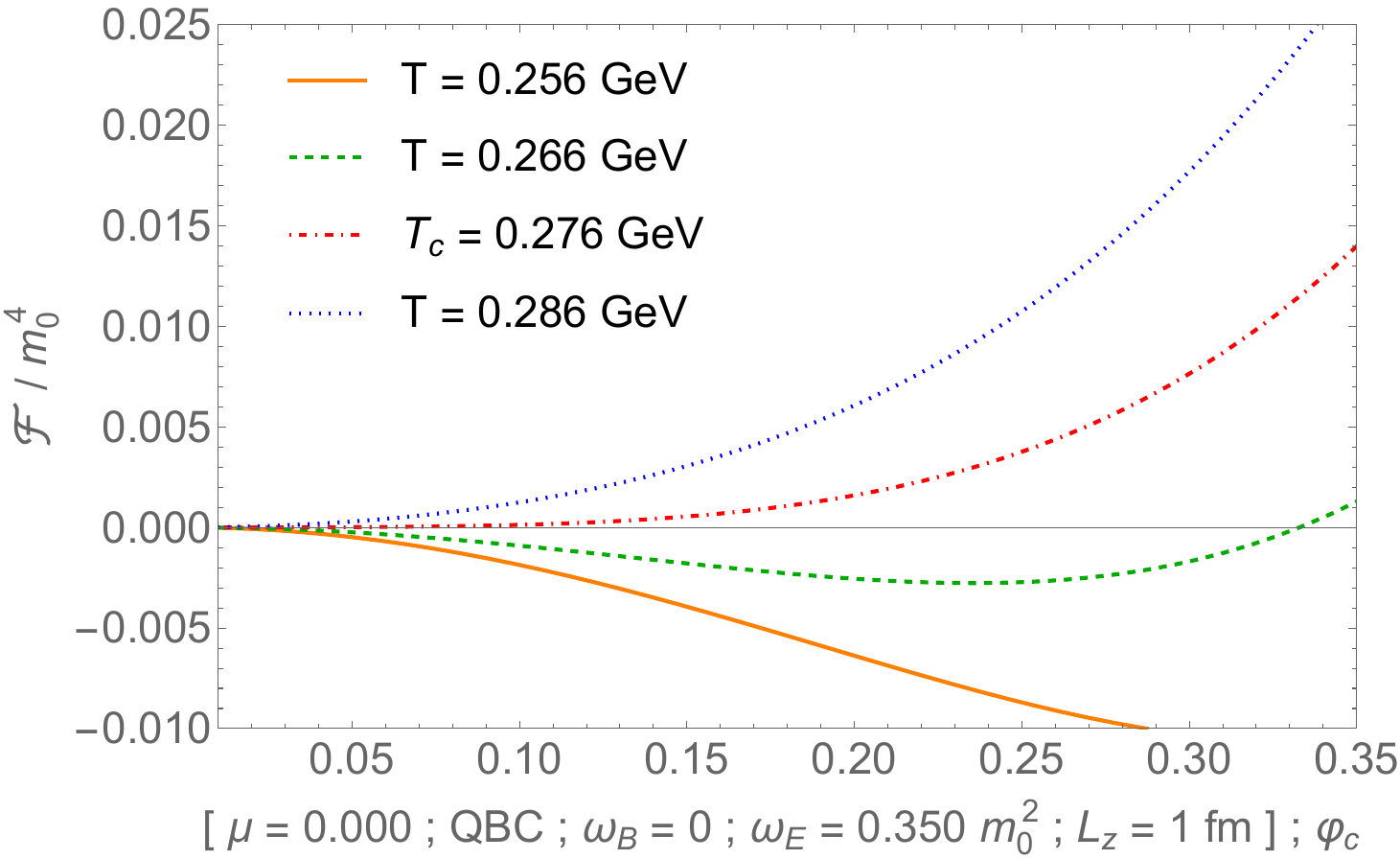} \\
\includegraphics[{width=6.49cm}]{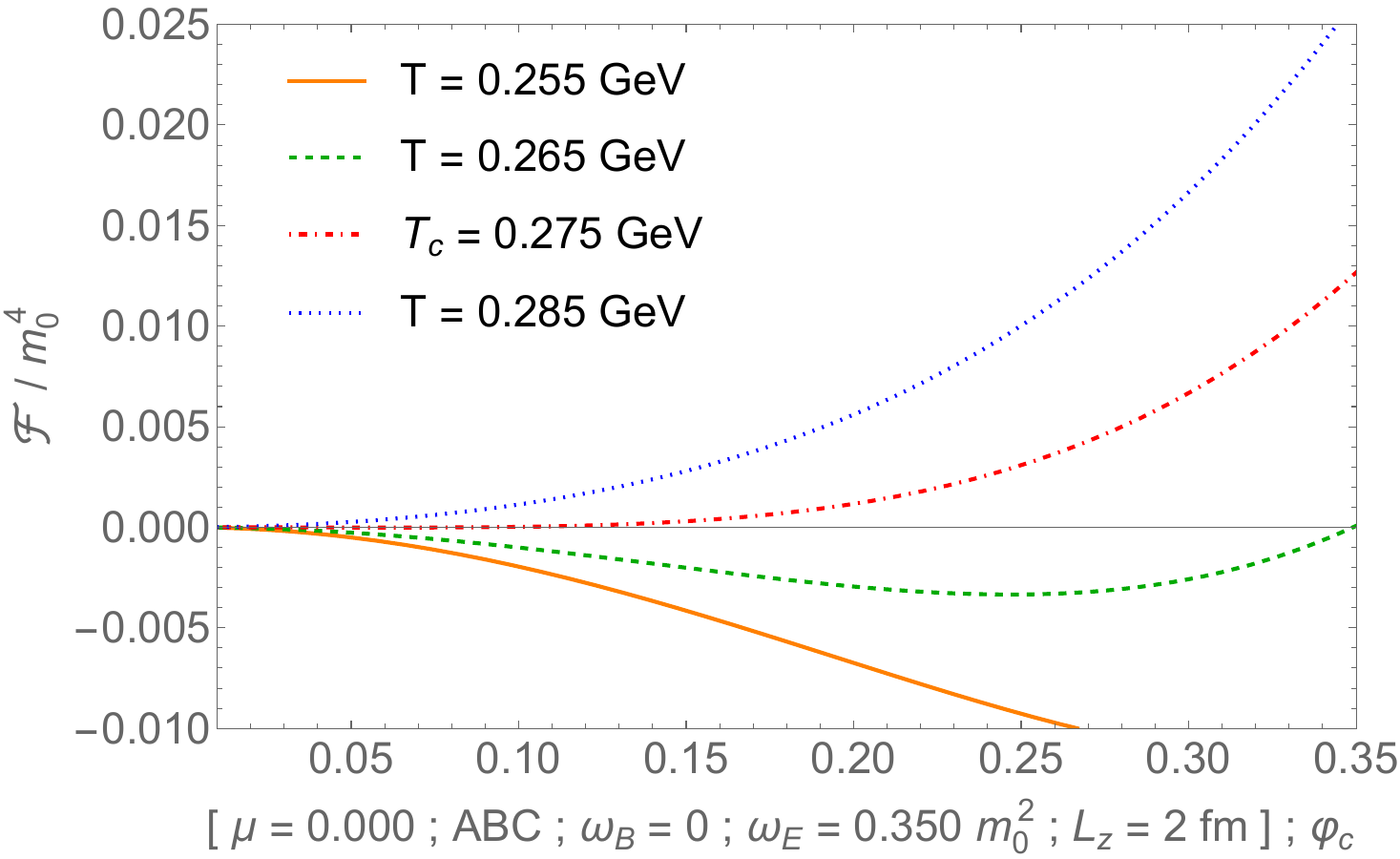}
\includegraphics[{width=6.49cm}]{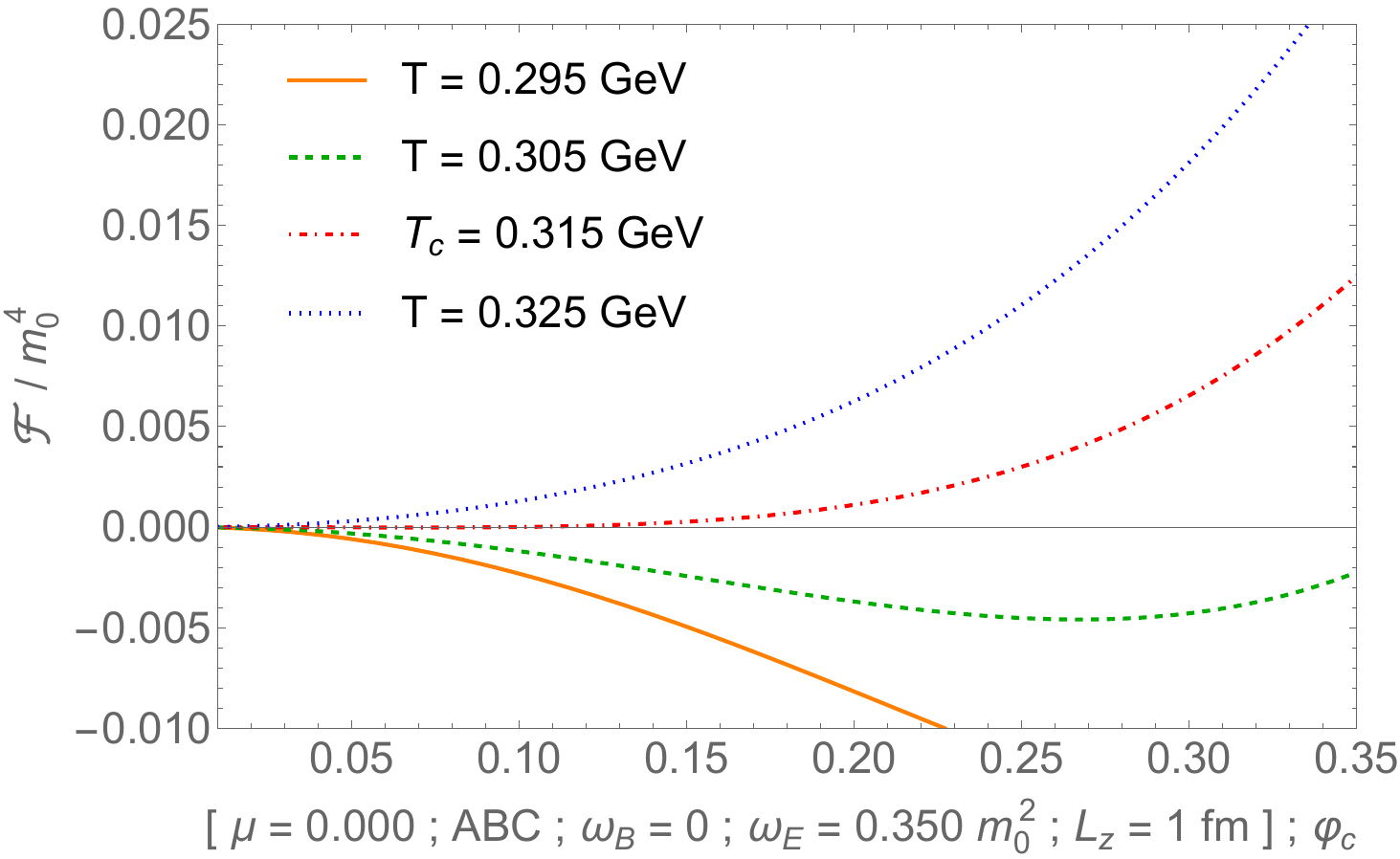} \\
\caption{~Influence of boundary conditions and external electric field over the phase transition of the system. On the left, we fix $L_{z}=2~\mathrm{fm}$. On the right, we have $L_{z}=1~\mathrm{fm}$.}
\label{Fig10}
\end{figure}
\begin{figure}
\centering
\includegraphics[{width=6.49cm}]{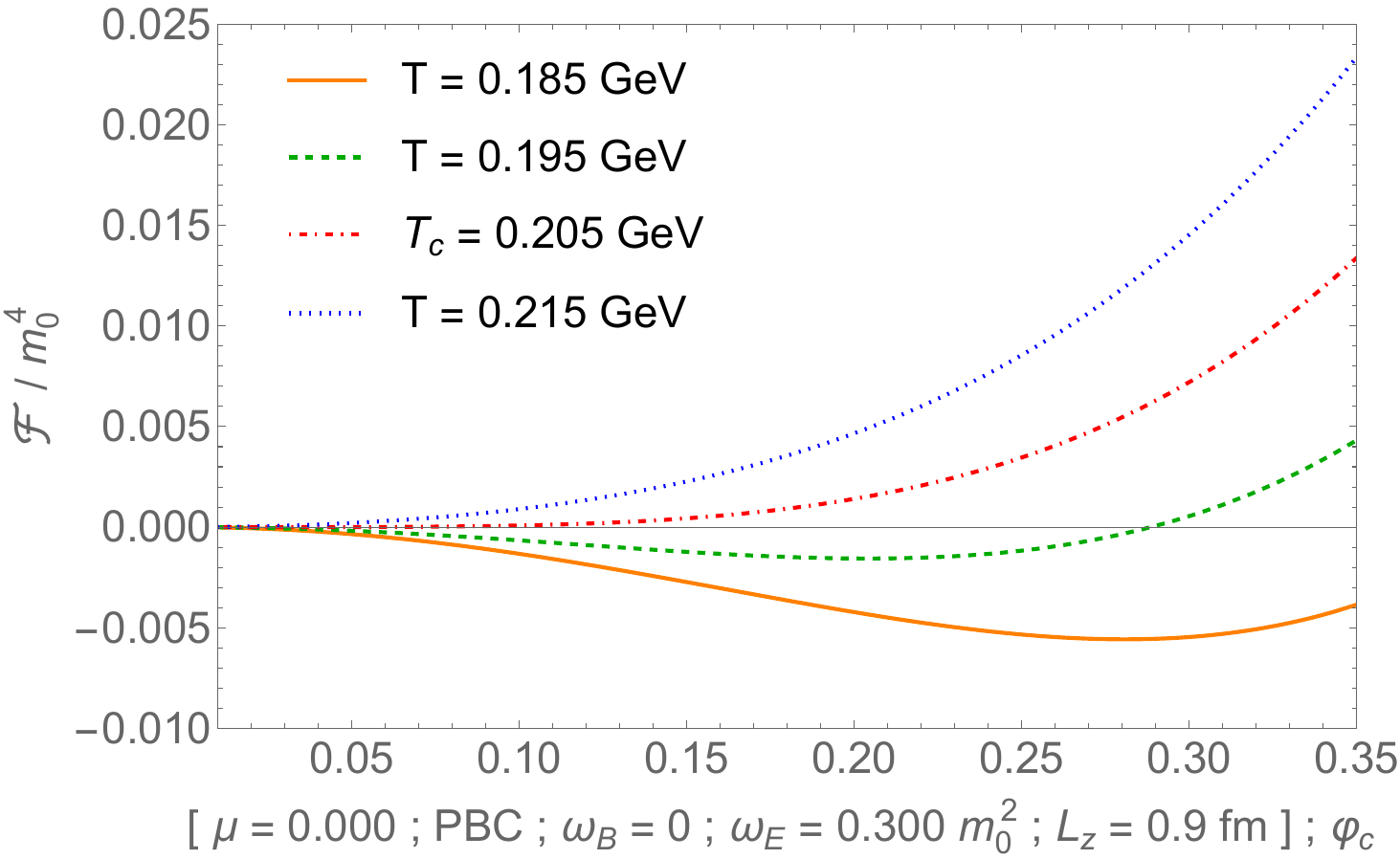}
\includegraphics[{width=6.49cm}]{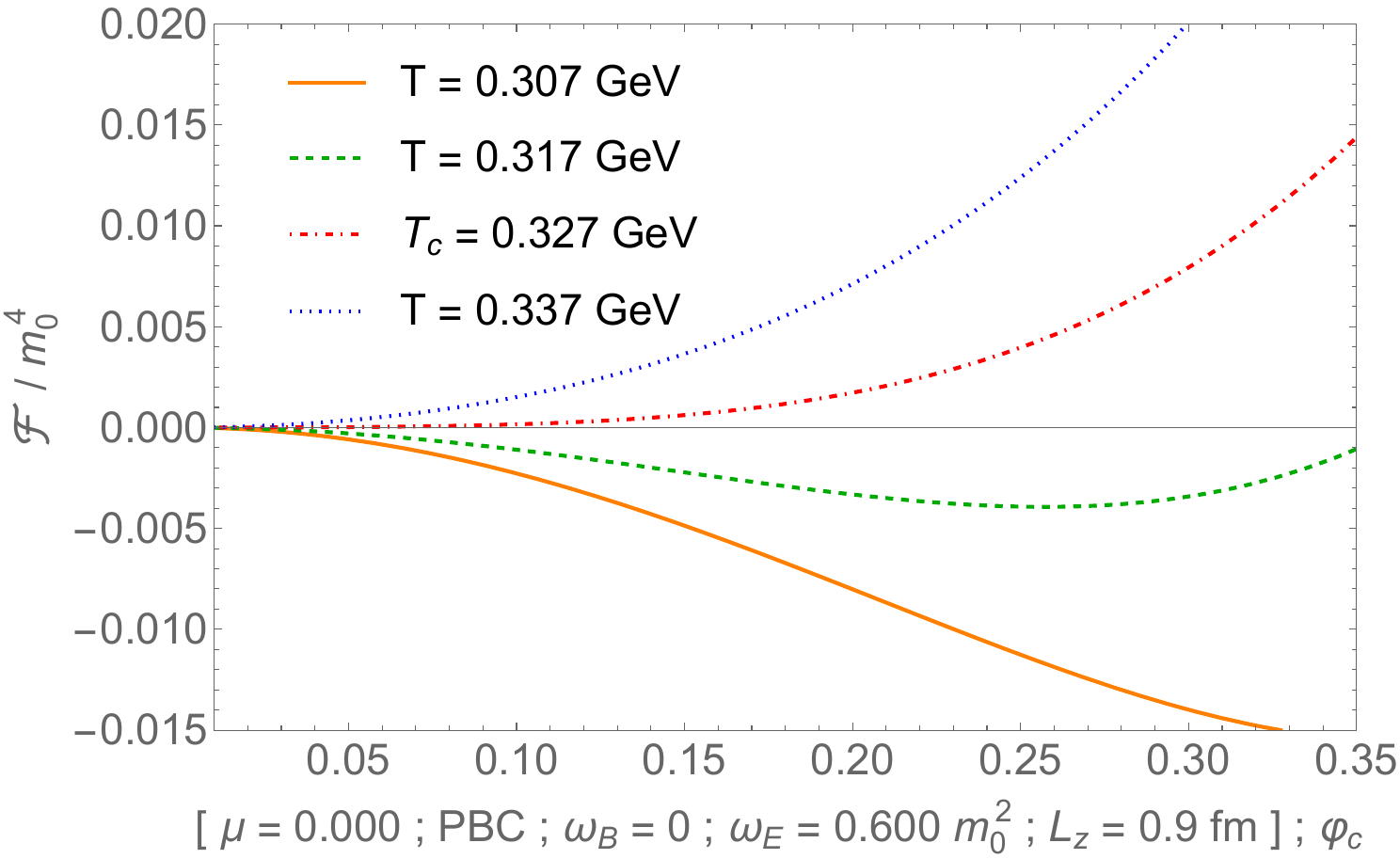} \\
\includegraphics[{width=6.49cm}]{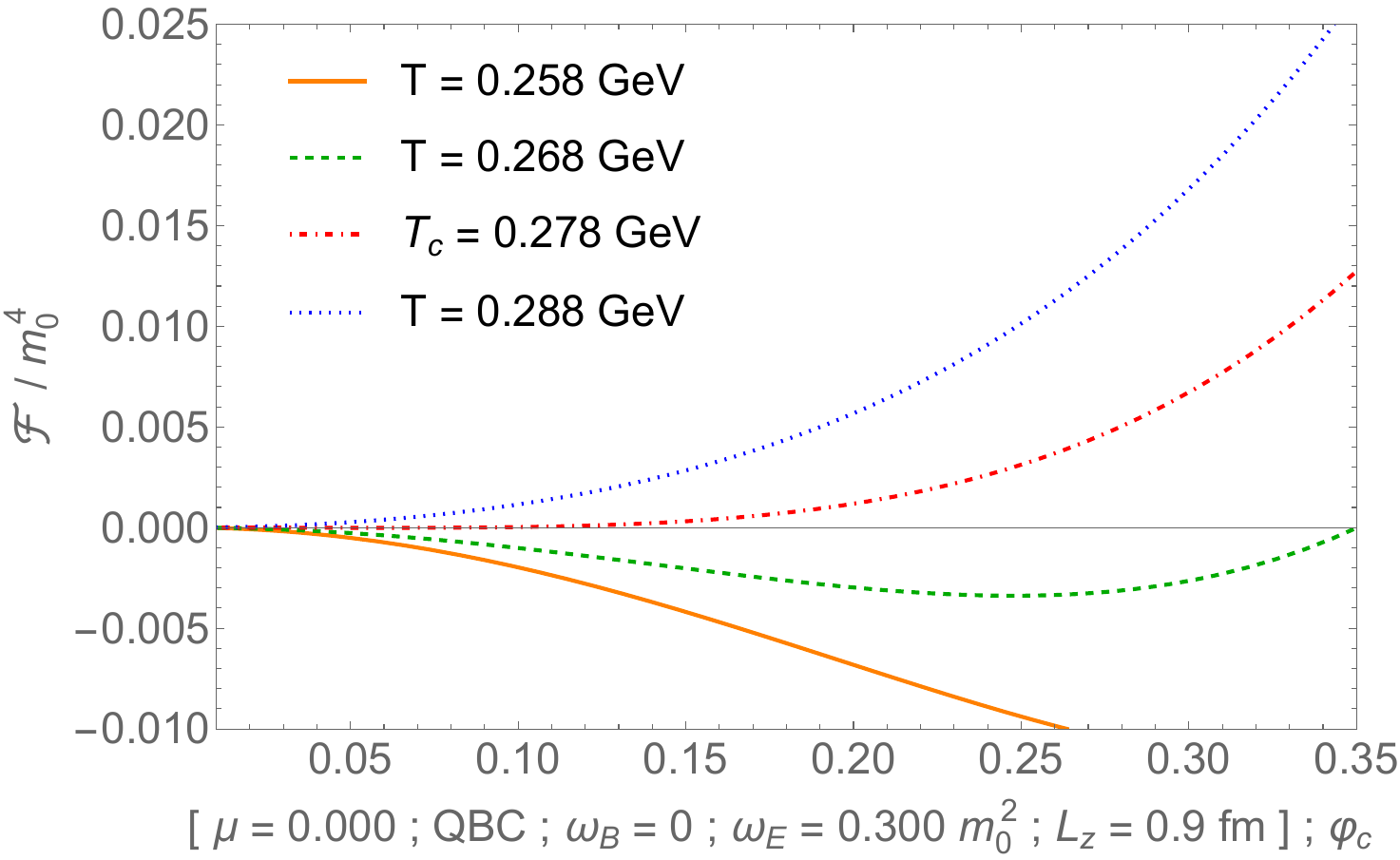}
\includegraphics[{width=6.49cm}]{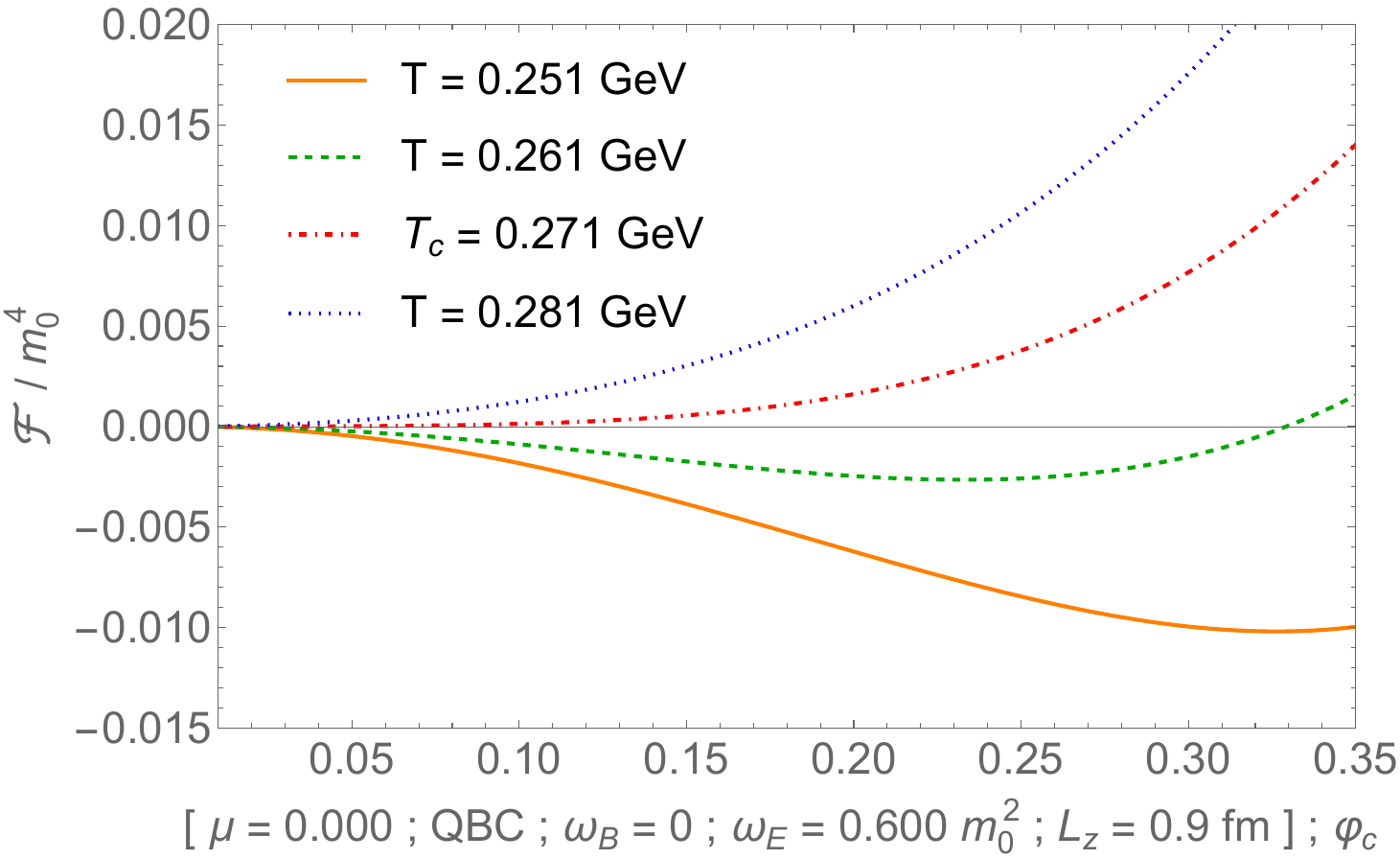} \\
\includegraphics[{width=6.49cm}]{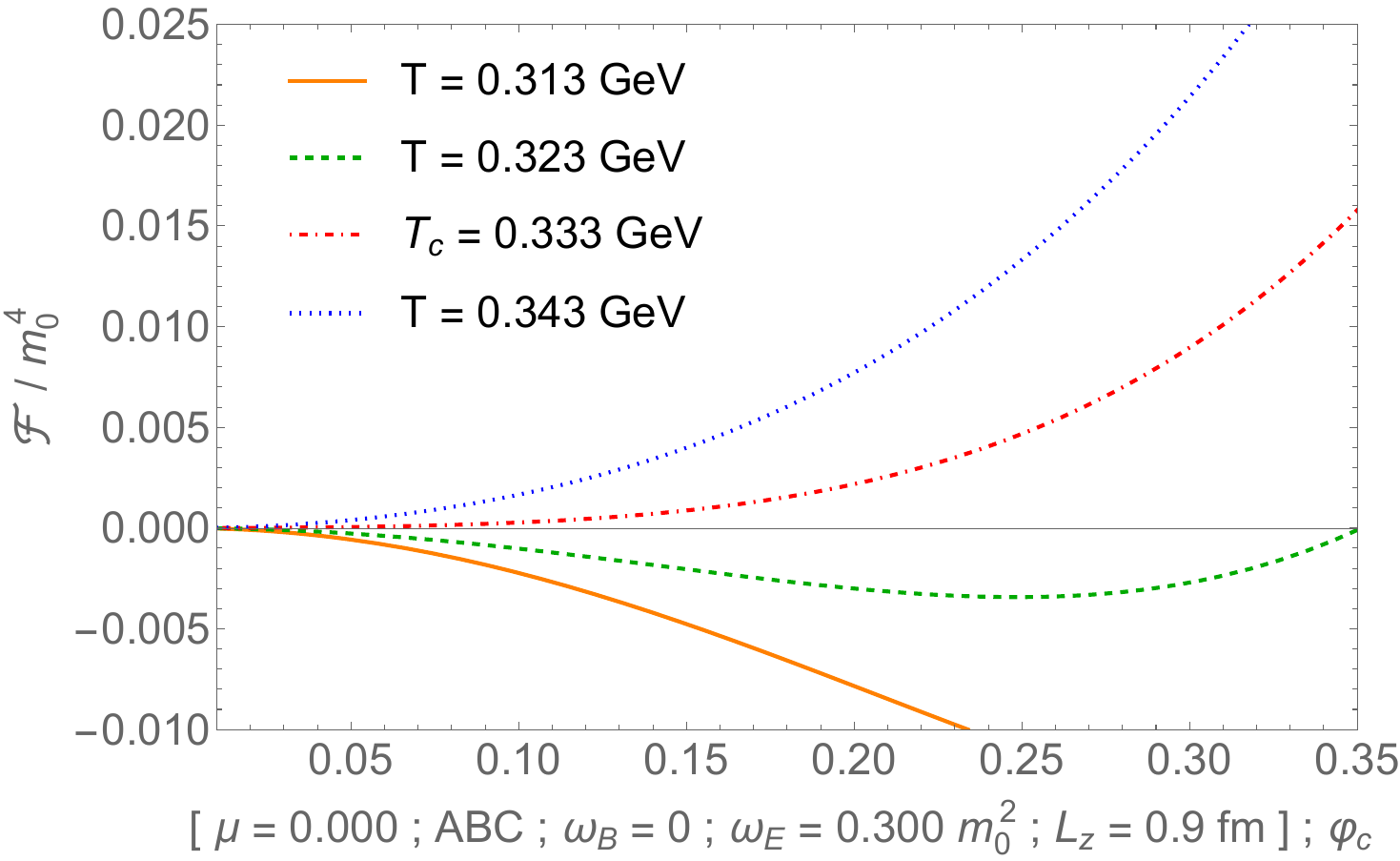}
\includegraphics[{width=6.49cm}]{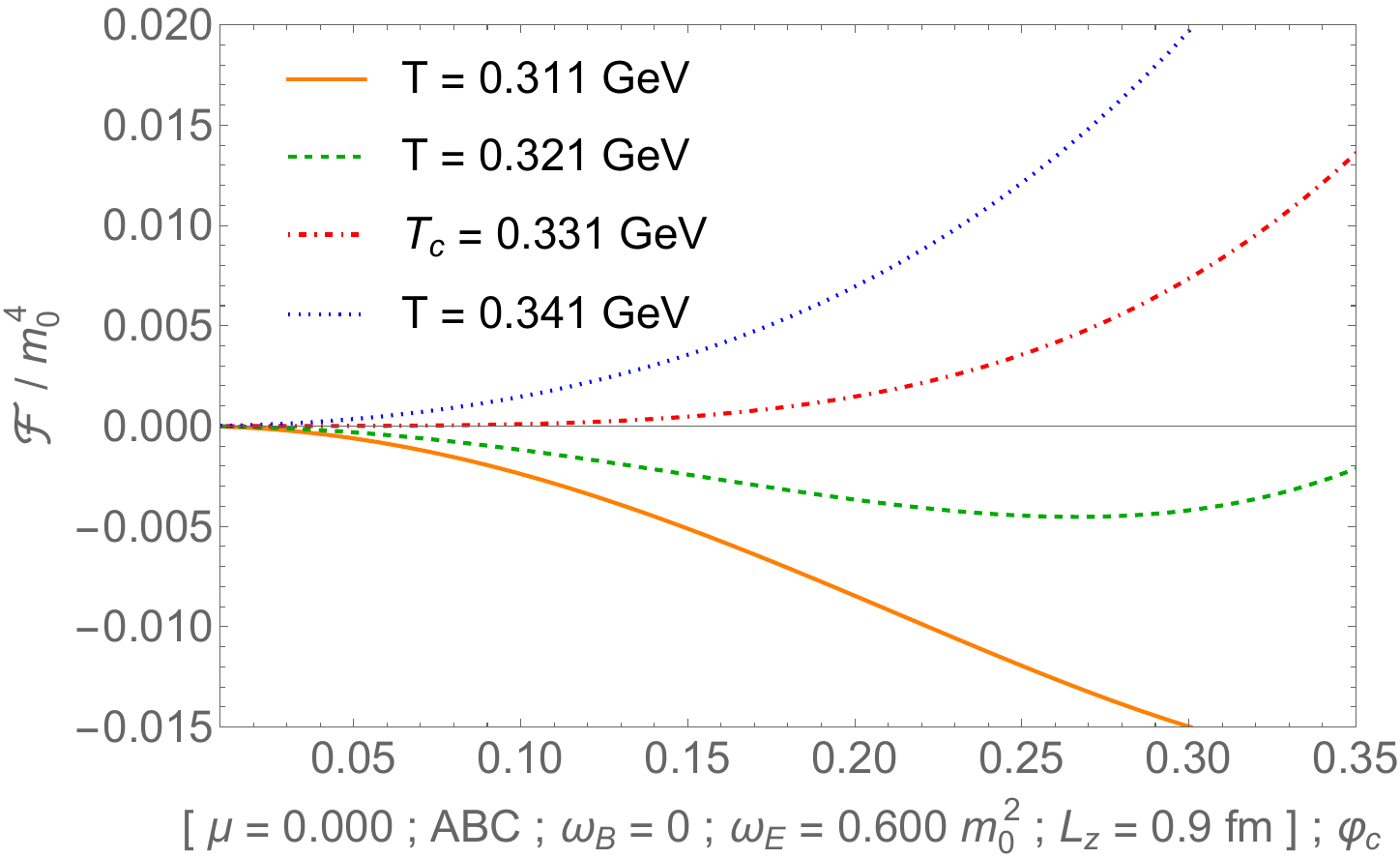} \\
\caption{~Influence of boundary conditions over the phase transition of the system. On the left, we fix $\omega_{E}=0.300~\mathrm{m_{0}^{2}}$. On the right, we have $\omega_{E}=0.600~\mathrm{m_{0}^{2}}$.}
\label{Fig11}
\end{figure}
\begin{figure}
\centering
\includegraphics[{width=6.49cm}]{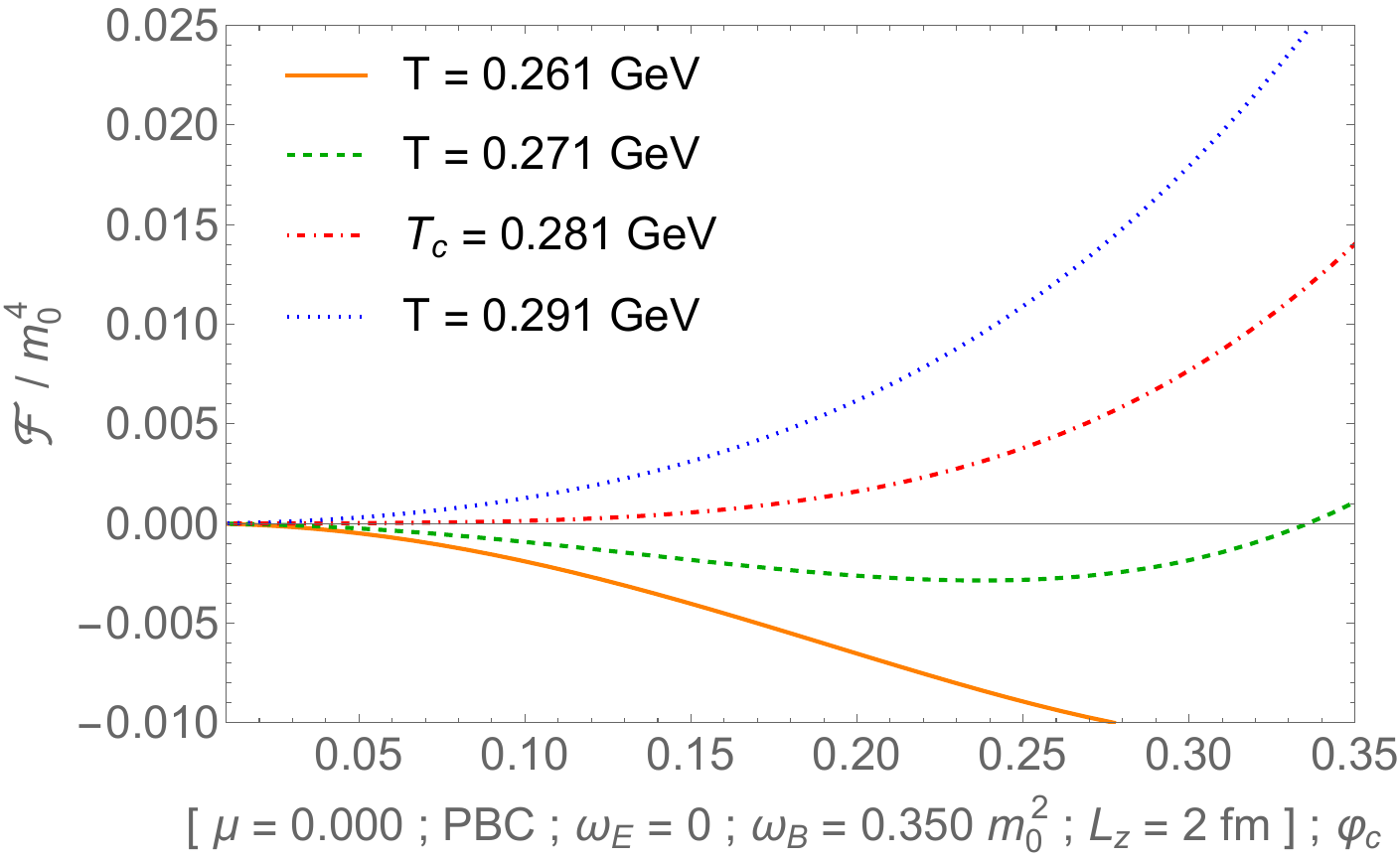}
\includegraphics[{width=6.49cm}]{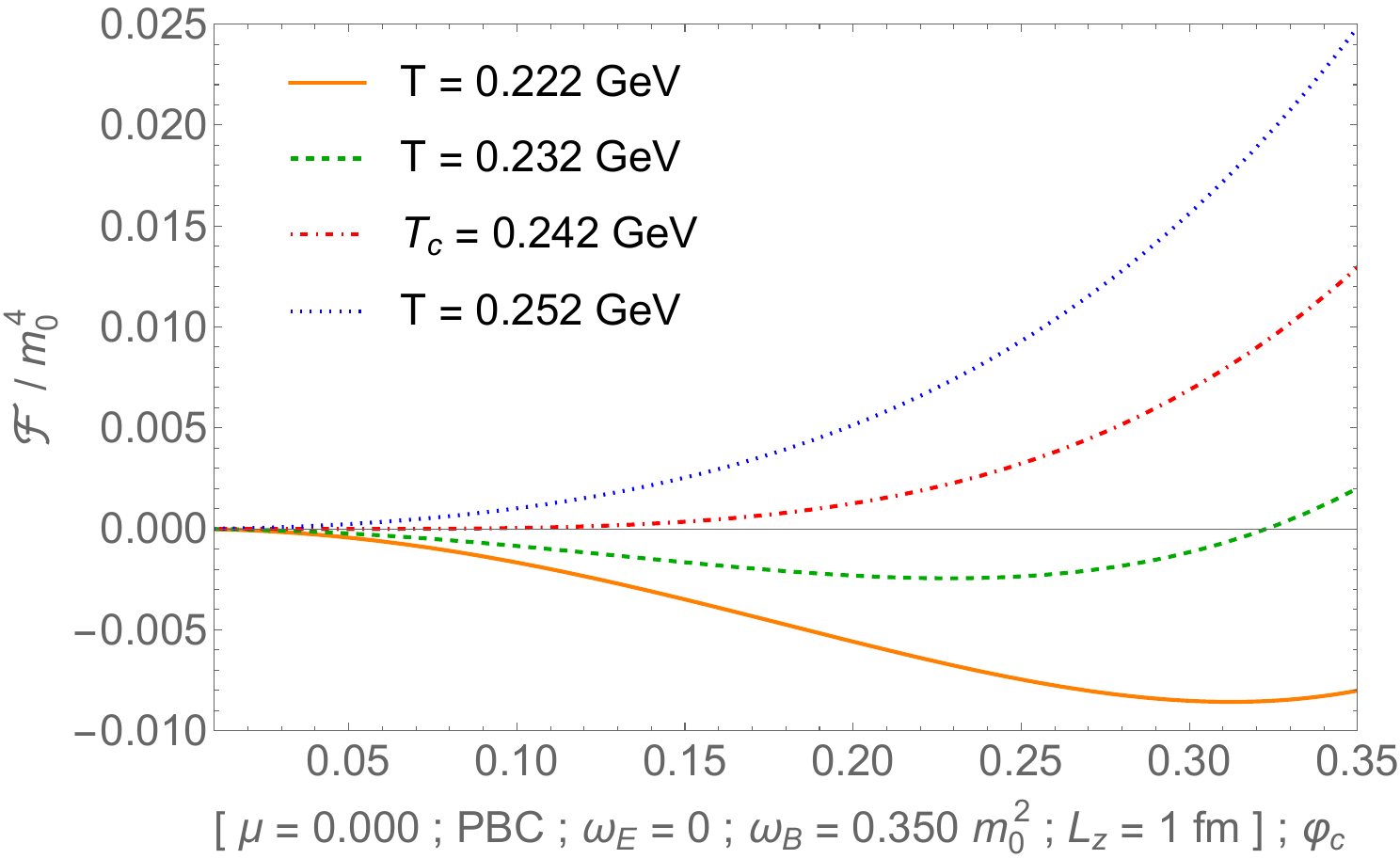} \\
\includegraphics[{width=6.49cm}]{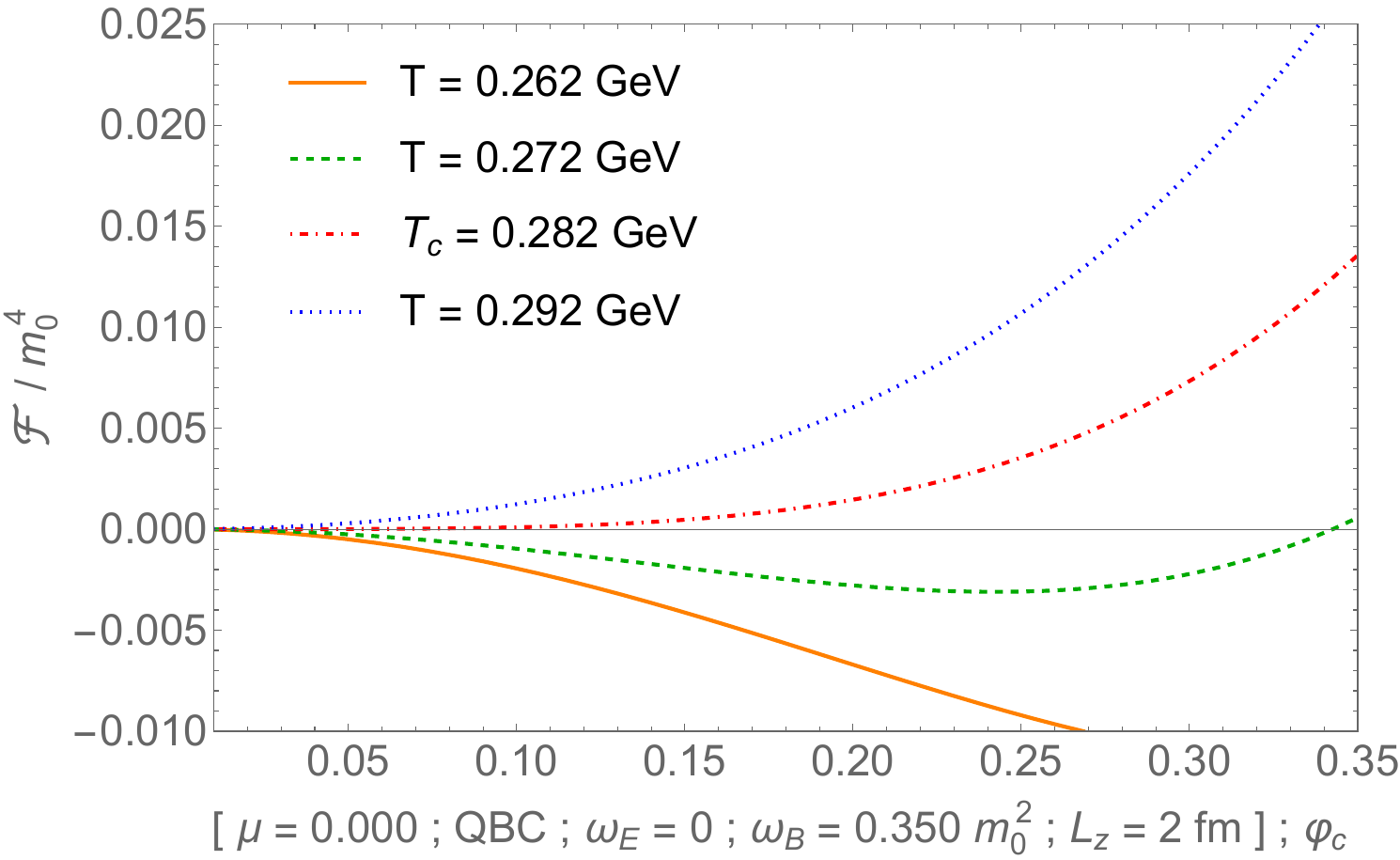}
\includegraphics[{width=6.49cm}]{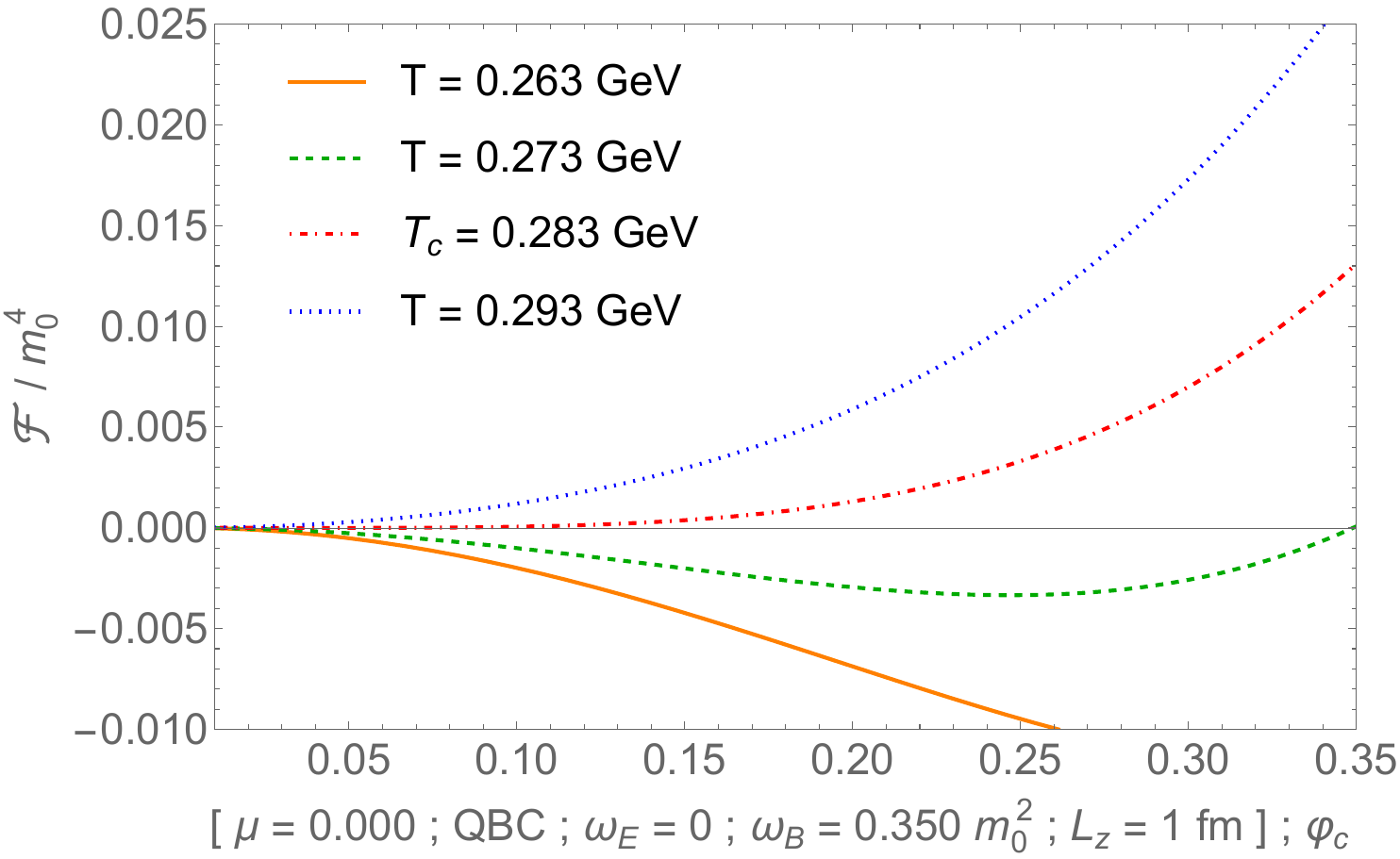} \\
\includegraphics[{width=6.49cm}]{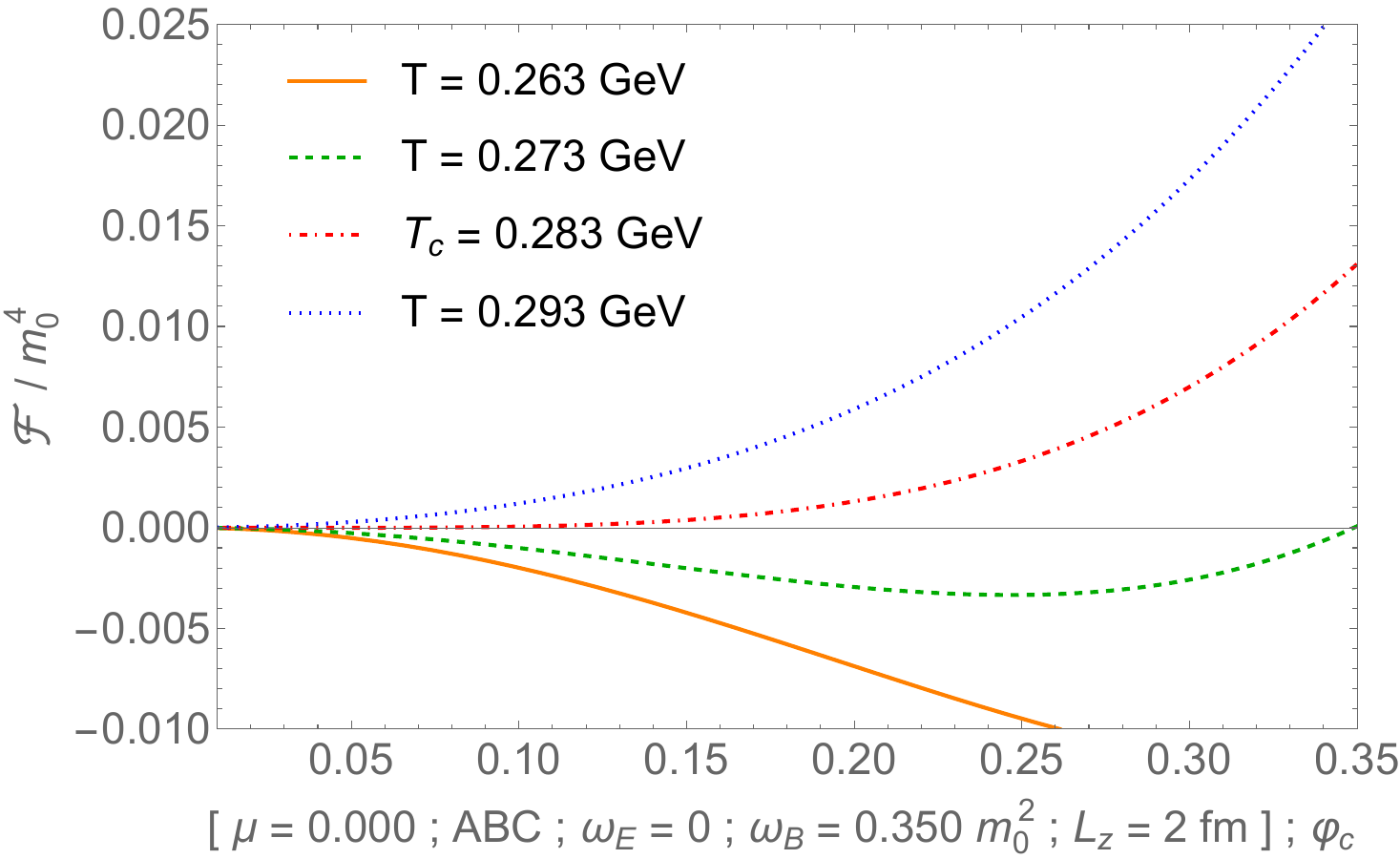}
\includegraphics[{width=6.49cm}]{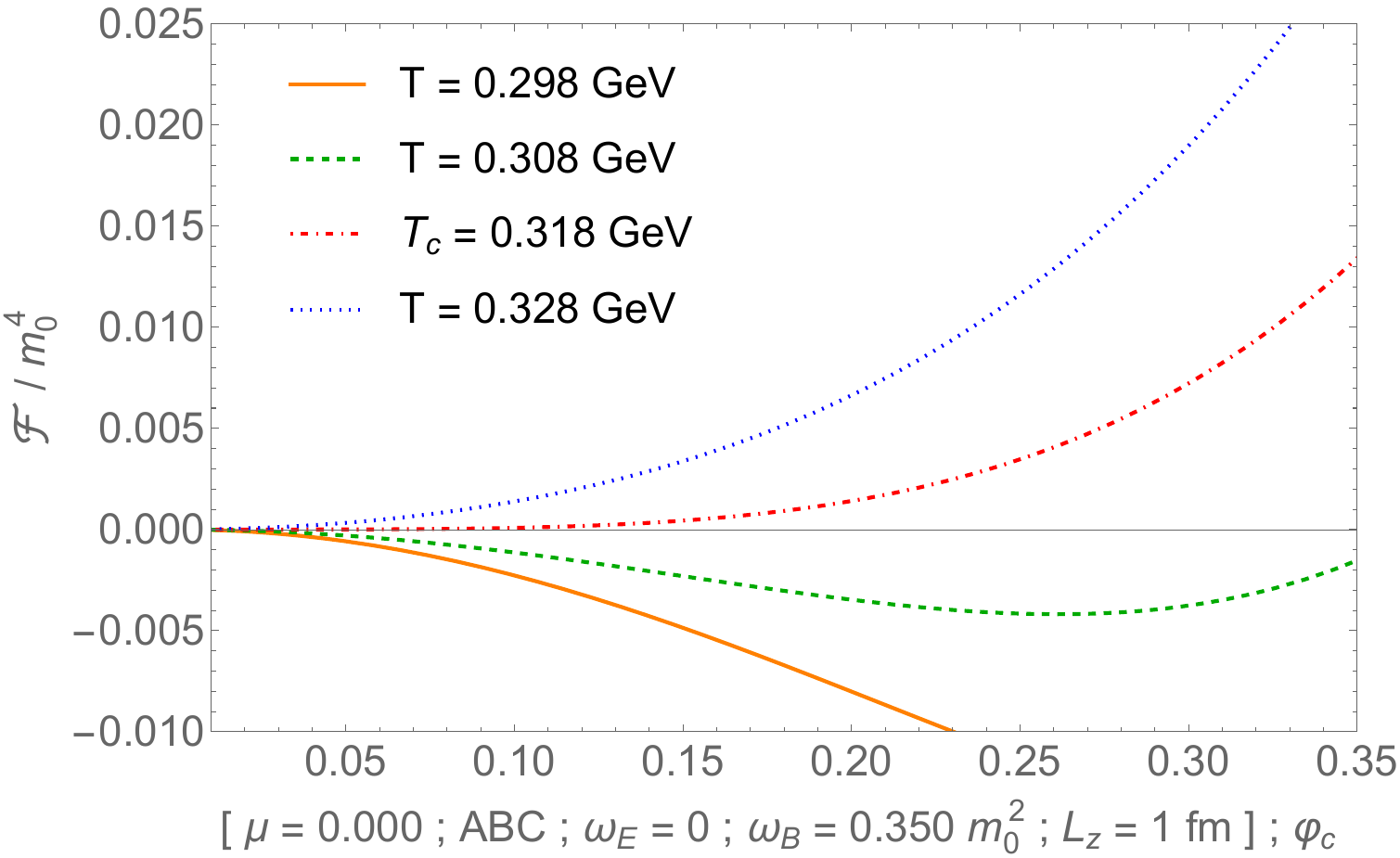} \\
\caption{~
Influence of boundary conditions and external magnetic field over the phase transition of the system. On the left, we fix $L_{z}=2~\mathrm{fm}$. On the right, we have $L_{z}=1~\mathrm{fm}$.}
\label{Fig12}
\end{figure}
\begin{figure}
\centering
\includegraphics[{width=6.49cm}]{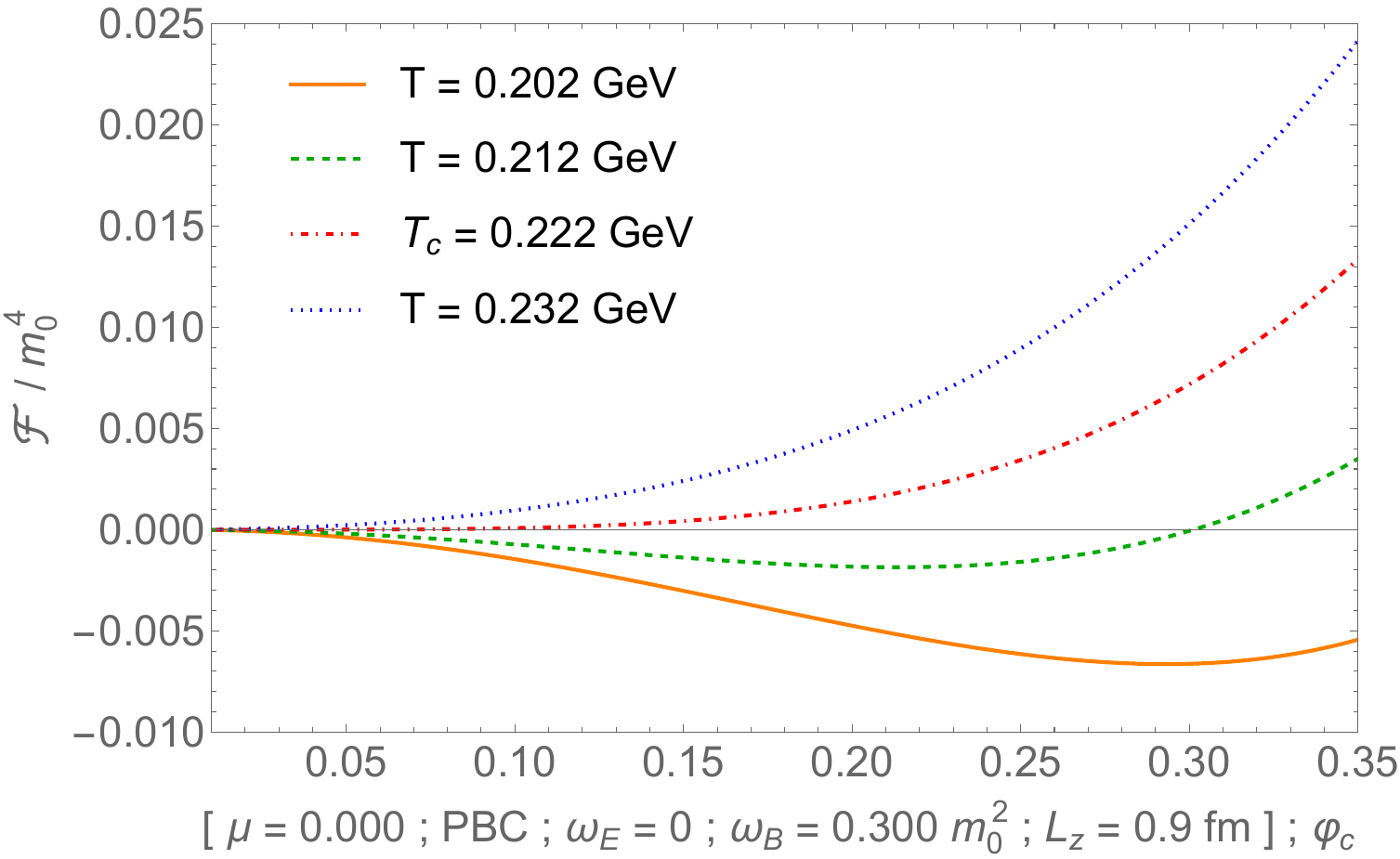}
\includegraphics[{width=6.49cm}]{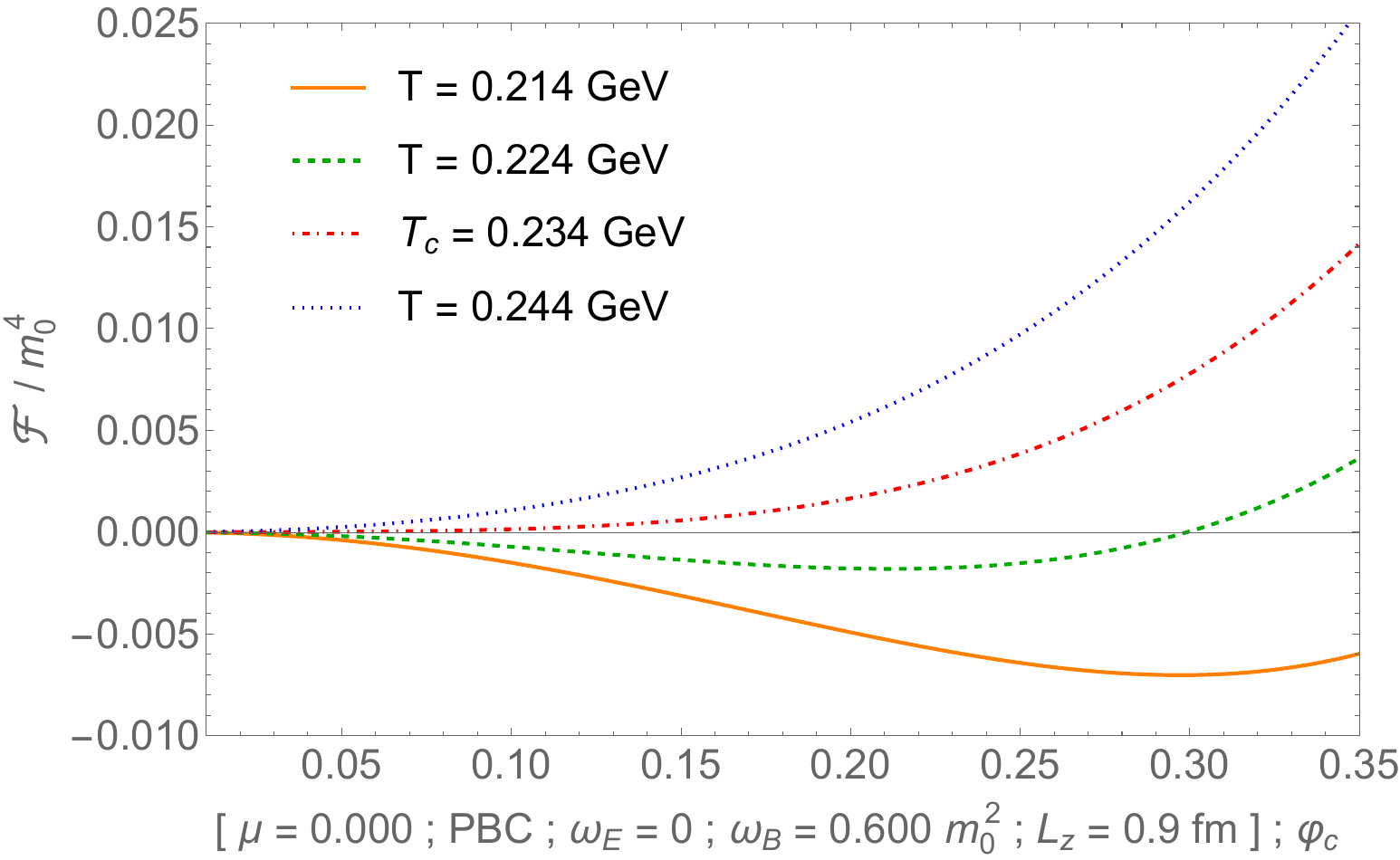} \\
\includegraphics[{width=6.49cm}]{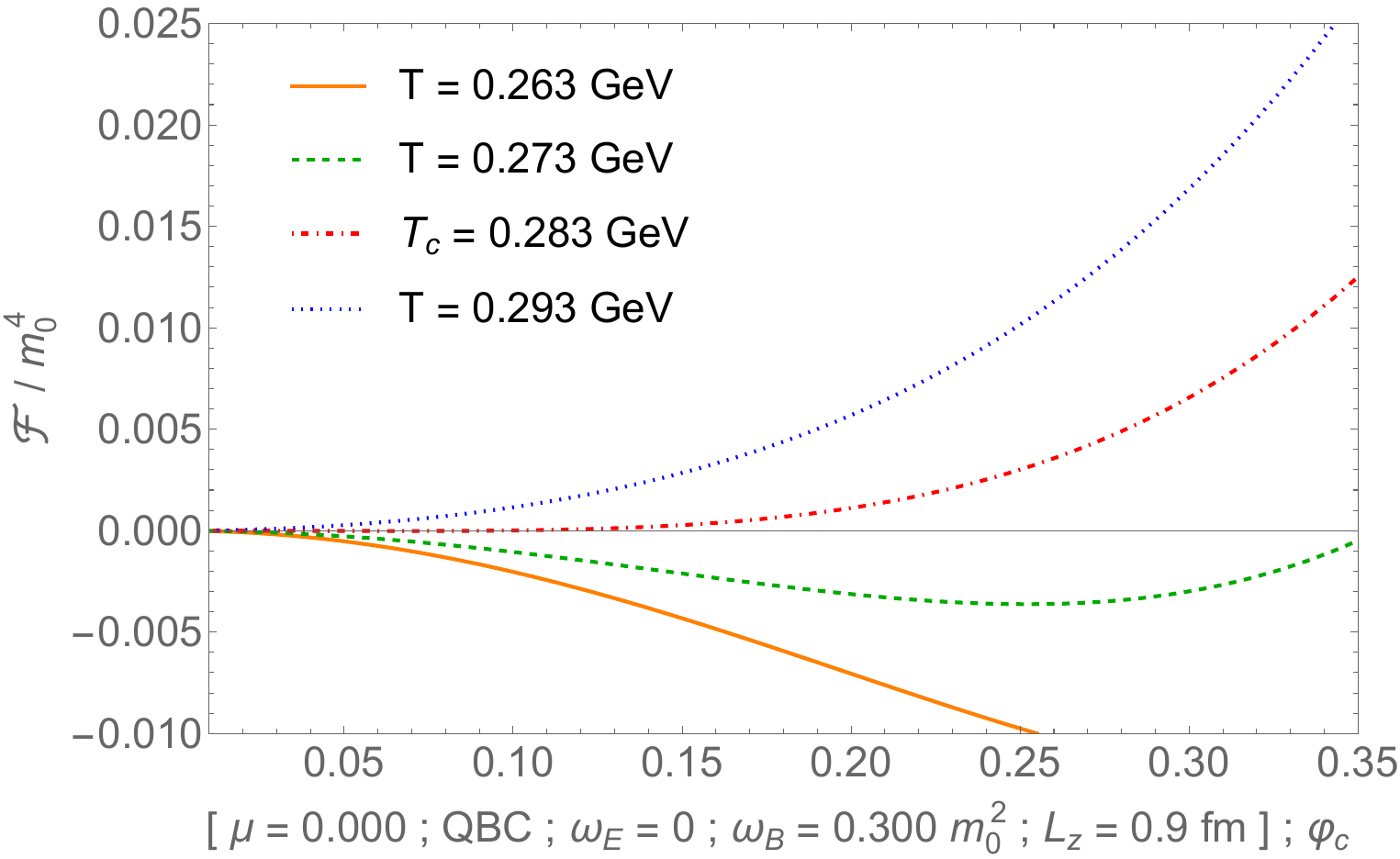}
\includegraphics[{width=6.49cm}]{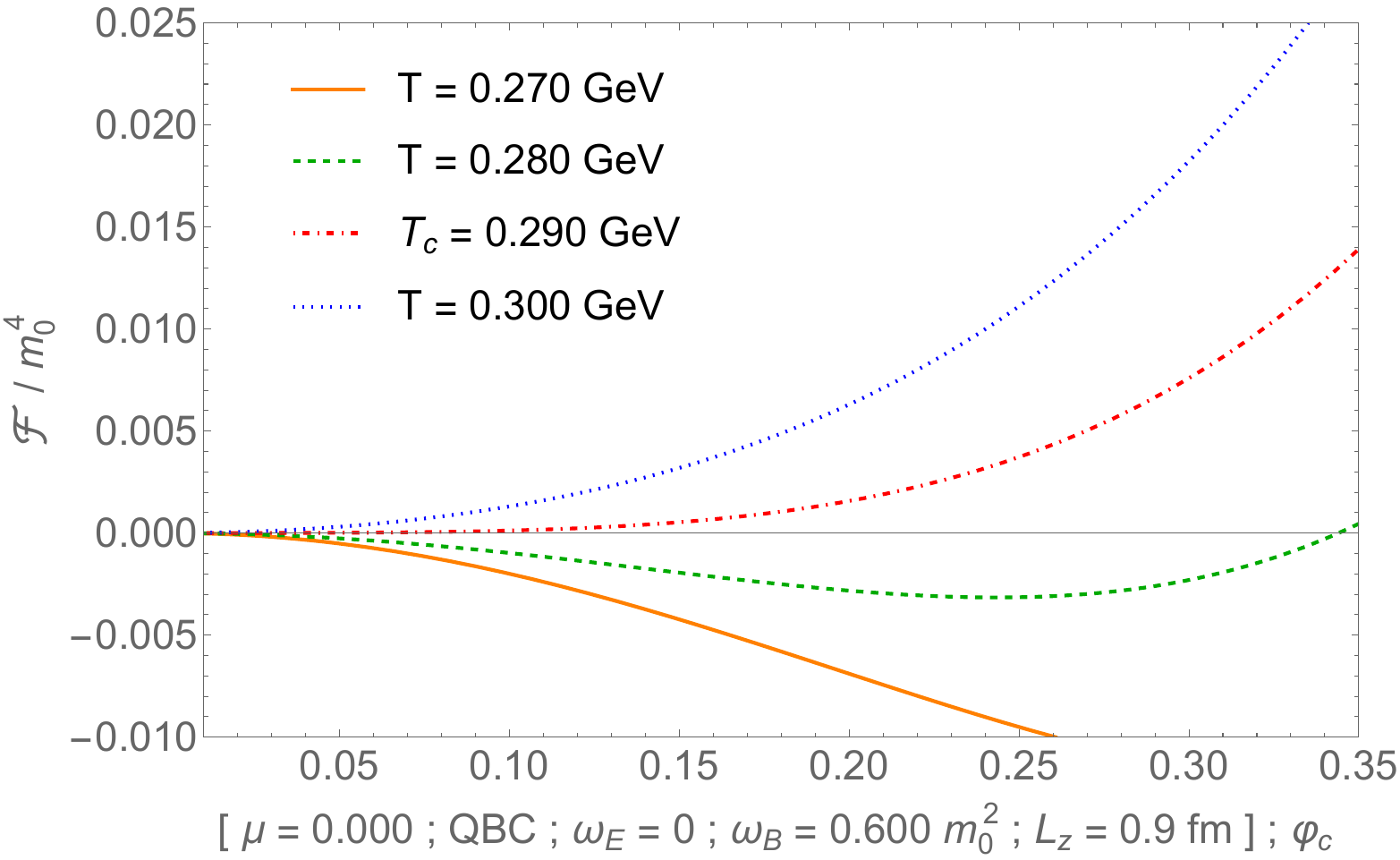} \\
\includegraphics[{width=6.49cm}]{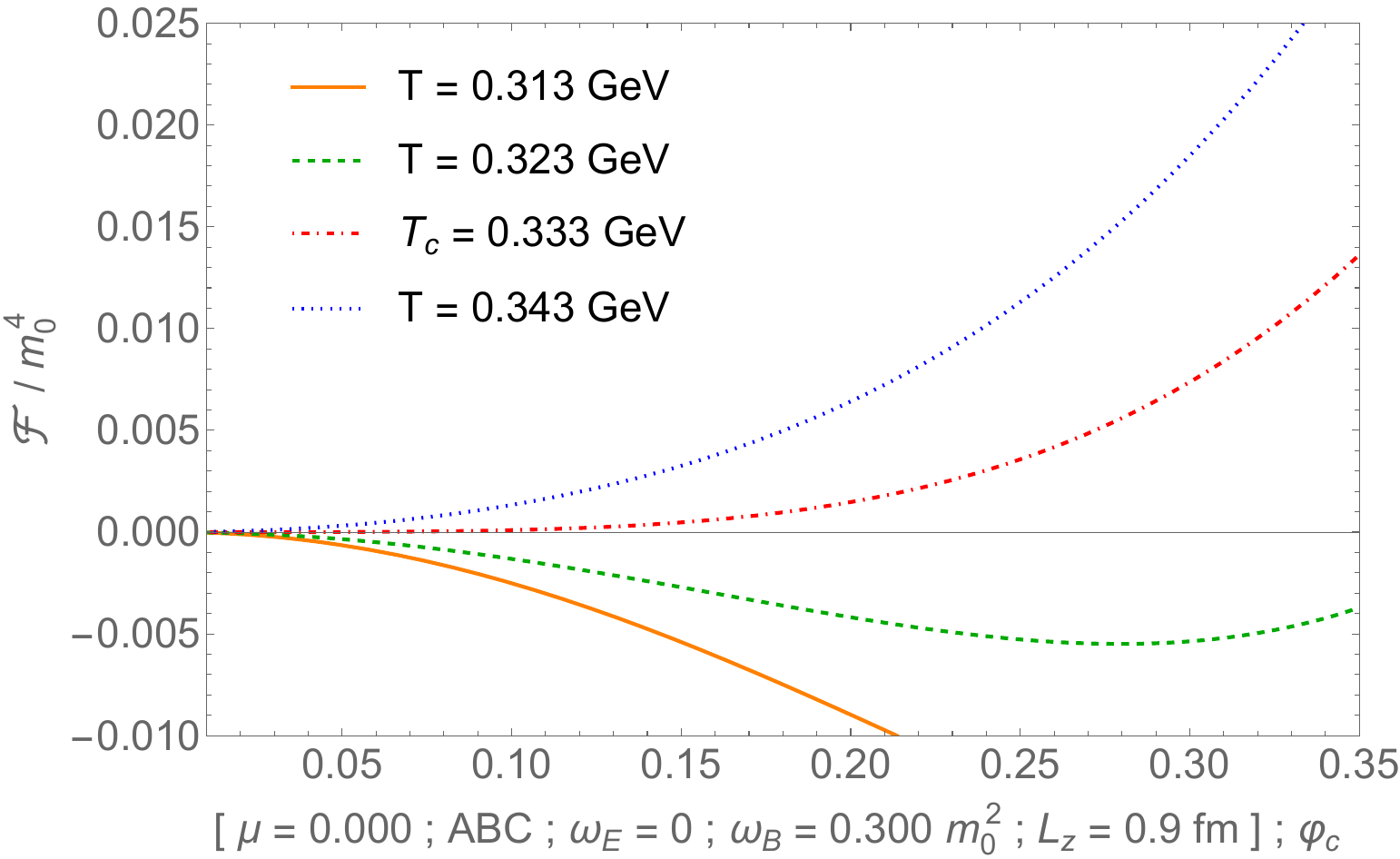}
\includegraphics[{width=6.49cm}]{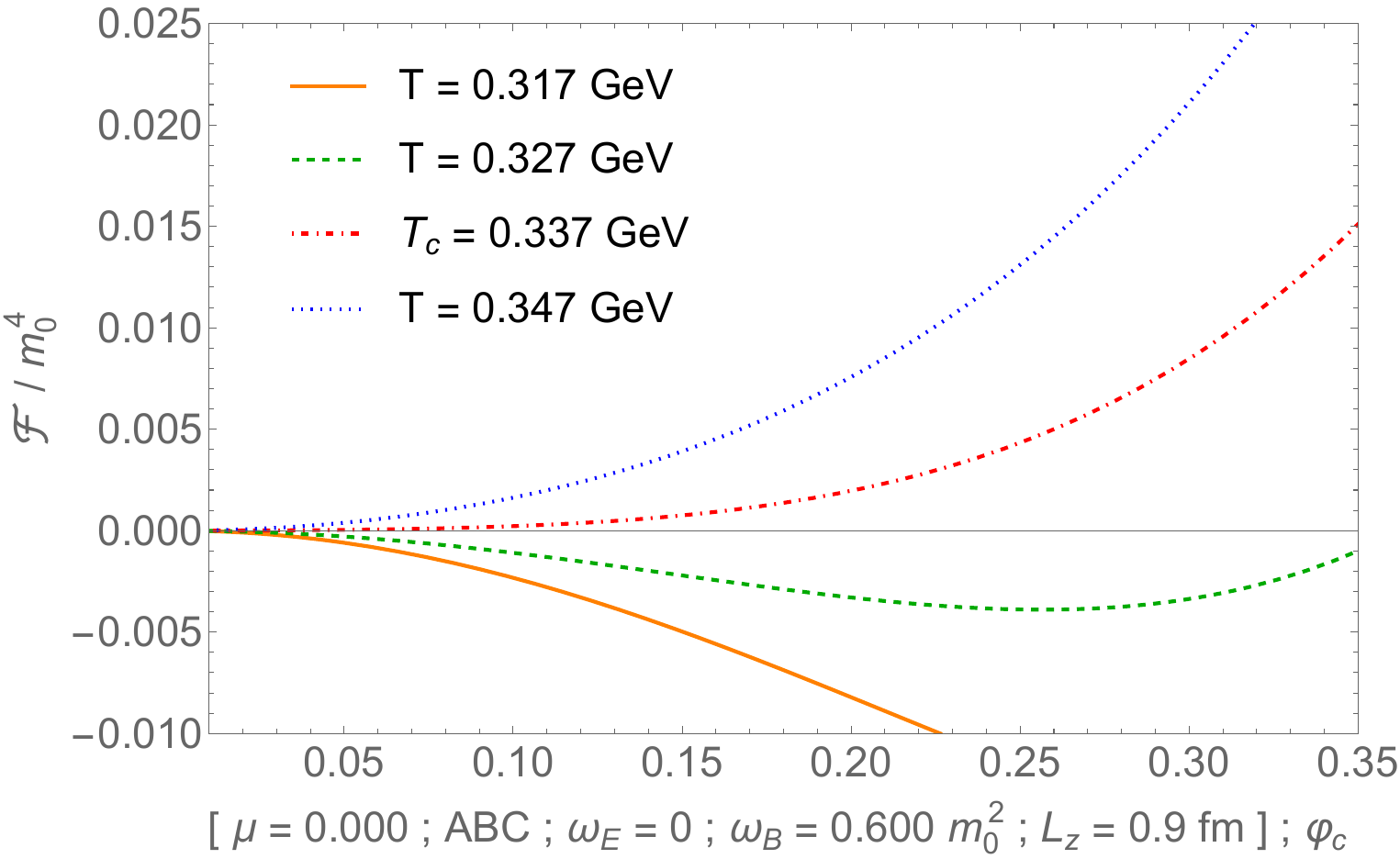} \\
\caption{~Influence of boundary conditions over the phase transition of the system. On the left, we fix $\omega_{B}=0.300~\mathrm{m_{0}^{2}}$. On the right, we have $\omega_{B}=0.600~\mathrm{m_{0}^{2}}$.}
\label{Fig13}
\end{figure}
\section{Summary and outlook}

\hspace{0.53cm}Along this paper, we have calculated radiatively corrections on the mass parameter due to both finite temperature and compactification of one spatial coordinate (in the $z$-direction) taking into account the existence of an electromagnetic background on the zero-spin bosonic system. Through a hybrid method that mixes Ritus' eigenfunctions and Schwinger's proper time, we were able to provide a closed expression
for the charged scalar propagator using all Landau levels. The propagator calculated in this note does not have any approximations. By using several kinds of boundary conditions at the planes that define the geometry of the system, we perform an analysis of the phase structure of the model by application of quantum field theory in a toroidal topology. 

From a practical point of view, we applied the Matsubara generalized formalism for a toy model with quartic-interaction of type $\lambda \Phi^{4}$ under electric and magnetic external fields. In virtue of ultraviolet divergence, we regularized the model by a technique that makes evident the vacuum contribution (where the divergence is located) and separates apart the thermal, spatial, and electromagnetic contributions. We have found the inverse electric catalysis for weaker values of the electric external field and the electric catalysis for stronger values of $\textbf{E}$ when the frontier of the system is under PBC conditions. The cases, QBC and ABC showed just inverse electric catalysis. However, the IEC phenomenon becomes negligible for ABC conditions at thicknesses smaller than $1~\mathrm{fm}$. Thus, the electric external field action on the system contributes to reinforcing the restored phase as follows: QBC and ABC cases, in all sectors of the electric field considered; PBC case, in the weak sector. Still in the PBC case, but for strong intensities of the electric field, we notice this external field tends to increase the broken phase region due to the augmentation of corrected mass. 

Also, we got the magnetic catalysis effect at fixed film thickness for all kinds of boundary conditions used in the paper. In this toy model, the magnetic external field always contributes to the rising broken phase regions. 

Even not using the fermionic theory, our results are according to, at least qualitatively, $\sigma$ meson mass behavior under a magnetic external field, see for instance ~\cite{PRDEmerson2022,MesonsMag}. The proximity with results that come from more appropriate theories (using quarks degrees of freedom) in the description of the effective mass $m_{\sigma}$ under changes in the thermodynamic variables, gives a reasonable status for the toy model treated here. 

We hope to continue investigating phase transitions under more general backgrounds, as in external fields with spatial dependencies. 

\section*{Acknowledgements}

This paper is dedicated to Emerson Filho and Maria Clara. 

\section*{Authorship contribution statement}
{\textbf{Emerson B. S. Corr\^ea:}} Design, implementation of the research, analysis of the results, writing - original draft.
\vspace{0.5cm}

\hspace{-0.50cm}{\textbf{Michelli S. R. Sarges:}} Design, implementation of the research, analysis of the results, writing - original draft.

\section*{Data Availability Statement}
No data associated in the manuscript.



%

\end{document}